\documentclass[12pt]{article}

\usepackage{geometry} 

\usepackage{geometry}
\usepackage{amsthm}
\usepackage{amssymb}
\usepackage{amsmath}
\usepackage{bm}
\usepackage{amsbsy}
\usepackage{bigints} 
\usepackage{relsize}
\usepackage{mathrsfs}
\usepackage{mathtools}
\usepackage{accents}
\usepackage{enumitem}
\usepackage{amsfonts}
\allowdisplaybreaks

\def\sigsqeps{\sigma^2_{\epsilon}}
\def\aeps{a_{\epsilon}}
\def\Asqeps{A_{\epsilon}^2}
\def\Sigmanu{\bSigma_{\nu}}
\def\munu{\bmu_{\nu}}
\def\sigsqmu{\sigma^2_{\mu}}
\def\amu{a_{\mu}}
\def\Asqmu{A_{\mu}^2}
\def\const{\text{const.}}

\def\mumu{\bmu_\mu}
\def\betamu{\bbeta_\mu}
\def\umu{\bu_\mu}
\def\numu{\bnu_\mu}
\def\Vpsi{\bV_\psi}

\newcommand\betapsi[1]{\bbeta_{\psi_{#1}}}
\newcommand\upsi[1]{\bu_{\psi_{#1}}}
\newcommand\nupsi[1]{\bnu_{\psi_{#1}}}
\newcommand\sigsqpsi[1]{\sigma^2_{\psi_{#1}}}
\newcommand\apsi[1]{a_{\psi_{#1}}}
\newcommand\Asqpsi[1]{A_{\psi_{#1}}^2}
\newcommand\hmu[1]{h_{\mu, i}}
\newcommand\hmupsi[1]{\bh_{\mu \psi, i}}
\newcommand\Hpsi[1]{\bH_{\psi, i}}
\newcommand\mupsi[1]{\bmu_{\psi_#1}}
\newcommand\tni[1]{\text{terms not involving $#1$}}

\usepackage{setspace}
\usepackage{epigraph}

\usepackage[T1]{fontenc}
\usepackage[utf8]{inputenc}
\usepackage{authblk}

\usepackage[english]{babel}
\makeatletter
\def\thmhead@plain#1#2#3{%
  \thmname{#1}\thmnumber{\@ifnotempty{#1}{ }\@upn{#2}}%
  \thmnote{ {\the\thm@notefont#3}}}
\let\thmhead\thmhead@plain
\makeatother

\theoremstyle{plain}
\newtheorem{theorem}{Theorem}[section]
\newtheorem{lemma}[theorem]{Lemma}
\newtheorem{proposition}[theorem]{Proposition}

\newtheorem{assumption}{Assumption}

\theoremstyle{definition}

\theoremstyle{remark}
\newtheorem*{remark}{Remark}

\usepackage[font={small}]{caption}
\usepackage{graphicx}
\usepackage{subcaption}
\usepackage{cleveref}
\usepackage[labelformat=simple]{subcaption}

\usepackage{tikz}
\usetikzlibrary{arrows,positioning,calc,fit}
\usetikzlibrary{shapes.geometric}
\usetikzlibrary{automata}
\usepackage{tkz-euclide}
\usepackage[bottom]{footmisc} 
\usepackage{booktabs}
\usepackage{siunitx}
\usepackage{tabularx}
\usepackage{colortbl}
\usepackage{rotating}
\usepackage{tcolorbox}
\usepackage{pgffor}

\usepackage{xcolor}

\usepackage{cite}
\usepackage{pslatex,apacite} 
\usepackage{etoolbox}
\pretocmd{\chapter}{}{}{}

\usepackage{import}
\usepackage{amsmath,amssymb,amsfonts}
\usepackage{bm}
\usepackage{algorithm}
\usepackage{algorithmicx}
\usepackage[noend]{algpseudocode}
\usepackage{varwidth}
\usepackage{xspace}
\makeatletter
\def\BState{\State\hskip-\ALG@thistlm}
\makeatother
\def\bzero{\boldsymbol{0}}
\def\bone{\boldsymbol{1}}
\usepackage{amsmath,amssymb,amsfonts}
\def\ba{\boldsymbol{a}}

\def\bc{\boldsymbol{c}}

\def\bh{\boldsymbol{h}}

\def\bm{\boldsymbol{m}}

\def\bq{\boldsymbol{q}}

\def\bt{\boldsymbol{t}}
\def\bu{\boldsymbol{u}}

\def\bx{\boldsymbol{x}}
\def\by{\boldsymbol{y}}

\def\bA{\boldsymbol{A}}

\def\bC{\boldsymbol{C}}
\def\bD{\boldsymbol{D}}

\def\bH{\boldsymbol{H}}
\def\bI{\boldsymbol{I}}

\def\bQ{\boldsymbol{Q}}

\def\bT{\boldsymbol{T}}
\def\bU{\boldsymbol{U}}
\def\bV{\boldsymbol{V}}

\def\bX{\boldsymbol{X}}
\def\bY{\boldsymbol{Y}}

\def\thick#1{\hbox{\rlap{$#1$}\kern0.25pt\rlap{$#1$}\kern0.25pt$#1$}}

\def\bbeta{\boldsymbol{\beta}}

\def\bepsilon{\boldsymbol{\epsilon}}

\def\bzeta{\boldsymbol{\zeta}}
\def\bdeta{\boldsymbol{\eta}}
\def\btheta{\boldsymbol{\theta}}
\def\biota{\boldsymbol{\iota}}

\def\bmu{\boldsymbol{\mu}}
\def\bnu{\boldsymbol{\nu}}

\def\bpsi{\boldsymbol{\psi}}

\def\bTheta{\boldsymbol{\Theta}}

\def\bLambda{\boldsymbol{\Lambda}}

\def\bXi{\boldsymbol{\Xi}}

\def\bSigma{\boldsymbol{\Sigma}}

\def\bPsi{\boldsymbol{\Psi}}

\def\smbalpha{{\thick{\scriptstyle{\alpha}}}}

\def\fhat{{\widehat f}}

\def\yhat{{\widehat y}}

\def\zetahat{{\widehat\zeta}}

\def\muhat{{\widehat\mu}}

\def\psihat{{\widehat\psi}}

\def\zetatilde{{\widetilde\zeta}}

\def\psitilde{{\widetilde\psi}}

\def\bzetahat{{\widehat\bzeta}}

\def\bPsihat{{\widehat\bPsi}}
\def\smbalpha{\widehat{\smbalpha}}

\def\bzetatilde{{\widetilde\bzeta}}

\def\bpsitilde{{\widetilde\bpsi}}

\def\bXitilde{{\widetilde\bXi}}

\def\bPsitilde{{\widetilde\bPsi}}

\def\hbar{{\overline h}}

\def\Qsc{{\mathcal Q}}

\def\R{{\mathbb R}}

\def\normal{\text{N}}
\def\hc{\text{Half-Cauchy}}

\def\igw{\text{Inverse G-Wishart}}

\def\lpyq{\text{log} \ \underline{p}(\boldsymbol{y}; q)}

\def\invchisq{\text{Inverse}-\chi^2}

\DeclareMathOperator*{\argmin}{argmin}
\DeclareMathOperator*{\argmax}{argmax}
\DeclareMathOperator*{\blockdiag}{blockdiag}

\def\Cov{\mathbb{C}\text{ov}}
\def\indsim{\stackrel{\text{ind.}}{\sim}}
\DeclareMathOperator*{\diag}{diag}

\DeclareMathOperator*{\neighbors}{neighbors}

\DeclareMathOperator{\E}{\mathbb{E}}

\DeclareMathOperator{\Var}{\mathbb{V}ar}
\DeclareMathOperator{\ind}{\mathbb{I}}
\DeclareMathOperator*{\tr}{tr}
\DeclareMathOperator{\vect}{vec}
\DeclareMathOperator{\vech}{vech}

\newcommand{\DKL}{D_{\text{KL}}}
\newcommand{\T}[1]{#1^{\intercal}}

\newcommand{\msg}[2]{m_{#1 \rightarrow #2}}

\newcommand{\np}[2]{\boldsymbol\eta_{#1 \rightarrow #2}}
\newcommand{\npbf}[2]{\boldsymbol\eta_{#1 \leftrightarrow #2}}
\newcommand{\npq}[1]{\boldsymbol\eta_{q^*(#1)}}

\algnewcommand\algorithmicinput{\textbf{Data Inputs:}}
\algnewcommand\DataInputs{\item[\algorithmicinput]}
\algnewcommand\HyperparameterInputs{\item[{\textbf{Hyperparameter Inputs:}}]}
\algnewcommand\ParameterInputs{\item[{\textbf{Parameter Inputs:}}]}
\algnewcommand\ParameterOutputs{\item[{\textbf{Parameter Outputs:}}]}
\algnewcommand\Inputs{\item[{\textbf{Inputs:}}]}
\algnewcommand\Outputs{\item[{\textbf{Outputs:}}]}
\algnewcommand\Updates{\item[{\textbf{Updates:}}]}
\algnewcommand\Initialise{\item[{\textbf{Initialise:}}]}
\algnewcommand\Initialize{\item[{\textbf{Initialize:}}]}
\title{Bayesian Functional Principal Components Analysis via Variational Message Passing}
\author[1,2]{Tui H. Nolan \thanks{Corresponding author: tui.nolan@uts.edu.au}}
\author[3]{Jeff Goldsmith}
\author[1,4]{David Ruppert}
\affil[1]{School of Operations Research and Information Engineering, Cornell University}
\affil[2]{School of Mathematical and Physical Sciences, University of Technology Sydney}
\affil[3]{Department of Biostatistics, Mailman School of Public Health, Columbia University}
\affil[4]{Department of Statistics and Data Science, Cornell University}

\begin{document}

\maketitle

\section*{\centering Abstract}

Functional principal components analysis is a popular tool for inference on functional data. Standard approaches
rely on an eigendecomposition of a smoothed covariance surface in order to extract the orthonormal functions
representing the major modes of variation. This approach can be a computationally intensive procedure, especially
in the presence of large datasets with irregular observations. In this article, we develop a Bayesian approach,
which aims to determine the Karhunen-Lo\`{e}ve decomposition directly without the need to smooth and estimate a
covariance surface. More specifically, we develop a variational Bayesian algorithm via message passing over a
factor graph, which is more commonly referred to as variational message passing.
Message passing algorithms are a powerful tool for compartmentalizing the algebra and coding required
for inference in hierarchical statistical models. Recently, there has been much focus on formulating variational
inference algorithms in the message passing framework because it removes the need for rederiving approximate
posterior density functions if there is a change to the model. Instead, model changes are handled by changing
specific computational units, known as fragments, within the factor graph. We extend the notion of variational message
passing to functional principal components analysis. Indeed, this is the first article to address a functional data model
via variational message passing. Our approach introduces two new fragments that are necessary for Bayesian
functional principal components analysis. We present the computational details, a set of simulations for assessing
accuracy and speed and an application to United States temperature data.


\section{Introduction}
\label{sec:intro}

Functional principal components analysis (FPCA) is the methodological extension of classical principal
components analysis (PCA) to functional data. Within the overarching framework of functional data analysis,
FPCA is a central technique. The advantages of using FPCA for functional data are derived
from analogous advantages that PCA affords for multivariate data analysis. For instance, PCA in the multivariate
data setting is used to reduce dimensionality and identify the major modes of variation of the
data set. The modes of variation are determined by the eigenvectors of the sample covariance matrix of the data
set, while dimension reduction is achieved by identifying the eigenvectors that maximize variation in the data.
In the functional setting, response curves are interpreted as independent realisations of an underlying
stochastic process. A covariance operator and its eigenfunctions play the analogous
role that the covariance matrix and its eigenvectors play in the multivariate data setting. By identifying the
eigenfunctions with the largest eigenvalues, one can reduce the
dimensionality of the entire data set by approximating each curve as a linear combination of the reduced set
of eigenfunctions.

There are technical issues that arise in the functional setting that are not present for multivariate data.
The domain of the functional curves is typically a compact interval $[0, T]$ of the real line.
Despite having a continuous domain, the curves are only observed at discrete points over this interval.
Furthermore, the points of observation, as well as the total number of observations, need not be the
same for each curve. Therefore, approaches that are used in PCA require
modifications to extend to the functional framework.
In FPCA, we often rely on nonparametric regression to smooth the eigenfunctions and employ
an appropriate step to ensure that they are orthonormal from the perspective of integration, rather
than inner products of vectors.

There have been numerous developments in FPCA methodology throughout the statistical literature.
A thorough introduction to the statistical framework and applications can be found in \citeA[Chapter~8]{ramsay05}
and \citeA[Section~2]{wang16}. Much of this work mirrors the eigendecomposition approach to PCA, in that an
eigenbasis is obtained from a covariance surface. \citeA{yao05} focused on the case of sparsely observed functional
data, and estimate
principal component scores through conditional expectations. \citeA{xiao16} developed a fast covariance estimation method for densely observed functional data. \citeA{di09} extended FPCA to multilevel functional data, extracting within and
between subject sources of variability, and \citeA{greven2011} developed methods for longitudinal functional data. However,
\citeA{goldsmith13} noted that these approaches implicitly condition on an estimated eigenbasis to estimate scores, meaning that inference on
individual curve estimates can be inaccurate.

Meanwhile, other approaches have built on or are similar to the probabilistic PCA framework that was introduced by \citeA{tipping99} and \citeA{bishop99}. Rather than first obtaining eigenfunctions from a covariance and then estimating scores, all quantities are considered unknown and are estimated jointly.  \citeA{james2000} used an EM algorithm for estimation and inference in the context of sparsely observed curves. Variational Bayes for FPCA was introduced by \citeA{vanderlinde08} via a generative model with a factorized approximation of the full posterior density function. \citeA{Goldsmith15} introduced a fully Bayes framework for multilevel function-on-scalar
regression models, and also considered observed values that arise from exponential family distributions. 

In frequentist versions of FPCA, the covariance function is determined through bivariate smoothing of the raw
covariances. Eigenfunctions and eigenvalues are then determined from the smoothed covariance function.
The key advantage in the Bayesian approach is that the covariance function is not estimated, meaning that
complex bivariate smoothing is not required. Indeed, the eigenfunctions and eigenvalues are computed directly
as part of a Bayesian hierarchical model. Furthermore, it is unnecessary to compute or store large covariance matrices for dense functional data, and for sparse, irregular functional data -- where estimating the raw covariance is difficult or impossible -- direct estimation of eigenfunctions in a Bayesian model is straightforward. For these reasons, we pursue a Bayesian approach to FPCA.

Although there have been numerous contributions to Bayesian implementations of FPCA, we argue that there are
additional considerations that should be addressed. First, MCMC modeling of FPCA is a computationally expensive
procedure and, in some biostatistical applications \cite{Goldsmith15}, the computational time can reach several
hours. Second, current versions of variational Bayes for FPCA, despite being a much faster computational alternative,
are difficult to extend to more complex likelihood specifications, such as multilevel data models and binary response
outcomes.

\citeA{minka05} presents a unifying view of approximate Bayesian inference under a message passing framework
that relies on the notion of messages passed between nodes of a factor graph. Mean field variational Bayes (MFVB)
can be incorporated into this framework through an alternate scheme known as variational message passing (VMP)
\cite{winn05}. \citeA{wand17} introduced computational units, known as fragments, that compartmentalize
the algebraic derivations that are necessary for approximate Bayesian inference in VMP. The notion of fragments
within a factor graph is essential for efficient extensions of variational Bayes-based FPCA to arbitrarily large statistical
models.

In this article, we propose an FPCA extension of the VMP framework for variational Bayesian inference set out in
\citeA{wand17}. Our novel methodology includes the introduction of two fragments  that are necessary for
computing approximate posterior density functions under an MFVB scheme,
as well as a sequence of post-processing steps for estimating the orthonormal eigenfunctions.
We provide an introduction to
variational Bayesian inference in Section \ref{sec:vbi}, with an overview of VMP in Section \ref{sec:vmp}.
Section \ref{sec:fpca} introduces the Bayesian hierarchical model for FPCA and its extensions under a
VMP formulation. In Section \ref{sec:post_vmp_steps}, we outline the post-VMP steps that are required for
producing orthonormal eigenfunctions. Simulations, including speed and accuracy comparisons with MCMC
algorithms, are presented in Section \ref{sec:sims}, and an application to United States temperature data is
provided in Section \ref{sec:us_weather_data}.


\section{Variational Bayesian Inference}
\label{sec:vbi}

The overarching aim of this article is the identification and derivation of fragments that are necessary for VMP
implementations of FPCA. VMP represents a class of methodologies, derived from MFVB approaches,
for approximate Bayesian inference over a factor graph.
In this section, we provide a brief introduction to the MFVB and VMP frameworks. For an in-depth
explanation of MFVB, we refer the reader to \citeA{ormerod10} and \citeA{blei17}; for a comprehensive review
of VMP, we refer the reader to \citeA{minka05} and \citeA{wand17}.

Variational Bayesian inference is based on the notion of minimal Kullback-Leibler divergence to approximate a
posterior density function. For arbitrary density functions $p_1$
and $p_2$ on $\R^d$, the Kullback-Leibler divergence of $p_1$ from $p_2$ is

\[
	\DKL (p_1, p_2) \equiv \bigintsss_{\R^d} \log \left\{ \frac{p_1 (\bx)}{p_2 (\bx)} \right\} p_1 (\bx) d\bx.
\]

\noindent Note that

\begin{equation}
	\DKL (p_1, p_2) \ge 0.
\label{nonneg_dkl}
\end{equation}

Consider a generic Bayesian model with observed data vector $\by$ and parameter vector $\btheta \in \Theta$,
where $\bTheta$ is a parameter space. We make the
assumption that $\by$ and $\btheta$ are continuous random variables with density functions $p(\by)$ and $p(\btheta)$.
For the case where some components are discrete, a similar treatment applies with summations replacing integrals.
Next, let $q (\btheta)$ represent an arbitrary density function over the parameter space $\Theta$. The essence of
variational Bayesian inference is to restrict $q$ to a class of density functions $\Qsc$ and use the optimal $q$-density
function, defined by

\begin{equation}
	q^* (\btheta) \equiv \argmin_{q \in \Qsc} \ \DKL \{ q(\btheta), p (\btheta | \by) \},
\label{vb_dkl}
\end{equation}

\noindent as an approximation to the true posterior density function $p (\btheta | \by)$.

Simple algebraic arguments \cite<e.g.>[Section~2.1]{ormerod10} show that the marginal log-likelihood satisfies:

\[
	\log p (\by) = \DKL \{ q(\btheta), p (\btheta | \by) \} + \log \underline{p} (\by; q),
\]

\noindent where

\[
	\underline{p} (\by; q) \equiv
		\exp \left[
			\bigintsss \log \left\{
				\frac{p (\by, \btheta)}{q (\btheta)}
			\right\} q (\btheta) d\btheta
		\right].
\]

\noindent From the non-negativity condition of \eqref{nonneg_dkl}, we have

\[
	\underline{p} (\by; q) \le p (\by)
\]

\noindent showing that $\underline{p} (\by; q)$ is a lower-bound on the marginal likelihood. This leads to an
equivalent form for the optimisation problem in \eqref{vb_dkl}:

\begin{equation}
	q^* (\btheta) \equiv \argmax_{q \in \Qsc} \{ \log \underline{p} (\by; q) \}.
\label{vb_elbo}
\end{equation}

\noindent As stated in \citeA{rohde16}, this alternate expression has the advantage of representing the
optimal $q$-density function as maximising the lower-bound on the marginal log-likelihood. For the remainder
of this article, we will address variational Bayesian inference with \eqref{vb_elbo}, rather than \eqref{vb_dkl}.


\subsection{Mean Field Variational Bayes}
\label{sec:mfvb}

MFVB is a class of variational Bayesian inference methods that uses a product density (or mean field) restriction
in the optimal $q$-density function. The mean field approximation, which has its roots in statistical physics
\cite{parisi88}, imposes the factorization

\begin{equation}
	q (\btheta) = \prod_{i=1}^N q (\btheta_i),
\label{mf_rest}
\end{equation}

\noindent for all $q \in \Qsc$, where $\{ \btheta_1, \dots, \btheta_N \}$ is some partition of $\btheta$. The
optimal $q$-density functions that satisfy \eqref{vb_elbo} are given by \cite<e.g.>{minka05, ormerod10}

\begin{equation}
	q^*_i (\btheta_i) = \frac{
		\exp \{
			\E_{q (\btheta \setminus \btheta_i)} \log p (\by, \btheta)
		\}
	} {
		\int \exp \{
			\E_{q (\btheta \setminus \btheta_i)} \log p (\by, \btheta)
		\} d\btheta_i
	}, \quad \text{for $i = 1, \dots, N$,}
\label{opt_q}
\end{equation}

\noindent where $\E_{q (\btheta \setminus \btheta_i)}$ denotes expectation with respect to the optimal posterior density
functions of all elements in the partition of $\btheta$, defined by \eqref{mf_rest}, except for the optimal posterior
density function of $\btheta_i$.

The parameter vectors that define each of the optimal $q$-density
functions in \eqref{opt_q} are interrelated and are updated
by a coordinate ascent algorithm \cite[Algorithm~1]{ormerod10}. However, the resulting parameter vector updates
are problem-specific and must be rederived if there is a change to the model. For instance, the updates for the
optimal posterior density functions of the coefficients in a linear regression model will differ from those in a
linear logistic regression model.


\subsection{Variational Message Passing}
\label{sec:vmp}

VMP is an alternate computational framework for variational Bayesian inference with a mean field product restriction.
The VMP infrastructure is a factor graph representation of the Bayesian model. \citeA{wand17} advocates for
the use of fragments, a sub-graph of a factor graph, as a means of compartmentalizing the algebra and computer
coding required for variational Bayesian inference. Posterior density estimation is achieved by messages passed
within and between factor graph fragments. Here, we give a brief description of the foundations of VMP. For a
thorough exposition of the VMP framework, we refer the reader to \citeA{wand17}.

Consider a generic Bayesian model with observed data vector $\by$ and parameter vectors $\btheta_1, \dots,
\btheta_5$. Suppose that the joint density function factorizes according to

\begin{equation}
	p (\by, \btheta_1, \dots, \btheta_5) =
		p (\by | \btheta_1, \btheta_2, \btheta_3) p (\btheta_1) p (\btheta_2 | \btheta_4, \btheta_5)
		p (\btheta_3 | \btheta_4) p (\btheta_4) p (\btheta_5).
\label{arb_bayes_mod}
\end{equation}

\begin{figure}
	\centering
	\begin{tikzpicture} [
		auto, node distance=1.5cm, every loop/.style={},
		thick,stochastic node/.style={circle,draw,font=\sffamily\Large\bfseries},
		factor/.style={regular polygon,regular polygon sides=4}
	]
		
		\node[fill, factor, draw, scale=1.5, label=above:{$p \left( \by | \btheta_1, \btheta_2, \btheta_3 \right)$}] (py) {};
		\node[stochastic node, scale=1.5, label=above:$\btheta_1$] (theta1) [right of=py] {};
		\node[stochastic node, scale=1.5, draw = blue, label=right:$\btheta_2$] (theta2) [below of=py] {};
		\node[stochastic node, scale=1.5, label=above:$\btheta_3$] (theta3) [left of=py] {};
		
		\node[fill, factor, draw, scale=1.5, label=right:{$p \left( \btheta_1 \right)$}] (ptheta1) [below of=theta1] {};
		
		\node[
			fill = blue, factor, draw = blue, scale=1.5,
			label=above:{$p \left( \btheta_2 | \btheta_4, \btheta_5 \right)$}
		] (ptheta2) [left of=theta2] {};
		\node[stochastic node, scale=1.5, draw = blue, label=left:$\btheta_4$] (theta4) [left of=ptheta2] {};
		\node[stochastic node, scale=1.5, draw = blue, label=left:$\btheta_5$] (theta5) [below of=ptheta2] {};
		
		\node[
			fill, factor, draw, scale=1.5,
			label=above:{$p \left( \btheta_3 | \btheta_4 \right)$}
		] (ptheta3) [left of=theta3] {};
		
		\node[fill, factor, draw, scale=1.5, label=left:{$p \left( \btheta_4 \right)$}] (ptheta4) [below of=theta4] {};
		
		\node[fill, factor, draw, scale=1.5, label=right:{$p \left( \btheta_5 \right)$}] (ptheta5) [right of=theta5] {};
		
		\path[every node/.style={font=\sffamily\small}]
		
		(py)
			edge node {} (theta1)
			edge node {} (theta2)
			edge node {} (theta3)
		
		(ptheta1)
			edge node {} (theta1)
		
		(ptheta2)
			edge [blue] node {} (theta2)
			edge [blue] node {} (theta4)
			edge [blue] node {} (theta5)
		
		(ptheta3)
			edge node {} (theta3)
			edge node {} (theta4)
		
		(ptheta4)
			edge node {} (theta4)
		
		(ptheta5)
			edge node {} (theta5);
		
	\end{tikzpicture}
\caption{
	A factor graph representation of the Bayesian model described by \eqref{arb_bayes_mod}. As an example,
	the factor graph fragment for $p (\btheta_2 | \btheta_4, \btheta_5)$ is highlighted in blue.
}
\label{fig:arb_bayes_mod_fg}
\end{figure}
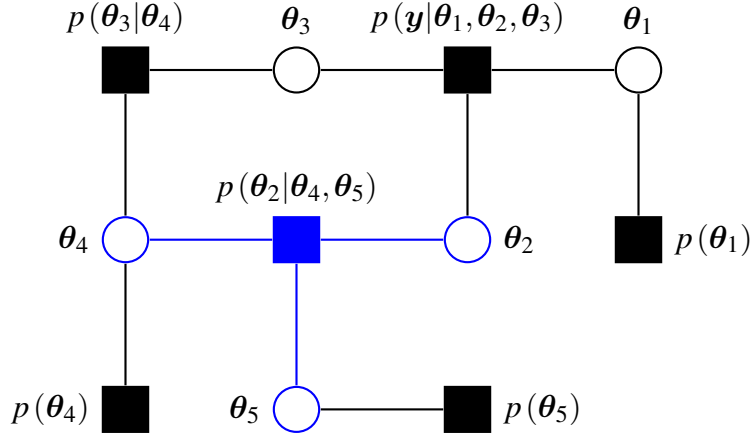

\noindent A factor graph representation of the Bayesian model expressed in \eqref{arb_bayes_mod} is presented
in Figure \ref{fig:arb_bayes_mod_fg}. The square nodes are the factors, which represent the distributional
specifications of the model, and the circular nodes are called stochastic nodes, which represent the parameter
vectors of the model. Furthermore, the graph is bipartite, meaning that stochastic nodes can only share an
edge with a factor and vice versa. Additionally, notice that the edges of the factor graph respect the distributional
dependencies of \eqref{arb_bayes_mod}. For instance, the factor for $p (\btheta_2 | \btheta_4, \btheta_5)$
shares edges with the stochastic nodes for $\btheta_2$, $\btheta_4$ and $\btheta_5$ only.

The VMP construction of variational Bayesian inference relies on messages passed between factors and stochastic
nodes. Consider the factor for $p (\btheta_2 | \btheta_4, \btheta_5)$ and the messages that it will pass to its
neighboring stochastic nodes $\btheta_2$, $\btheta_4$ and $\btheta_5$. The messages passed from this
factor are updated according to

\begin{equation}
	\msg{p (\btheta_2 | \btheta_4, \btheta_5)}{\btheta_i} (\btheta_i)
		\longleftarrow
			\frac{
				\E_{q(\btheta \setminus \btheta_i)} \log p (\btheta_2 | \btheta_4, \btheta_5)
			} {
				\int \E_{q(\btheta \setminus \btheta_i)} \log p (\btheta_2 | \btheta_4, \btheta_5) d\btheta_i
			},
	\quad \text{for $i = 2, 4, 5$.}
\label{msg_fact_st}
\end{equation}

\noindent Before continuing, we will make a brief note on our notation for computational updates. In \eqref{msg_fact_st},
we use a left arrow to indicate a computational update, rather than an equality symbol. This reflects the fact that at each
iteration of the VMP algorithm, the algebraic form of the message remains the same, whereas the parameter values of the
message will change between iterations as the Kullback-Leibler divergence is minimized. In addition, the subscript
of the message in \eqref{msg_fact_st} indicates the direction of the message. Finally, each message is a function of
the parameter vector that it is sent to or sent from.

Next, consider the stochastic node $\btheta_4$, which receives messages from the factors
$p (\btheta_2 | \btheta_4, \btheta_5)$, $p (\btheta_3 | \btheta_4)$ and $p (\btheta_4)$. The messages that $\btheta_4$
returns to these factors are

\begin{align}
\begin{split}
	\msg{\btheta_4}{p (\btheta_2 | \btheta_4, \btheta_5)} (\btheta_4)
		&\longleftarrow
			\msg{p (\btheta_3 | \btheta_4)}{\btheta_4} (\btheta_4) \
			\msg{p(\btheta_4)}{\btheta_4} (\btheta_4) \\
	\msg{\btheta_4}{p (\btheta_3 | \btheta_4)} (\btheta_4)
		&\longleftarrow
			\msg{p (\btheta_2 | \btheta_4, \btheta_5)}{\btheta_4} (\btheta_4) \
			\msg{p (\btheta_4)}{\btheta_4} (\btheta_4) \\
	\msg{\btheta_4}{p (\btheta_4)} (\btheta_4)
		&\longleftarrow
			\msg{p (\btheta_2 | \btheta_4, \btheta_5)}{\btheta_4} (\btheta_4) \
			\msg{p (\btheta_3 | \btheta_4)}{\btheta_4} (\btheta_4).
\end{split}
\label{examp_st_f_msg}
\end{align}

\noindent In general, the message that a stochastic node passes to a neighbouring factor is simply the product
of the messages that it has received from its other neighbouring factors. Note that the form of the message
updates in \eqref{msg_fact_st} are such that the range of each message is a subset of $\R$.
Therefore, the computations in \eqref{examp_st_f_msg} simply involve multiplication of scalars.

Now, consider a general Bayesian model with $M$ factors $p_1, \dots, p_M$ and $N$ parameter vectors
$\btheta_1, \dots, \btheta_N$. We define the set of stochastic nodes that are connected to the factor
$p_j$ as

\[
	\neighbors (j) \equiv \{ i = 1, \dots, N: \text{$\btheta_i$ shares an edge with $p_j$} \}.
\]

\noindent Then, the message from factor
$p_j$ to the stochastic node $\btheta_i$ is

\begin{equation}
	\msg{p_j}{\btheta_i} (\btheta_i)
		\longleftarrow
			\frac{
				\E_{q(\btheta \setminus \btheta_i)} \log p_j
			} {
				\int \E_{q(\btheta \setminus \btheta_i)} \log p_j d\btheta_i
			}.
\label{msg_fact_sn}
\end{equation}

\noindent The message from $\btheta_i$ to $p_j$ is

\begin{equation}
	\msg{\btheta_i}{p_j} (\btheta_i)
		\longleftarrow
			\prod_{j' \neq j : i \in \neighbors(j')} \msg{p_{j'}}{\btheta_i} (\btheta_i).
\label{msg_sn_fact}
\end{equation}

\noindent Upon convergence of the messages, the optimal $q$-density functions, which satisfy \eqref{vb_elbo}
take the form

\begin{equation}
	q^* (\btheta_i) \propto \prod_{j : i \in \neighbors(j)} \msg{p_j}{\btheta_i} (\btheta_i), \quad i = 1, \dots, N.
\label{opt_q_vmp}
\end{equation}

\noindent The VMP iterative loop has the following generic steps \cite{minka05, wand17}:

\begin{enumerate}
	\item Choose a factor.
	\item Update the messages passed from the factor's neighbouring stochastic nodes to the factor.
	\item Update the messages passed from the factor to its neighbouring stochastic nodes.
\end{enumerate}


\subsubsection{Exponential Family Form}
\label{sec:exp_fam_form}

A key step in deriving and implementing VMP algorithms is the representation of probability density functions
in exponential family form. In particular, the messages in \eqref{msg_fact_sn} and \eqref{msg_sn_fact}
are typically in the exponential family of density functions with vector of sufficient statistics $\bT (\btheta_i)$. We have

\[
	\msg{p_j}{\btheta_i} (\btheta_i)
		\propto
			\exp \{ \T{\bT (\btheta_i)} \np{p_j}{\btheta_i} \}, \quad
	\text{and} \quad
	\msg{\btheta_i}{p_j} (\btheta_i)
		\propto
			\exp \{ \T{\bT (\btheta_i)} \np{\btheta_i}{p_j} \},
\]

\noindent where $\np{p_j}{\btheta_i}$ and $\np{\btheta_i}{p_j}$ are the message natural parameter vectors.
\citeA{wand17} explains how natural parameter vectors play a central role in the messages that are
passed within and between factor graph fragments. Furthermore, the natural parameter vectors for the
optimal $q$-density functions in \eqref{opt_q_vmp} take the form

\begin{equation}
	\npq{\btheta_i} = \sum_{j : i \in \neighbors(j)} \np{p_j}{\btheta_i}, \quad i = 1, \dots, N.
\label{etaq}
\end{equation}

\noindent In addition, we adopt the notation

\begin{equation}
	\npbf{p_j}{\btheta_i} \equiv \np{p_j}{\btheta_i} + \np{\btheta_i}{p_j}.
\label{npbf}
\end{equation}

Before introducing the exponential family forms for key distributions in the VMP setting, we outline some
matrix and vector operators. We define the $\vect$ and $\vech$ operators,
which are well-established \cite<e.g.>{gentle07}.
For a $d_1 \times d_2$ matrix, the $\vect$ operator concatenates the columns of the matrix from left to right.
For a $d_1 \times d_1$ matrix, the $\vech$ operator concatenates the columns of the matrix after removing
the above diagonal elements. For example, suppose that

\[
	\bA = \begin{bmatrix}
		2 & -1 \\
		-3 & 1
	\end{bmatrix}.
\]

\noindent Then $\vect (\bA) = \T{(2, -3, -1, 1)}$ and $\vech (\bA) = \T{(2, -3, 1)}$.
For a $d^2 \times 1$ vector
$\ba$, $\vect^{-1} (\ba)$ is the $d \times d$ matrix such that $\vect \{\vect^{-1} (\ba)\} = \ba$. Additionally, the matrix
$\bD_d$ is the duplication matrix of order $d$, and it is such that $\bD_d \vech (\bA) = \vect (\bA)$ for a
$d \times d$ symmetric matrix $\bA$. Furthermore, $\bD_d^{+} \equiv (\T{\bD_d} \bD_d)^{-1} \T{\bD_d}$ is the
Moore-Penrose inverse of $\bD_d$, where $\bD_d^+ \vect (\bA) = \vech(\bA)$.

The normal distribution is one of the most important distributions within the exponential family, and it plays a
major role in VMP versions of variational Bayesian inference. Consider the $d \times 1$ multivariate normal
random vector $\bx \sim \normal (\bmu, \bSigma)$. The probability density function of $\bx$ can be shown to satisfy

\begin{equation}
	p(\bx) = \exp \left\{ \T{\bT_{\vect} (\bx)} \bdeta_{\vect} - A_{\vect} (\bdeta_{\vect}) - \frac{d}{2} \log (2 \pi) \right\}
\label{vec_repn}
\end{equation}

\noindent where

\[
	\bT_{\vect} (\bx) \equiv \begin{bmatrix}
		\bx \\
		\vect (\bx \T{\bx})
	\end{bmatrix} \quad
	\text{and} \quad
	\bdeta_{\vect} \equiv \begin{bmatrix}
		\bdeta_{\vect, 1} \\
		\bdeta_{\vect, 2}
	\end{bmatrix} \equiv \begin{bmatrix}
		\bSigma^{-1} \bmu \\
		-\frac12 \vect (\bSigma^{-1})
	\end{bmatrix}
\]

\noindent are, respectively, the vector of sufficient statistics and the natural parameter vector. The function

\[
	A_{\vect} (\bdeta_{\vect}) =
		-\frac14 \T{\bdeta}_{\vect, 1} \left\{ \vect^{-1} (\bdeta_{\vect, 2}) \right\}^{-1} \bdeta_{\vect, 1}
		- \frac12 \log \left| -2 \vect^{-1} (\bdeta_{\vect, 2}) \right|
\]

\noindent is the log-partition function. The inverse mapping of the natural parameter vector is
\cite[equation~S.4]{wand17}

\begin{equation}
	\bmu = -\frac12 \left\{ \vect^{-1} (\bdeta_{\vect, 2}) \right\}^{-1} \bdeta_{\vect, 1} \quad
	\text{and} \quad
	\bSigma = -\frac12 \left\{ \vect^{-1} (\bdeta_{\vect, 2}) \right\}^{-1}.
\label{gauss_vec_comm_params}
\end{equation}

\noindent We will refer to the representation of the multivariate normal probability density function in
\eqref{vec_repn} as the \emph{vec-based representation}. Alternatively, a more storage-economical
representation of the multivariate normal probability density function is the \emph{vech-based representation}:

\[
	p (\bx) = \exp \left\{ \T{\bT_{\vech} (\bx)} \bdeta_{\vech} - A_{\vech} (\bdeta_{\vech}) - \frac{d}{2} \log (2 \pi) \right\},
\]

\noindent where the vector of sufficient statistics, the natural parameter vector and the log-partition function are,
respectively,

\[
	\bT_{\vech} (\bx) \equiv \begin{bmatrix}
		\bx \\
		\vech (\bx \T{\bx})
	\end{bmatrix}, \quad
	\bdeta_{\vech} \equiv \begin{bmatrix}
		\bdeta_{\vech, 1} \\
		\bdeta_{\vech, 2}
	\end{bmatrix} \equiv \begin{bmatrix}
		\bSigma^{-1} \bmu \\
		-\frac12 \T{\bD_d} \vect(\bSigma^{-1})
	\end{bmatrix}
\]

\noindent and

\[
	A_{\vech} (\bdeta_{\vech}) =
		-\frac14 \T{\bdeta}_{\vech, 1} \left\{
			\vect^{-1} (\bD_d^{+ \intercal} \bdeta_{\vech, 2})
		\right\}^{-1} \bdeta_{\vech, 1}
		- \frac12 \log \left| -2 \vect^{-1} (\bD_d^{+ \intercal} \bdeta_{\vech, 2}) \right|.
\]

\noindent The inverse mapping of the natural parameter vector under the vech-based representation is

\begin{equation}
	\bmu = -\frac12 \left\{ \vect^{-1} (\bD_d^{+ \intercal}\bdeta_{\vech, 2}) \right\}^{-1} \bdeta_{\vech, 1} \quad
	\text{and} \quad
	\bSigma = -\frac12 \left\{ \vect^{-1} (\bD_d^{+ \intercal}\bdeta_{\vech, 2}) \right\}^{-1}.
\label{gauss_vech_comm_params}
\end{equation}

The other major distribution within the exponential family that is pivotal for this article is the inverse-$\chi^2$ distribution.
A random variable $x$ has an inverse-$\chi^2$ distribution with shape parameter $\xi > 0$ and scale parameter
$\lambda > 0$ if the probability density function of $x$ is

\[
	p(x) = 
		\frac{(\lambda/2)^{\xi/2}}{\Gamma (\xi/2)}
		x^{-(\xi + 2)/2} \exp \left( -\frac{\lambda}{2 x} \right) \ind (x > 0),
\]

\noindent where $\Gamma (\cdot)$ is the gamma function defined by $\Gamma (z) \equiv \int_0^\infty u^{z - 1} e^u
du$. The exponential family representation of the inverse-$\chi^2$ density function is

\[
	p(x) = \exp \left\{ \T{\bT (x)} \bdeta - A(\bdeta) \right\} \ind (x > 0),
\]

\noindent where the vector of sufficient statistics, the natural parameter vector and the log-partition function are,
respectively,

\[
	\bT (x) \equiv \begin{bmatrix}
		\log (x) \\
		1/x
	\end{bmatrix} \quad
	\text{and} \quad
	\bdeta \equiv \begin{bmatrix}
		\eta_1 \\
		\eta_2
	\end{bmatrix} \equiv \begin{bmatrix}
		-\frac12 (\xi + 2) \\
		-\frac{\lambda}{2}
	\end{bmatrix}
\]

\noindent and

\[
	A(\bdeta) \equiv \log \{ \Gamma (\xi/2) \} - \frac{\xi}{2} \log (\lambda/2).
\]

\noindent The inverse mapping of the natural parameter vector is \cite[equation~8]{maestrini20}

\[
	\xi = -2 \eta_1 - 2 \quad
	\text{and} \quad
	\lambda = -2 \eta_2.
\]

The generalization of the inverse-$\chi^2$ distribution is the inverse G-Wishart distribution, written $\bX \sim \igw (G, \xi, \Lambda)$,
for a symmetric and positive definite $d \times d$ matrix $\bX$.
The parameter $G$ is an undirected graph with $d$ nodes and edge
set $E$ with pairs of nodes connected by an edge. Furthermore, $\xi > 0$ and
$\bLambda$ is a symmetric and positive definite $d \times d$ matrix.
We say that, the symmetric matrix $\bA$ respects $G$ if $\bA_{ij} = 0$
for all $( i, j ) \notin E$. Now, impose the additional constraint that  $\bX^{-1}$ respects $G$. Furthermore, let $G_{\text{full}}$ represent a
complete graph where each node is connected
to every other node by an edge, and let $G_{\text{diag}}$ represent a sparse graph where the edge set is empty.
The justification for this notation is that a matrix with no zero constraints respects $G_{\text{full}}$ and a diagonal
matrix respects $G_{\text{diag}}$.
The two graph constructions generate two definitions for the inverse G-Wishart distribution \cite[Definition~3]{maestrini20}:

\begin{enumerate}[label={(\alph*)}]
	\item If $G = G_{\text{full}}$ and $\xi$ is restricted such that $\xi > 2d - 2$ then we say that $\bX$ has an inverse G-Wishart
	distribution with graph $G$, shape parameter $\xi$ and scale matrix $\bLambda$ if and only if the non-zero
	values of the density function of $\bX$ satisfy
		\[
			p (\bX) \propto |\bX|^{-(\xi + 2)/2} \exp \left\{ -\frac12 \tr (\bLambda \bX^{-1}) \right\}
		\]
	\item If $G = G_{\text{diag}}$, then we say the $\bX$ has an inverse G-Wishart
	distribution with graph $G$, shape parameter $\xi > 0$ and scale matrix $\bLambda$ if and only if the non-zero
	values of the density function of $\bX$ satisfy
		\[
			p (\bX) \propto |\bX|^{-(\xi + 2)/2} \exp \left\{ -\frac12 \tr (\bLambda \bX^{-1}) \right\}.
		\]
\end{enumerate}

The exponential family form of the inverse G-Wishart distribution is not important for this article; we refer the reader to
\citeA[Section~2.2]{maestrini20}. Instead, our interest lies in the role of the graphical parameter, which must be
incorporated as an argument for variational message updates involving inverse G-Wishart random matrices, including
inverse-$\chi^2$ random variables. We follow the advice in \citeA[Sections~5 \& 6]{maestrini20} for message passing
updates involving graphical parameters.


\subsubsection{Factor Graph Fragments}
\label{sec:fg_frag}

A factor graph fragment (or fragment, for short) is the computational unit of VMP.
A fragment, as defined by \citeA{wand17},
is a subgraph of a factor graph
consisting of a single factor and each of its neighboring stochastic nodes.
An example of a factor graph fragment is the fragment for $p (\btheta_2 | \btheta_4, \btheta_5)$ in Figure
\ref{fig:arb_bayes_mod_fg}, which is highlighted in blue. The factor representing $p (\btheta_2 | \btheta_4, \btheta_5)$,
the neighboring stochastic nodes $\btheta_2$, $\btheta_4$ and $\btheta_5$
and the connecting edges comprise the fragment.

The identification and derivation of the algebraic computations in fragments is the key step for extending VMP
to arbitrarily large Bayesian statistical models. As explained at the end of Section \ref{sec:mfvb}, the form of each of the
optimal $q$-density functions is problem-specific, requiring rederivations for any changes in the model. However,
the fragment based approach in VMP means that updates are localized within the fragment.
This represents enormous savings in algebraic derivations and coding because once a fragment has
been coded and stored, it can simply be called as a function without the need to rederive the updates.
Any change in the model is simply handled by removing the fragment that does not fit the new model and replacing
it with an appropriate alternative.

The following is a list of some of the major contributions to fragment updates for VMP:

\begin{itemize}
	\item \citeA{wand17} identified five fundamental fragments for Gaussian response semiparametric regression.
	This includes the Gaussian prior fragment, inverse G-Wishart prior fragment and the iterated inverse
	G-Wishart fragment, which are necessary for a VMP construction of Bayesian FPCA. The author also
	identified several other fragments for direct implementation in various extensions for semiparametric
	regression.
	\item \citeA{nolan17} developed fast, stable and accurate numerical integration techniques for 
	the logistic likelihood fragment. This had been previously introduced in \citeA{wand17} using the
	variational lower bound of \citeA{jaakkola00}, however the performance of this approximation can be poor
	\cite{knowles11}. Instead, \citeA{nolan17} incorporated the normal scale mixture uniform approximation of
	the logistic function \cite{monahan92} into the logistic likelihood fragment for highly accurate inference.
	\item \citeA{maestrini18} derived algorithmic updates for the skew $t$ likelihood fragment with all skew $t$
	parameters inferred, rather than being held fixed.
	\item \citeA{mclean19} built on previous VMP constructions by focusing on regression models where
	the response variable is modeled to have an elaborate distribution, such as Negative Binomial and
	$t$ likelihoods.
	\item \citeA{nolanmw20} used a set of solutions to sparse multilevel matrix problems \cite{nolan19} to
	streamline the computations of VMP for Gaussian response linear mixed models. This involved the
	introduction of four new fragments.
	\item \citeA{maestrini20} provide corrections for the matrix versions of the inverse G-Wishart prior fragment
	and the iterated inverse G-Wishart fragment from \citeA{wand17}. In particular, Section 4.1.3 of
	\citeA{wand17}, which introduces the iterated inverse G-Wishart fragment, only addresses the case where
	the graph parameter is $G = G_{\text{full}}$. However, it does not account for the case where $G = G_{\text{diag}}$.
	The necessary corrections are outlined in Section 6.1 of \citeA{maestrini20}. Although the scalar versions of these
	fragments from \citeA{wand17} are sufficient for the current article, we will use the updated fragments
	from \citeA{maestrini20} since they are the new standard for approximate Bayesian inference on variance
	and covariance matrix parameters.
\end{itemize}


\section{Functional Principal Components Analysis}
\label{sec:fpca}

Consider a random sample of i.i.d.\ smooth random functions $y_1, \dots, y_n \in L^2 [0, 1]$. We will assume the
existence of a continuous mean function $\mu = \E y_i$ and continuous covariance surface
$\sigma (t, s) = \E [ \{ y_i (t) - \mu (t) \} \{ y_i (s) - \mu (s) \} ]$, $i = 1, \dots, n$.
Then, the covariance operator $\Sigma$ of $y_i$ is defined as

\begin{equation}
	(\Sigma f) (t) \equiv \int_0^1 \sigma (t, s) f(s) ds, \quad f \in L^2 [0, 1].
\label{cov_op}
\end{equation}

\noindent Mercer's Theorem implies that the spectral decomposition of $\Sigma$ satisfies $\sigma (s, t) =
\sum_{l=1}^\infty \gamma_l \ \psi^*_l (s) \ \psi^*_l (t)$, where the $\gamma_l$ are the eigenvalues of
$\Sigma$ in descending
order and $\psi^*_l$ are the corresponding orthonormal eigenfunctions. The Karhunen-Lo\`{e}ve decomposition
is the basis for the FPCA expansion \cite{yao05}:

\begin{equation}
	y_i (t) = \mu (t) + \sum_{l=1}^\infty \zeta^*_{il} \ \psi^*_l (t), \quad i = 1, \dots, n,
\label{kl_expansion}
\end{equation}

\noindent where $\zeta^*_{il} = \int_0^1 \{ y_i (t) - \mu(t) \} \psi^*_l(t) dt$ are the principal components
scores. The $\zeta^*_{il}$ are independent across $i$ and uncorrelated across $l$, with $\E (\zeta^*_{il}) = 0$
and $\Var (\zeta^*_{il}) = \gamma_l$.
The asterisk is used as a reminder that the eigenfunctions in
\eqref{kl_expansion} are orthonormal and that the scores are independent.

Expansion \eqref{kl_expansion} facilitates dimension reduction by providing a best approximation for each
curve $y_1, \dots, y_n$ in terms of the truncated sums involving the first $L$ orthonormal eigenfunctions
$\psi^*_1, \dots, \psi^*_L$. That is, for any choice of $L$ orthonormal eigenfunctions $\psi_1, \dots, \psi_L$, the
minimum of

\[
	\sum_{i=1}^n || y_i - \mu - \sum_{l=1}^L \langle y_i - \mu , \psi_l \rangle \psi_l ||^2
\]

\noindent is achieved for $\psi_l = \psi^*_l$, $l = 1, \dots, L$, where $|| \cdot ||$ denotes the $L^2$ norm and
$\langle \cdot, \cdot \rangle$ denotes the $L^2$ inner product. For this reason, we define the best estimate of
$y_i$ as

\begin{equation}
	\yhat_i (t) \equiv \mu (t) + \sum_{l=1}^L \zeta^*_{il} \ \psi^*_l (t), \quad i = 1, \dots, n.
\label{yhat}
\end{equation}

Next, we make some observations involving the scores and the orthonormal eigenfunctions in \eqref{kl_expansion}
and \eqref{yhat}:

\begin{enumerate}
	\item If $\gamma_l = \gamma_k$, $l \neq k$, then the corresponding orthonormal
	eigenfunctions $\psi^*_l$ and $\psi^*_k$
	are not unique. We will simply assume that the eigenvalues are unique, which is a reasonable assumption
	for most applications (e.g. climate data
	and biostatistical problems).
	\item If all the eigenvalues are unique, the corresponding orthonormal eigenfunctions are only unique up to
	a change of sign. Issues of identifiability are always present when one attempts to infer eigenfunctions or
	eigenvectors. However, choosing one eigenfunction over its opposite sign has no effect on the resulting fits,
	although one choice of sign may provide more natural interpretation of the eigenfunction.
	Here, we simply assume that the inner product of the Bayesian estimate of an
	eigenfunction, with the desired sign, and the eigenfunction itself is positive.
\end{enumerate}

\noindent We state these assumptions formally.

\begin{assumption}
	
	The eigenvalues of the covariance operator in \eqref{cov_op} are unique.
	
\label{asspn:scores}
\end{assumption}

\begin{assumption}
	
	The signs of the orthonormal eigenfunctions $\psi^*_1, \dots, \psi^*_L$ are such that if $\psihat_l$ is an
	estimator of $\psi^*_l$, then $\langle \psi^*_l , \psihat_l \rangle > 0$.
	
\label{asspn:signs}
\end{assumption}

Expansions similar to \eqref{yhat} are also possible, where

\begin{equation}
	\yhat_i (t) \equiv \mu (t) + \sum_{l=1}^L \zeta_{il} \ \psi_l (t), \quad i = 1, \dots, n,
\label{yhat_not_orthogonal}
\end{equation}

\noindent where $\zeta_{il}$ are correlated across $l$, but remain independent across $i$, and the $\psi_l$ are not
orthonormal. Theorem \ref{thm:orth_basis} shows that an orthogonal decomposition of the resulting basis functions
and scores is sufficient for establishing the appropriate estimator \eqref{yhat}.
Its proof is provided in Appendix \ref{app:proof_thm_orth_basis}.

\begin{theorem}
	
	Consider Assumptions \ref{asspn:scores} and \ref{asspn:signs} and the approximations of the response
	curves $y_1, \dots, y_n$ in \eqref{yhat_not_orthogonal}. Then, there exists a unique set of orthonormal
	eigenfunctions $\psi^*_1, \dots, \psi^*_L$ and an uncorrelated set of scores $\zeta^*_{i1}, \dots, \zeta^*_{iL}$,
	$i = 1, \dots, n$, such that
	
	\[
		\yhat_i (t) = \mu (t) + \sum_{l=1}^L \zeta^*_{il} \ \psi^*_l (t), \quad i = 1, \dots, n.
	\]
	
\label{thm:orth_basis}
\end{theorem}

Theorem \ref{thm:orth_basis} motivates estimation of the Karhunen-Lo\`{e}ve decomposition directly to infer
the eigenfunctions and scores. In this approach, all components of the Karhunen-Lo\`{e}ve decomposition are
viewed as unknown so that scores and eigenfunctions are estimated jointly.
The other class of methods use covariance decompositions to obtain the eigenfunctions
and subsequently estimates the scores given the eigenfunctions using the Karhunen-Lo\`{e}ve decomposition
\cite<e.g.>{yao05, di09, xiao16}. There are several advantages in the former method in that it does not require
estimation or smoothing of a large covariance and can more directly handle sparse or irregular functional data.
The Bayesian model is described in Section \ref{sec:bayes_mod}, while new fragments that are relevant
for FPCA via VMP are introduced in Sections \ref{sec:fpca_gauss_lik_frag} and \ref{sec:mean_fpc_gauss_pen_frag}.


\subsection{Bayesian Model Construction}
\label{sec:bayes_mod}

In practice, the curves $y_1, \dots, y_n$ are indirectly observed as noisy observations at discrete points in time.
Furthermore, the observation times are not necessarily the same for each curve.
Let the set of design points for the $i$th curve be summarized by the vector

\begin{equation}
	\bt_i \equiv \T{(t_{i1}, \dots, t_{iT_i})}, \quad i = 1, \dots, n,
\label{t_i}
\end{equation}

\noindent where $T_i$ is the number of observations on the $i$th curve. In addition, we represent the
observations for the $i$th curve, $y_i (t)$, by the vector

\begin{equation}
	\by_i \equiv \T{\{ y_i (t_{i1}) + \epsilon_{i1}, \dots, y_i (t_{iT_i}) + \epsilon_{iT_i} \}} \quad i = 1, \dots, n,
\label{disc_obs}
\end{equation}

\noindent where $\epsilon_{ij}$ are i.i.d.\ noise terms with $\E (\epsilon_{ij}) = 0$ and $\Var (\epsilon_{ij}) = \sigsqeps$.
The finite decomposition in \eqref{yhat} takes the form:

\begin{equation}
	\by_i = \bmu_i + \sum_{l=1}^L \zeta_{il} \bpsi_{il} + \bepsilon_{i}, \quad i = 1, \dots, n,
\label{resp_mod}
\end{equation}

\noindent where $\bmu_i \equiv \T{\{ \mu (t_{i1}), \dots, \mu (t_{iT_i}) \}}$,
$\bpsi_{il} \equiv \T{\{ \psi_l (t_{i1}), \dots, \psi_l (t_{iT_i}) \}}$, for $l = 1, \dots, L$, and
$\bepsilon_{i} \equiv \T{(\epsilon_{i1}, \dots, \epsilon_{iT_i})}$ is a vector of measurement errors
for the observations on curve $y_i (t)$.

We model continuous curves from discrete observations via nonparametric regression \cite{ruppert03, ruppert09},
using the mixed model-based penalized spline basis function representation, as in \citeA{durban05}. The
representation for the mean function and the FPCA eigenfunctions are:

\[
	\mu (t) \approx \beta_{\mu, 0} + \beta_{\mu, 1} t + \sum_{k=1}^K u_{\mu, k} z_k (t) \quad
	\text{and} \quad
	\psi_l (t) \approx \beta_{\psi_l, 0} + \beta_{\psi_l, 1} t + \sum_{k=1}^K u_{\psi_l, k} z_k (t) \quad
	\text{for $l = 1, \dots, L$},
\]

\noindent where $\{ z_k (\cdot) \}_{1 \le k \le K}$ is a suitable set of
basis functions. Splines and wavelet families are the most common choices for the $z_k$. In our simulations, we
use O'Sullivan penalized splines, which are described in Section 4 of \citeA{wand08}.

In order to avoid notational clutter, we incorporate the following definitions:

\[
\begin{gathered}
	\betamu \equiv \T{(\beta_{\mu, 0}, \beta_{\mu, 1})} \quad
		\umu \equiv \T{(u_{\mu, 1}, \dots, u_{\mu, K})} \quad
		\numu \equiv \T{(\T{\betamu}, \T{\umu})} \\
	\betapsi{l} \equiv \T{(\beta_{\psi_l, 0}, \beta_{\psi_1, 1})}, \quad
		\upsi{l} \equiv \T{(u_{\psi_l, 1}, \dots, u_{\psi_l, K})} \quad
		\text{and} \quad
		\nupsi{l} \equiv \T{(\T{\betapsi{l}}, \T{\upsi{l}})} \quad
		\text{for $l = 1, \dots, L$}.
\end{gathered}
\]

\noindent Then simple derivations that stem from \eqref{resp_mod} show that the vector of observations on
each of the response curves satisfies the representation:

\[
	\by_i = \bC_i \left(
		\numu + \sum_{l=1}^L \zeta_{il} \nupsi{l}
	\right) + \bepsilon_{i}, \quad i = 1, \dots, n,
\]

\noindent where

\begin{equation}
	\bC_i \equiv \begin{bmatrix}
		1 & t_{i1} & z_1 (t_{i1}) & \dots & z_K (t_{i1}) \\
		\vdots & \vdots & \vdots & \ddots & \vdots \\
		1 & t_{iT_i} & z_1 (t_{iT_i}) & \dots & z_K (t_{iT_i})
	\end{bmatrix}.
\label{C_mat}
\end{equation}

\noindent In addition, we define:

\begin{equation}
	\by \equiv \T{(\T{\by_1}, \dots, \T{\by_N})}, \quad
	\bnu \equiv \T{(\T{\numu}, \T{\nupsi{1}}, \dots, \T{\nupsi{L}})} \quad
	\text{and} \quad
	\bzeta_i \equiv \T{(\zeta_{i1}, \dots, \zeta_{iL})} \quad i = 1, \dots, n.
\label{partitioned_vectors}
\end{equation}

Next, we present the Bayesian FPCA Gaussian response model:

\begin{equation}
\begin{gathered}
	\by_i | \bnu, \bzeta_i, \sigsqeps \indsim \normal \left\{
		\bC_i \left( \numu + \sum_{l=1}^L \zeta_{il} \nupsi{l} \right), \sigsqeps \bI_{T_i}
	\right\}, \quad
	\bzeta_i \indsim \normal (\bzero, \bSigma_{\zeta_i}), \quad
	i = 1, \dots, n, \\
	$$
	\left.\begin{bmatrix}
		\numu \\
		\nupsi{l}
	\end{bmatrix} \ \right| \ \sigsqmu, \sigsqpsi{l}
		\indsim
			\normal \left(
				\begin{bmatrix}
					\bmu_\mu \\
					\bmu_{\psi_l} \\
				\end{bmatrix},
				\begin{bmatrix}
					\bSigma_\mu & \T{\textbf{O}} \\
					\textbf{O} & \bSigma_{\psi_l}
				\end{bmatrix}
			\right), \quad
	\sigsqpsi{l} | \apsi{l} \indsim \invchisq (1, 1/\apsi{l}), \\
	$$
	\apsi{l} \indsim \invchisq (1, 1/\Asqpsi{l}), \quad l = 1, \dots, L, \\
	$$
	\sigsqmu | \amu \sim \invchisq (1, 1/\amu), \quad \amu \sim \invchisq(1, 1/\Asqmu), \\
	$$
	\sigsqeps | \aeps \sim \invchisq (1, 1/\aeps), \quad \aeps \sim \invchisq(1, 1/\Asqeps),
\end{gathered}
\label{bayes_fpca_mod}
\end{equation}

\noindent where

\begin{equation}
\begin{gathered}
	\bmu_\mu \equiv \T{(\T{\bmu_{\beta_\mu}}, \T{\bzero_K})}, \quad
	\bSigma_\mu \equiv \begin{bmatrix}
		\bSigma_{\beta_\mu} & \T{\textbf{O}} \\
		\textbf{O} & \sigsqmu \bI_{K}
	\end{bmatrix}, \\
	$$
	\bmu_{\psi_l} \equiv \T{(\T{\bmu_{\beta_{\psi_l}}}, \T{\bzero_K})}, \quad
	\bSigma_{\psi_l} \equiv \begin{bmatrix}
		\bSigma_{\beta_{\psi_l}} & \T{\textbf{O}} \\
		\textbf{O} & \sigsqpsi{l} \bI_{K}
	\end{bmatrix}, \quad l = 1, \dots, L,
\end{gathered}
\label{sub_vecs_mats}
\end{equation}

\noindent and $\bmu_{\beta_\mu}$ ($2 \times 1$), $\bmu_{\beta_{\psi_l}}$ ($2 \times 1$, $l = 1, \dots, L$),
$\bSigma_{\beta_\mu}$ ($2 \times 2$, positive definite), $\bSigma_{\beta_{\psi_l}}$ ($2 \times 2$, positive definite,
$l = 1, \dots, L$), $\bSigma_{\zeta_i}$ ($L \times L$, positive definite, $i = 1, \dots, n$), $A_\nu > 0$,
$A_{\psi_l} > 0$ ($l = 1, \dots, L$) are the model hyperparameters.
Note that the iterated inverse-$\chi^2$ distributional specification on $\sigsqeps$,
which involves an inverse-$\chi^2$ prior specification on the auxiliary variable $\aeps$, is equivalent to $\sigsqeps \sim
\hc (A_{\epsilon})$. This auxiliary variable-based hierarchical construction facilitates arbitrarily non-informative
priors on standard deviation parameters \cite{gelman06}. Similar comments also apply to the iterated inverse-$\chi^2$
distributional specifications for $\sigsqmu$ and $\sigsqpsi{1}, \dots, \sigsqpsi{L}$.

Full Bayesian inference for the parameter set $\bnu$, $\bzeta_1, \dots, \bzeta_n$, $\sigsqeps$, $\aeps$,
$\sigsqmu$, $\amu$, $\sigsqpsi{1}, \dots, \sigsqpsi{L}$ and $\apsi{1}, \dots, \apsi{L}$ requires the determination
of the posterior density function

\[
	p (
		\bnu, \bzeta_1, \dots, \bzeta_n, \sigsqeps, \aeps, \sigsqmu, \amu,
		\sigsqpsi{1}, \dots, \sigsqpsi{L}, \apsi{1}, \dots, \apsi{L} | \by
	),
\]

\noindent but it is typically analytically intractable. The standard approach for overcoming this deficiency is to
employ MCMC approaches. However, we propose two major arguments against this approach. First, MCMC
simulations are very slow for model \eqref{bayes_fpca_mod}, even for moderate dimensions of $\bnu$, which
depends on the number of eigenfunctions ($L$) and O'Sullivan penalized spline basis functions ($K$).
Second, the mean function $\mu (t)$ and the eigenfunctions $\psi_1 (t), \dots,
\psi_L (t)$ are typically highly correlated, which is expected to lead to poor mixing.
A possible remedy for this is to use an inverse G-Wishart prior structure that permits
correlations amongst the spline coefficients \cite{goldsmith16}. However, this is beyond the scope of this article,
which is not concerned with improving MCMC methods for FPCA.

Alternatively, variational approximate inference for model \eqref{bayes_fpca_mod} involves
the mean field restriction:

\begin{align}
\begin{split}
	p (
		\bnu, \bzeta_1, \dots, \bzeta_n, \sigsqeps, \aeps, &\sigsqmu, \amu,
		\sigsqpsi{1}, \dots, \sigsqpsi{L}, \apsi{1}, \dots, \apsi{L} | \by
	) \approx \\
		&\left\{ \prod_{i=1}^N q (\bzeta_i) \right\} q (\bnu, \aeps, \amu, \apsi{1}, \dots, \apsi{L})
		q(\sigsqeps, \sigsqmu, \sigsqpsi{1}, \dots, \sigsqpsi{L}).
\end{split}
\label{fpca_mf_min_restrn}
\end{align}

\noindent The approximation in \eqref{fpca_mf_min_restrn} represents the minimal mean-field restriction that is
required for approximate variational inference. Here, we have assumed posterior independence between
global parameters and response curve-specific parameters, as well as incorporating the notion of
\emph{asymptotic independence} between regression coefficients and variance parameters
\cite[Section~3.1]{menictas13}.
However, induced factorizations, based on graph theoretic
results \cite[Section~10.2.5]{bishop06}, admit further factorizations, and the right-hand side of
\eqref{fpca_mf_min_restrn} becomes

\begin{equation}
	\left\{ \prod_{i=1}^N q (\bzeta_i) \right\} q (\bnu) q(\sigsqeps) q(\aeps)
	q(\sigsqmu) q (\amu) \left\{ \prod_{l=1}^L q(\sigsqpsi{l}) q(\apsi{l}) \right\}.
\label{fpca_mf_restrn}
\end{equation}

\begin{figure}
	\centering
	\begin{tikzpicture} [
		auto, node distance=1.2cm, every loop/.style={},
		thick,stochastic node/.style={circle,draw,font=\sffamily\Large\bfseries},
		factor/.style={regular polygon,regular polygon sides=4}
	]
	
	\node [
		fill = blue, factor, draw = blue, scale = 1,
		label = right:{$p ( \by | \bnu, \bzeta_1, \dots, \bzeta_n, \sigsqeps )$}
	] (py) {};
	\node [
		stochastic node, draw = blue, scale = 1,
		label = right:$\bnu$
	] (nu) [below of = py] {};
	\node [
		stochastic node, draw = blue, scale = 1,
		label = above:$\bzeta_1$
	] (zeta1) [above left = 0.2cm and 1.5cm of py] {};
	\node [
		stochastic node, draw = blue, scale=1,
		label = below:$\bzeta_n$
	] (zetaN) [below left = 0.2cm and 1.5cm of py] {};
	\node [font = \Large, rotate=-90] (zeta_dots) at ($(zeta1)!0.5!(zetaN)$) {$\dots$};
	\node [
		stochastic node, draw = blue, scale=1,
		label = above:$\sigsqeps$
	] (sigsqeps) [above of = py] {};
	
	\node [
		fill = red, factor, draw = red, scale = 1,
		label = right:{$p ( \bnu | \sigsqmu, \sigsqpsi{1}, \dots, \sigsqpsi{L} )$}
	] (pnu) [below of = nu] {};
	\node[stochastic node, draw = red, scale = 0.7] (nu_inner) at (nu.center) {};
	\node [
		stochastic node, draw = red, scale=1,
		label = left:$\sigsqpsi{1}$
	] (sigsqpsi1) [below = 1.3cm of pnu] {};
	\node [
		stochastic node, draw = red, scale=1,
		label = right:$\sigsqpsi{L}$
	] (sigsqpsiL) [right = 2.5cm of sigsqpsi1] {};
	\node [font = \huge] (sigsqpsi_dots) at ($(sigsqpsi1)!0.5!(sigsqpsiL)$) {$\dots$};
	\node [
		stochastic node, draw = red, scale=1,
		label = left:$\sigsqmu$
	] (sigsqmu) [left = 2.5cm of sigsqpsi1] {};
	
	\node [
		fill, factor, draw, scale = 1,
		label = above:{$p ( \bzeta_1 )$}
	] (pzeta1) [left = 1.3cm of zeta1] {};
	\node [
		fill, factor, draw, scale = 1,
		label = below:{$p ( \bzeta_n )$}
	] (pzetaN) [left = 1.3cm of zetaN] {};
	\node [font = \Large, rotate=-90] (pzeta_dots) at ($(pzeta1)!0.5!(pzetaN)$) {$\dots$};
	
	\node [
		fill, factor, draw, scale = 1,
		label = above:{$p ( \sigsqeps | \aeps )$}
	] (psigsqeps) [right = 1.3cm of sigsqeps] {};
	\node [
		stochastic node, draw, scale=1,
		label = above:$\aeps$
	] (aeps) [right = 1.3cm of psigsqeps] {};
	
	\node [
		fill, factor, draw, scale = 1,
		label = above:{$p ( \aeps )$}
	] (paeps) [right = 1.3cm of aeps] {};
	
	\node [
		fill, factor, draw, scale = 1,
		label = left:{$p ( \sigsqmu | \amu )$}
	] (psigsqmu) [below of = sigsqmu] {};
	\node [
		stochastic node, draw, scale=1,
		label = left:$\amu$
	] (amu) [below of = psigsqmu] {};
	
	\node [
		fill, factor, draw, scale = 1,
		label = left:{$p ( \amu )$}
	] (pamu) [below of = amu] {};
	
	\node [
		fill, factor, draw, scale = 1,
		label = left:{$p ( \sigsqpsi{1} | \apsi{1} )$}
	] (psigsqpsi1) [below of = sigsqpsi1] {};
	\node [
		stochastic node, draw, scale = 1,
		label = left:{$ \apsi{1} $}
	] (apsi1) [below of = psigsqpsi1] {};
	\node [
		fill, factor, draw, scale = 1,
		label = right:{$p ( \sigsqpsi{L} | \apsi{L} )$}
	] (psigsqpsiL) [below of = sigsqpsiL] {};
	\node [
		stochastic node, draw, scale = 1,
		label = right:{$ \apsi{L} $}
	] (apsiL) [below of = psigsqpsiL] {};
	\node [font = \huge] (psigsqpsi_dots) at ($(psigsqpsi1)!0.5!(psigsqpsiL)$) {$\dots$};
	\node [font = \huge] (apsi_dots) at ($(apsi1)!0.5!(apsiL)$) {$\dots$};
	
	\node [
		fill, factor, draw, scale = 1,
		label = left:{$p ( \apsi{1} )$}
	] (papsi1) [below of = apsi1] {};
	\node [
		fill, factor, draw, scale = 1,
		label = right:{$p ( \apsi{L} )$}
	] (papsiL) [below of = apsiL] {};
	\node [font = \huge] (papsi_dots) at ($(papsi1)!0.5!(papsiL)$) {$\dots$};
	
	\path[every node/.style={font=\sffamily\small}]
	
	(py)
		edge [blue] node {} (nu)
		edge [blue] node {} (zeta1)
		edge [blue] node {} (zetaN)
		edge [blue] node {} (sigsqeps)
	
	(pnu)
		edge [red] node {} (nu_inner)
		edge [red] node {} (sigsqmu)
		edge [red] node {} (sigsqpsi1)
		edge [red] node {} (sigsqpsiL)
	
	(pzeta1) edge node {} (zeta1)
	(pzetaN) edge node {} (zetaN)
	
	(psigsqeps)
		edge node {} (sigsqeps)
		edge node {} (aeps)
	
	(paeps)
		edge node {} (aeps)
	
	(psigsqmu)
		edge node {} (sigsqmu)
		edge node {} (amu)
	
	(pamu)
		edge node {} (amu)
	
	(psigsqpsi1)
		edge node {} (sigsqpsi1)
		edge node {} (apsi1)
	(psigsqpsiL)
		edge node {} (sigsqpsiL)
		edge node {} (apsiL)
	
	(papsi1)
		edge node {} (apsi1)
	(papsiL)
		edge node {} (apsiL);
	
	\end{tikzpicture}
\caption{The factor graph for the Bayesian FPCA model in \eqref{bayes_fpca_mod}.}
\label{fig:fg_fpca}
\end{figure}
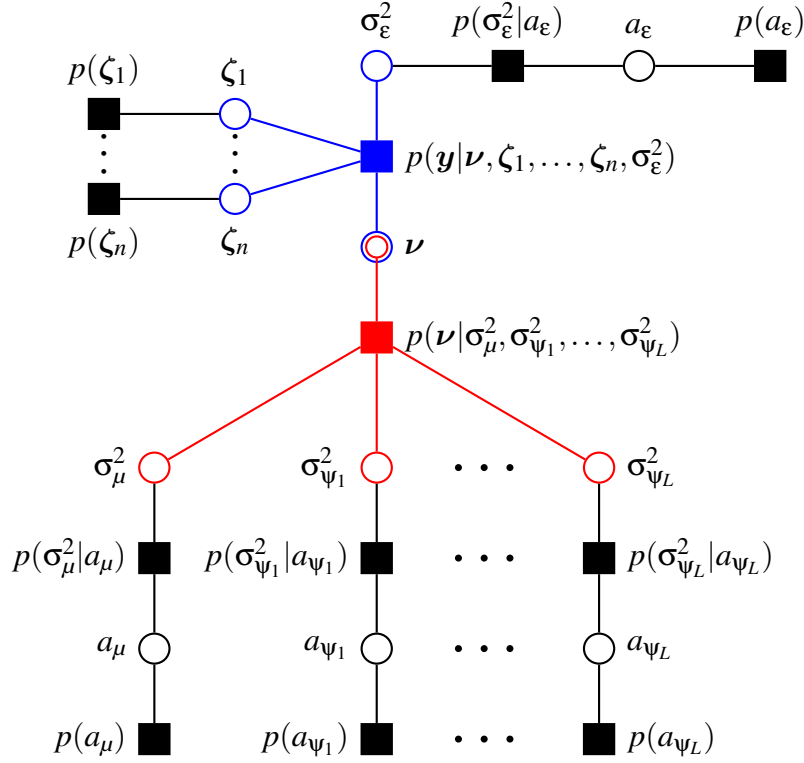

\noindent From here, we work with the factorization in \eqref{fpca_mf_restrn} to minimize the Kullback-Leibler
divergence  of the right-hand side of \eqref{fpca_mf_min_restrn} from its left-hand side.
The factor graph for model \eqref{bayes_fpca_mod} that represents the factorization in \eqref{fpca_mf_restrn}
is presented in Figure \ref{fig:fg_fpca}.

From Figure \ref{fig:fg_fpca}, we identify the following factor graph fragments that are involved in the Bayesian
FPCA model \eqref{bayes_fpca_mod}:

\begin{itemize}
	\item The fragments for $p (\bzeta_1), \dots, p(\bzeta_n)$ are Gaussian prior fragments. The updates for
	this fragment are presented in Section 4.1.1 of \citeA{wand17}, where a vec-based representation of the
	multivariate normal density function is used.
	\item The fragments for $p (\aeps)$, $p (\amu)$ and $p (\apsi{1}), \dots, p (\apsi{L})$ are scalar versions
	of the inverse G-Wishart prior fragment, which is presented as Algorithm 1 of \citeA{maestrini20}.
	\item The fragments for $p (\sigsqeps | \aeps)$, $p (\sigsqmu | \amu)$ and $p (\sigsqpsi{1} | \apsi{1}), \dots,
	p (\sigsqpsi{L} | \apsi{L})$ are scalar versions of the iterated inverse G-Wishart fragment, which is presented
	as Algorithm 2 of \citeA{maestrini20}.
	\item The fragments for $p (\by | \bnu, \bzeta_1, \dots, \bzeta_n, \sigsqeps)$ and $p (\bnu | \sigsqmu,
	\sigsqpsi{1}, \dots, \sigsqpsi{L})$ are two new fragments that have not been addressed in previous literature
	on VMP, but are crucial for FPCA modelling.
	We name the fragment for $p (\by | \bnu, \bzeta_1, \dots, \bzeta_n, \sigsqeps)$ the \emph{functional
	principal component
	Gaussian likelihood fragment} and the fragment for $p (\bnu | \sigsqmu, \sigsqpsi{1}, \dots, \sigsqpsi{L})$
	the \emph{functional principal component Gaussian penalization fragment}.
\end{itemize}


\subsection{Functional Principal Component Gaussian Likelihood Fragment}
\label{sec:fpca_gauss_lik_frag}

The Functional Principal Component Gaussian likelihood fragment, shown in blue in Figure
\ref{fig:fg_fpca}, is defined by the factor

\begin{equation}
	p (\by | \bnu, \bzeta_1, \dots, \bzeta_n, \sigsqeps) =
		\prod_{i=1}^n p (\by_i | \bnu, \bzeta_i, \sigsqeps),
\label{fpc_gauss_lik_factorized}
\end{equation}

\noindent where

\begin{equation}
	\by_i | \bnu, \bzeta_i, \sigsqeps \indsim \normal \left\{
		\bC_i \left( \numu + \sum_{l=1}^L \bzeta_{il} \nupsi{l} \right), \sigsqeps \bI_{T_i}
	\right\}, \quad \text{for $i = 1, \dots, n$}.
\label{fpc_gauss_lik_factor}
\end{equation}

\noindent The purpose of this fragment is to provide message updates for the variational posterior density functions
of $\bnu$, $\bzeta_i, \dots, \bzeta_n$ and $\sigsqeps$ at every iteration of the VMP algorithm.
Here, we outline the construction of the messages that are passed from the factor representing
the likelihood $p (\by | \bnu, \bzeta_1, \dots, \bzeta_n, \sigsqeps)$ to each of its neighbouring stochastic nodes.
On the other hand, the
derivations of these messages and the derivations of expected values of random variables, random vectors
and random matrices that these messages depend on are deferred to Appendix \ref{app:fpca_gauss_lik_frag}.

The message from $p (\by | \bnu, \bzeta_1, \dots, \bzeta_n, \sigsqeps)$ to $\bnu$ can be shown to be
proportional to a multivariate
normal density function, with vec-based exponential density function representation:

\begin{equation}
	\msg{p (\by | \bnu, \bzeta_1, \dots, \bzeta_n, \sigsqeps)}{\bnu} (\bnu) \propto
		\exp \left\{
			\T{\begin{bmatrix}
				\bnu \\
				\vect (\bnu \T{\bnu})
			\end{bmatrix}}
			\np{p (\by | \bnu, \bzeta_1, \dots, \bzeta_n, \sigsqeps)}{\bnu}
		\right\}.
\label{msg_lik_nu}
\end{equation}

\noindent The update for the natural parameter vector in \eqref{msg_lik_nu} is

\begin{equation}
	\np{p (\by | \bnu, \bzeta_1, \dots, \bzeta_n, \sigsqeps)}{\bnu}
		\longleftarrow
			\begin{bmatrix}
				\E_q (1/\sigsqeps) \displaystyle\sum_{i=1}^n \T{\left\{
					\T{\E_q (\bzetatilde_i)} \otimes \bC_i
				\right\}} \by_i \\
				-\frac12 \E_q (1/\sigsqeps) \displaystyle\sum_{i=1}^n \vect \left\{
					\E_q (\bzetatilde_i \T{\bzetatilde_i}) \otimes (\T{\bC_i} \bC_i)
				\right\}
			\end{bmatrix},
\label{np_lik_nu}
\end{equation}

\noindent where, 

\begin{equation}
	\bzetatilde_i \equiv \T{(1, \T{\bzeta_i})}, \quad \text{for $i = 1, \dots, n$.}
\label{zeta_tilde}
\end{equation}

\noindent Before proceeding to the other messages in this fragment, we make a brief comment on using the vec-based
representation of the message in \eqref{msg_lik_nu}, as opposed to the storage-economical vech-based representation.
In preliminary simulations, we found that computations using the vech-based representation were enormously hindered
by the need to use a huge Moore-Penrose inverse matrix. For instance, consider the case where
there are two basis functions ($L = 2$) and 25 O'Sullivan penalized spline basis functions ($K = 25$) for nonparametric
regression. In this instance, the vector $\bnu$ is $81 \times 1$ ($d = 81$) and the Moore-Penrose inverse matrix
$\bD_{81}^+$ has dimension $3321 \times 6561$, inhibiting the computational speed. For this reason, we have
decided to use the vec-based representation of the message in \eqref{msg_lik_nu}, which does not require the
use of a Moore-Penrose inverse matrix.

For each $i = 1, \dots, n$, the message from $p (\by | \bnu, \bzeta_1, \dots, \bzeta_n, \sigsqeps)$ to $\bzeta_i$ is

\begin{equation}
	\msg{p (\by | \bnu, \bzeta_1, \dots, \bzeta_n, \sigsqeps)}{\bzeta_i} (\bzeta_i) \propto
		\exp \left\{
			\T{\begin{bmatrix}
				\bzeta_i \\
				\vech (\bzeta_i \T{\bzeta_i})
			\end{bmatrix}}
			\np{p (\by | \bnu, \bzeta_1, \dots, \bzeta_n, \sigsqeps)}{\bzeta_i}
		\right\},
\label{msg_lik_zeta}
\end{equation}

\noindent which is proportional to a multivariate normal density function.
The update for the natural parameter vector in \eqref{msg_lik_zeta} is

\begin{equation}
	\np{p (\by | \bnu, \bzeta_1, \dots, \bzeta_n, \sigsqeps)}{\bzeta_i}
		\longleftarrow
			\begin{bmatrix}
				\E_q (1/\sigsqeps) \left\{
					\T{\E_q (\Vpsi)} \T{\bC_i} \by_i - \E_q (\hmupsi{i})
				\right\} \\
				-\frac12 \E_q (1/\sigsqeps) \T{\bD_L} \vect \{ \E_q (\Hpsi{i}) \}
			\end{bmatrix},
\label{np_lik_zeta}
\end{equation}

\noindent where

\begin{equation}
	\Vpsi \equiv \begin{bmatrix}
		\nupsi{1} & \dots & \nupsi{L}
	\end{bmatrix} \quad
	\text{and} \quad
	\hmupsi{i} \equiv \T{\Vpsi} \T{\bC_i} \bC_i \numu, \quad
	\Hpsi{i} \equiv \T{\Vpsi} \T{\bC_i} \bC_i \Vpsi, \quad
	\text{for $i = 1, \dots, n$}.
\label{psi_related_defns}
\end{equation}

The message from $p (\by | \bnu, \bzeta_1, \dots, \bzeta_n, \sigsqeps)$ to $\sigsqeps$ is

\begin{equation}
	\msg{p (\by | \bnu, \bzeta_1, \dots, \bzeta_n, \sigsqeps)}{\sigsqeps} (\sigsqeps) \propto
		\exp \left\{
			\T{\begin{bmatrix}
				\log (\sigsqeps) \\
				1/\sigsqeps
			\end{bmatrix}}
			\np{p (\by | \bnu, \bzeta_1, \dots, \bzeta_n, \sigsqeps)}{\sigsqeps}
		\right\},
\label{msg_lik_sigsqeps}
\end{equation}

\noindent and it is proportional to an inverse-$\chi^2$ density function. The update for the natural parameter vector in
\eqref{msg_lik_sigsqeps} is

\begin{equation}
	\np{p (\by | \bnu, \bzeta_1, \dots, \bzeta_n, \sigsqeps)}{\sigsqeps}
		\longleftarrow
			\begin{bmatrix}
				-\frac12 \displaystyle\sum_{i=1}^n T_i \\
				-\frac12 \displaystyle\sum_{i=1}^n \E_q \left\{ \T{\left(
					\by_i - \bC_i \bV \bzetatilde_i
				\right)} \left(
					\by_i - \bC_i \bV \bzetatilde_i
				\right) \right\}
			\end{bmatrix},
\label{np_lik_sigsqeps}
\end{equation}

\noindent where

\begin{equation}
	\bV \equiv \begin{bmatrix}
		\numu & \nupsi{1} & \dots & \nupsi{L}
	\end{bmatrix}.
\label{V_mat}
\end{equation}

\noindent We must remember that the inverse-$\chi^2$ density function message that is passed
to $\sigsqeps$ is part of the inverse G-Wishart class of density functions for VMP. Within this class of messages,
a graph message is also required to specify whether the density function respects a full or a diagonal matrix. This
graphical message does not affect inverse-$\chi^2$ density function messages, however we will include
a graph message with the aim of providing fragments that are compatible with previously constructed inverse
G-Wishart fragments \cite[Algorithms 1 \& 2]{maestrini20}. According to Section 7.4 of \citeA{maestrini20},
the auxiliary-based hierarchical prior specification of $\sigsqeps$ in \eqref{bayes_fpca_mod} requires a
graphical message of the form

\begin{equation}
	G_{p (\by | \bnu, \bzeta_1, \dots, \bzeta_n, \sigsqeps) \rightarrow \sigsqeps}
		\longleftarrow
			G_{\text{full}}.
\label{G_lik_sigsqeps}
\end{equation}

\noindent That is, the univariate random variable $\sigsqeps$ is chosen to respect a ``full'' $1 \times 1$ matrix.

Pseudocode for the functional principal component
Gaussian Likelihood Fragment is presented in Algorithm \ref{alg:fpca_gauss_lik_frag}.
A derivation of all the relevant expectations and natural parameter vector updates is provided in Appendix
\ref{app:fpca_gauss_lik_frag}.

\begin{algorithm}
	\caption{
		Pseudocode for the functional principal component Gaussian likelihood fragment.
	}
	\label{alg:fpca_gauss_lik_frag}
	\begin{algorithmic}[1]
		\Inputs
			\begin{varwidth}[t]{\linewidth} $
				\np{\bnu}{p (\by | \bnu, \bzeta_1, \dots, \bzeta_n, \sigsqeps)}, \quad
				\{ \np{\bzeta_i}{p (\by | \bnu, \bzeta_1, \dots, \bzeta_n, \sigsqeps)} : i = 1, \dots, n \}
			$\par$
				\{
					\np{\sigsqeps}{p (\by | \bnu, \bzeta_1, \dots, \bzeta_n, \sigsqeps)}, \
					G_{\sigsqeps \rightarrow p (\by | \bnu, \bzeta_1, \dots, \bzeta_n, \sigsqeps)}
				\}
			$ \end{varwidth}
		\Updates
			\State Update all expectations with respect to the optimal posterior distribution.
				\Comment{see Appendix \ref{app:fpca_gauss_lik_frag}}
			\State Update $\np{p (\by | \bnu, \bzeta_1, \dots, \bzeta_n, \sigsqeps)}{\bnu}$
				\Comment{equation \eqref{np_lik_nu}}
			\For{$i = 1, \dots, n$}
				\State Update $\np{p (\by | \bnu, \bzeta_1, \dots, \bzeta_n, \sigsqeps)}{\bzeta_i}$
					\Comment{equation \eqref{np_lik_zeta}}
			\EndFor
			\State Update $\np{p (\by | \bnu, \bzeta_1, \dots, \bzeta_n, \sigsqeps)}{\sigsqeps}$
				\Comment{equation \eqref{np_lik_sigsqeps}}
			\State Update $G_{p (\by | \bnu, \bzeta_1, \dots, \bzeta_n, \sigsqeps) \rightarrow \sigsqeps}$
				\Comment{equation \eqref{G_lik_sigsqeps}}
		\Outputs
			\begin{varwidth}[t]{\linewidth} $
				\np{p (\by | \bnu, \bzeta_1, \dots, \bzeta_n, \sigsqeps)}{\bnu}, \quad
				\{ \np{p (\by | \bnu, \bzeta_1, \dots, \bzeta_n, \sigsqeps)}{\bzeta_i} : i = 1, \dots, n \}
			$\par$
				\{
					\np{p (\by | \bnu, \bzeta_1, \dots, \bzeta_n, \sigsqeps)}{\sigsqeps}, \
					G_{p (\by | \bnu, \bzeta_1, \dots, \bzeta_n, \sigsqeps) \rightarrow \sigsqeps}
				\}
			$ \end{varwidth}
	\end{algorithmic}
\end{algorithm}


\subsection{Functional Principal Component Gaussian Penalization Fragment}
\label{sec:mean_fpc_gauss_pen_frag}

The functional principal component Gaussian penalization fragment,
shown in red in Figure \ref{fig:fg_fpca}, is defined by the factor
$p (\bnu | \sigsqmu, \sigsqpsi{1}, \dots, \sigsqpsi{L})$,  where

\begin{equation}
	\left.\begin{bmatrix}
		\numu \\
		\nupsi{l}
	\end{bmatrix} \ \right| \ \sigsqmu, \sigsqpsi{1}, \dots, \sigsqpsi{L}
		\indsim
			\normal \left(
				\begin{bmatrix}
					\bmu_\mu \\
					\bmu_{\psi_l} \\
				\end{bmatrix},
				\begin{bmatrix}
					\bSigma_\mu & \T{\textbf{O}} \\
					\textbf{O} & \bSigma_{\psi_l}
				\end{bmatrix}
			\right), \quad \text{for $l = 1, \dots, L$.}
\label{mean_fpc_gauss_pen_factor}
\end{equation}

\noindent and all sub-vectors and sub-matrices are defined in \eqref{sub_vecs_mats}. The purpose of
this fragment is to provide message updates for the variational posterior density functions of $\bnu$, $\sigsqmu$
and $\sigsqpsi{1}, \dots, \sigsqpsi{L}$ at each iteration of the VMP algorithm. Here, as in Section
\ref{sec:fpca_gauss_lik_frag}, we outline the messages that are passed from this factor to its neighbouring
stochastic nodes. For detailed derivations of these messages and all relevant expectations, we defer the
reader to Appendix \ref{app:mean_fpc_gauss_pen_frag}.

First, let us introduce the vector and matrix

\begin{equation}
	\munu \equiv \T{(\T{\mumu}, \T{\mupsi{1}}, \dots, \T{\mupsi{L}})} \quad
	\text{and} \quad
	\Sigmanu \equiv \blockdiag (\bSigma_\mu, \bSigma_{\psi_1}, \dots, \bSigma_{\psi_L}).
\label{munu_Sigmanu}
\end{equation}

\noindent Then, the message from $p (\bnu | \sigsqmu, \sigsqpsi{1}, \dots, \sigsqpsi{L})$ to $\bnu$
can be shown to be proportional to a multivariate normal density function, with vec-based exponential density
function representation

\begin{equation}
	\msg{p (\bnu | \sigsqmu, \sigsqpsi{1}, \dots, \sigsqpsi{L})}{\bnu} (\bnu)
		\propto
			\exp \left\{
				\T{\begin{bmatrix}
					\bnu \\
					\vect (\bnu \T{\bnu})
				\end{bmatrix}}
				\np{p (\bnu | \sigsqmu, \sigsqpsi{1}, \dots, \sigsqpsi{L})}{\bnu}
			\right\},
\label{msg_pen_nu}
\end{equation}

\noindent where

\begin{equation}
	\np{p (\bnu | \sigsqmu, \sigsqpsi{1}, \dots, \sigsqpsi{L})}{\bnu}
		\longleftarrow
			\begin{bmatrix}
				\E_q (\Sigmanu^{-1}) \munu \\
				-\frac12 \vect \left\{ \E_q (\Sigmanu^{-1}) \right\}
			\end{bmatrix}.
\label{np_pen_nu}
\end{equation}

\noindent Once again, we have used a vec-based representation of the message to $\bnu$ as opposed to
a storage-economical vech-based representation. The major reason for this is outlined in the discussion following
\eqref{zeta_tilde}.

The message from $p (\bnu | \sigsqmu, \sigsqpsi{1}, \dots, \sigsqpsi{L})$ to $\sigsqmu$ is

\begin{equation}
	\msg{p (\bnu | \sigsqmu, \sigsqpsi{1}, \dots, \sigsqpsi{L})}{\sigsqmu} (\sigsqmu)
		\propto
			\exp \left\{
				\T{\begin{bmatrix}
					\log (\sigsqmu) \\
					1/\sigsqmu
				\end{bmatrix}} 
				\np{p (\bnu | \sigsqmu, \sigsqpsi{1}, \dots, \sigsqpsi{L})}{\sigsqmu}
			\right\},
\label{msg_pen_sigsqmu}
\end{equation}

\noindent which is an inverse-$\chi^2$ density function after normalization. The update for the natural parameter vector
in \eqref{msg_pen_sigsqmu} is

\begin{equation}
	\np{p (\bnu | \sigsqmu, \sigsqpsi{1}, \dots, \sigsqpsi{L})}{\sigsqmu}
		\longleftarrow
			\begin{bmatrix}
				-\frac{K}{2} \\
				-\frac12 \E_q (\T{\umu} \umu)
			\end{bmatrix}.
\label{np_pen_sigsqmu}
\end{equation}

For $l = 1, \dots, L$, the message from $p (\bnu | \sigsqmu, \sigsqpsi{1}, \dots, \sigsqpsi{L})$ to $\sigsqpsi{l}$
is similar to the message to $\sigsqmu$. The message is

\begin{equation}
	\msg{p (\bnu | \sigsqmu, \sigsqpsi{1}, \dots, \sigsqpsi{L})}{\sigsqmu} (\sigsqpsi{l})
		\propto
			\exp \left\{
				\T{\begin{bmatrix}
					\log (\sigsqpsi{l}) \\
					1/\sigsqpsi{l}
				\end{bmatrix}} 
				\np{p (\bnu | \sigsqmu, \sigsqpsi{1}, \dots, \sigsqpsi{L})}{\sigsqpsi{l}}
			\right\},
\label{msg_pen_sigsqpsi}
\end{equation}

\noindent which is an inverse-$\chi^2$ density function after normalization. The update for the natural parameter vector
in \eqref{msg_pen_sigsqpsi} is

\begin{equation}
	\np{p (\bnu | \sigsqmu, \sigsqpsi{1}, \dots, \sigsqpsi{L})}{\sigsqpsi{l}}
		\longleftarrow
			\begin{bmatrix}
				-\frac{K}{2} \\
				-\frac12 \E_q (\T{\upsi{l}} \upsi{l})
			\end{bmatrix}.
\label{np_pen_sigsqpsi}
\end{equation}

Finally, recall the discussion following \eqref{V_mat}. Each of the messages to the variance parameters $\sigsqmu,
\sigsqpsi{1}, \dots, \sigsqpsi{L}$ must be paired with a graph message. For the same reasons that were used to
justify the graphical message in \eqref{G_lik_sigsqeps}, the graph messages received by $\sigsqmu,
\sigsqpsi{1}, \dots, \sigsqpsi{L}$ are, respectively,

\begin{equation}
	G_{p (\bnu | \sigsqmu, \sigsqpsi{1}, \dots, \sigsqpsi{L}) \rightarrow \sigsqmu}
		\longleftarrow
			G_{\text{full}} \quad
	\text{and} \quad
	G_{p (\bnu | \sigsqmu, \sigsqpsi{1}, \dots, \sigsqpsi{L}) \rightarrow \sigsqpsi{l}}
		\longleftarrow
			G_{\text{full}}, \quad \text{for $l = 1, \dots, L$.}
\label{G_pen}
\end{equation}

Pseudocode for the functional principal component Gaussian penalization fragment is presented in Algorithm
\ref{alg:mean_fpc_gauss_pen_frag}.
A derivation of all the relevant expectations and natural parameter vector updates is provided in Appendix
\ref{app:mean_fpc_gauss_pen_frag}.

\begin{algorithm}
	\caption{
		Pseudocode for the functional principal component Gaussian penalization fragment.
	}
	\label{alg:mean_fpc_gauss_pen_frag}
	\begin{algorithmic}[1]
		\Inputs
			\begin{varwidth}[t]{\linewidth} $
				\np{\bnu}{p (\bnu | \sigsqmu, \sigsqpsi{1}, \dots, \sigsqpsi{L})}, \quad
				\{
					\np{\sigsqmu}{p (\bnu | \sigsqmu, \sigsqpsi{1}, \dots, \sigsqpsi{L})}, \
					G_{\sigsqmu \rightarrow p (\bnu | \sigsqmu, \sigsqpsi{1}, \dots, \sigsqpsi{L})}
				\}
			$\par$
				\{
					\np{\sigsqpsi{l}}{p (\bnu | \sigsqmu, \sigsqpsi{1}, \dots, \sigsqpsi{L})}, \
					G_{\sigsqpsi{l} \rightarrow p (\bnu | \sigsqmu, \sigsqpsi{1}, \dots, \sigsqpsi{L})} :
					l = 1, \dots, L
				\}
			$ \end{varwidth}
		\Updates
			\State Update all expectations with respect to the optimal posterior distribution.
				\Comment{see Appendix \ref{app:mean_fpc_gauss_pen_frag}}
			\State Update $\np{p (\bnu | \sigsqmu, \sigsqpsi{1}, \dots, \sigsqpsi{L})}{\bnu}$
				\Comment{equation \eqref{np_pen_nu}}
			\State Update $\np{p (\bnu | \sigsqmu, \sigsqpsi{1}, \dots, \sigsqpsi{L})}{\sigsqmu}$
				\Comment{equation \eqref{np_pen_sigsqmu}}
			\State Update $G_{p (\bnu | \sigsqmu, \sigsqpsi{1}, \dots, \sigsqpsi{L}) \rightarrow \sigsqmu}$
				\Comment{equation \eqref{G_pen}}
			\For{$l = 1, \dots, L$}
				\State Update $\np{p (\bnu | \sigsqmu, \sigsqpsi{1}, \dots, \sigsqpsi{L})}{\sigsqpsi{l}}$
					\Comment{equation \eqref{np_pen_sigsqpsi}}
				\State Update $G_{p (\bnu | \sigsqmu, \sigsqpsi{1}, \dots, \sigsqpsi{L}) \rightarrow \sigsqpsi{l}}$
					\Comment{equation \eqref{G_pen}}
			\EndFor
		\Outputs
			\begin{varwidth}[t]{\linewidth} $
				\np{p (\bnu | \sigsqmu, \sigsqpsi{1}, \dots, \sigsqpsi{L})}{\bnu}, \quad
				\{
					\np{p (\bnu | \sigsqmu, \sigsqpsi{1}, \dots, \sigsqpsi{L})}{\sigsqmu}, \
					G_{p (\bnu | \sigsqmu, \sigsqpsi{1}, \dots, \sigsqpsi{L}) \rightarrow \sigsqmu}
				\}
			$\par$
				\{
					\np{p (\bnu | \sigsqmu, \sigsqpsi{1}, \dots, \sigsqpsi{L})}{\sigsqpsi{l}}, \
					G_{p (\bnu | \sigsqmu, \sigsqpsi{1}, \dots, \sigsqpsi{L}) \rightarrow \sigsqpsi{l}} :
					l = 1, \dots, L
				\}
			$ \end{varwidth}
	\end{algorithmic}
\end{algorithm}


\section{Post-VMP Steps}
\label{sec:post_vmp_steps}

The FPCA model for curve estimation \eqref{yhat}, which has its genesis in the Karhunen-Lo\`{e}ve decomposition
\eqref{kl_expansion}, relies on orthogonal functional principal component
eigenfunctions and independent vectors of scores with uncorrelated entries.
However, the variational Bayesian FPCA resulting from a VMP treatment
does not enforce any orthogonality restrictions on the resulting eigenfunctions. Although curve estimation is still
valid without these constraints, interpretation of the analysis is more straightforward with orthogonal
eigenfunctions. Furthermore, the eigenfunctions are not guaranteed to be normalized.
In the following sections, we outline
a sequence of post-VMP steps that aid inference and interpretability for variational Bayes-based FPCA.


\subsection{Establishing the Optimal Posterior Density Functions}
\label{sec:opt_dens_funcs}

We are primarily concerned with the optimal posterior density functions for the vector of spline coefficients for
the mean function and eigenfunctions $\bnu$ and the vectors of principal component scores $\bzeta_1, \dots,
\bzeta_n$. Upon convergence of the algorithm, the natural parameter vectors for these optimal posterior density
functions are, according to \eqref{etaq},

\[
	\npq{\bnu} \longleftarrow
		\np{p (\by | \bnu, \bzeta_1, \dots, \bzeta_n, \sigsqeps)}{\bnu}
		+ \np{p (\bnu | \sigsqmu, \sigsqpsi{1}, \dots, \sigsqpsi{L})}{\bnu}
\]

\noindent and

\[
	\npq{\bzeta_i} \longleftarrow
		\np{p (\by | \bnu, \bzeta_1, \dots, \bzeta_n, \sigsqeps)}{\bzeta_i}
		+ \np{p (\bzeta_i)}{\bzeta_i}, \quad
	\text{for $i = 1, \dots, n$.}
\]

\noindent The optimal posterior density for each of these parameters is a Gaussian density
function, where the mean vector $\E_q (\bnu)$ and covariance matrix $\Cov_q (\bnu)$ for $q^* (\bnu)$
can be computed from
\eqref{gauss_vec_comm_params}, and the corresponding parameters $\E_q (\bzeta_i)$ and $\Cov_q (\bzeta_i)$
for $q^* (\bzeta_i)$, $i = 1, \dots, n$, can be
computed from \eqref{gauss_vech_comm_params}. Note that we partition $\E_q (\bnu)$ as

\[
	\E_q (\bnu) = \T{\{ \T{\E_q (\numu)}, \T{\E_q (\nupsi{1})}, \dots, \T{\E_q (\nupsi{L})} \}}
\]

\noindent in a similar fashion to \eqref{partitioned_vectors}.


\subsection{Establishing a Vector Version of the Karhunen-Lo\`{e}ve Decomposition}
\label{sec:biorthogonal}

In this section, we outline a sequence of steps to establish orthogonal
functional principal component eigenfunctions and uncorrelated scores.
Note that we will treat the estimated functional principal component eigenfunctions as fixed curves that
are estimated from the posterior mean of the spline coefficients $\E_q (\bnu)$. As a consequence, the pointwise
posterior variance in the response curve estimates result from the variance in the principal component scores
alone. This treatment is in line with standard approaches to FPCA, where the randomness in the model is
generated by the functional principal component scores  \cite<e.g.>{yao05, benko09}.

Now, we outline the steps to construct orthogonal functional principal component eigenfunctions and
uncorrelated scores. The existence and uniqueness of the eigenfunctions are justified by Theorem \ref{thm:orth_basis}.
First, set up an equidistant grid of design points $\bt_g = \T{(t_{g1}, \dots, t_{gn_g})}$,
where $t_{g1} = 0$, $t_{gn_g} = 1$ and $n_g$ is the length of the grid. Then define $\bC_g$ in an analogous fashion
to \eqref{C_mat}:

\[
	\bC_g \equiv \begin{bmatrix}
		1 & t_{g1} & z_1 (t_{g1}) & \dots & z_K (t_{g1}) \\
		\vdots & \vdots & \vdots & \ddots & \vdots \\
		1 & t_{gn_g} & z_1 (t_{gn_g}) & \dots & z_K (t_{gn_g})
	\end{bmatrix}.
\]

\noindent Establish the posterior estimates of the mean function $\E_q \{ \mu (\bt_g) \} = \bC_g \E_q (\numu)$
and the functional principal components eigenfunctions $\E_q \{ \psi_l (\bt_g) \} = \bC_g \E_q (\nupsi{l})$,
$l = 1, \dots, L$. Then define the matrix $\bPsi$ such that

\[
	\bPsi \equiv \begin{bmatrix} \E_q \{ \psi_1 (t_g) \} & \cdots & \E_q \{ \psi_L (t_g) \} \end{bmatrix}.
\]

\noindent Establish the singular value decomposition of $\bPsi$ such that $\bPsi = \bU_\psi \bD_\psi \T{\bV}_\psi$,
where $\bU_\psi$ is an $n_g \times L$ matrix consisting of the first $L$ left singular vectors of $\bPsi$,
$\bV_\psi$ is an $L \times L$ matrix consisting of the right singular vectors of $\bPsi$, and
$\bD_\psi$ is an $L \times L$ diagonal matrix consisting of the singular values of $\bPsi$.

Next, define

\[
	\bXi \equiv \T{\begin{bmatrix} \E_q (\bzeta_1) & \cdots & \E_q (\bzeta_n) \end{bmatrix}}
\]

\noindent Set $\bm_\zeta$ to be the $L \times 1$ sample mean vector of the columns of $\bD_\psi \T{\bV}_\psi \T{\bXi}$,
and set

\begin{equation}
	\muhat (\bt_g) \equiv \E_q \{ \mu (\bt_g) \} + \bU_\psi \bm_\zeta.
\label{mu_shift}
\end{equation}

\noindent Then set $\bC_\zeta$ to be the $L \times L$ sample covariance matrix of the
columns of $\bD_\psi \T{\bV}_\psi \T{\bXi} - \bm_\zeta \T{\bone_n}$ and establish its spectral decomposition
$\bC_\zeta = \bQ \bLambda \T{\bQ}$, where
$\bLambda$ is a diagonal matrix consisting of the eigenvalues of $\bC_\zeta$ in descending order along its
main diagonal and $\bQ$ is the orthogonal matrix consisting of the corresponding eigenvectors of $\bC_\zeta$ along
its columns.

Finally, define the matrices

\begin{equation}
	\bPsitilde \equiv \bU_\psi \bQ \bLambda^{1/2} \quad
	\text{and} \quad
	\bXitilde \equiv (\bXi \bV_\psi \bD_\psi - \bone_n \T{\bm_\zeta}) \bQ \bLambda^{-1/2}.
\label{psi_zeta_orth}
\end{equation}

\noindent Notice that $\bPsitilde$ is an $n_g \times L$ matrix and $\bXitilde$ is an $n \times L$ matrix. Next, partition
these matrices such that

\[
	\bPsitilde = \begin{bmatrix} \bpsitilde_1 (\bt_g) & \cdots & \bpsitilde_L (\bt_g) \end{bmatrix} \quad
	\text{and} \quad
	\bXitilde = \begin{bmatrix}
		\zetatilde_{11} & \cdots & \zetatilde_{1L} \\
		\vdots & \ddots & \vdots \\
		\zetatilde_{n1} & \cdots & \zetatilde_{nL}
	\end{bmatrix}
\]

\noindent The columns of $\bPsitilde$ are orthonormal vectors, but we require continuous curves that are orthonormal in
$L^2 [0, 1]$. We can adjust this by finding an approximation of $|| \bpsitilde_l ||$, $l = 1, \dots, L$, through numerical
integration. This allows us to establish estimates of the orthonormal functions $\psi^*_1, \dots, \psi^*_L$ in
\eqref{yhat} over the vector $\bt_g$ with

\begin{equation}
	\psihat_l (\bt_g) \equiv \frac{\bpsitilde_l (\bt_g)}{|| \bpsitilde_l ||}, \quad l = 1, \dots, L,
\label{psi_star}
\end{equation}

\noindent as well as estimates of the scores with

\[
	\zetahat_{il} \equiv || \bpsitilde_l || \zetatilde_{il}, \quad i = 1, \dots, n, \quad l = 1, \dots, L.
\]

\noindent Lemma \ref{lem:response_est} outlines the construction of posterior curve estimation for the
response vectors $y_1 (\bt_g),
\dots, y_n (\bt_g)$. Proposition \ref{prop:bi_orthogonal} shows that the form of the predicted response vectors
in Lemma \ref{lem:response_est} is a vector version of the Karhunen-Lo\`{e}ve decomposition. Here, we define
$\bzetahat_i \equiv \T{(\zetahat_{i1}, \dots, \zetahat_{iL})}$, $i = 1, \dots, n$.

\begin{lemma}
	
	The posterior estimate for the response vector $y_i (\bt_g)$ is given by
	
	\begin{equation}
		\yhat_i (\bt_g) = \muhat (\bt_g) + \sum_{l=1}^L \zetahat_{il} \psihat_l (\bt_g), \quad i = 1, \dots, n.
	\label{post_curve_est}
	\end{equation}
	
\label{lem:response_est}
\end{lemma}

\begin{remark}
	
	The posterior estimates $\yhat_1 (\bt_g), \dots, \yhat_n (\bt_g)$ in \eqref{post_curve_est} are
	the same as those prior to the post-processing steps outlined above. That is,
	
	\[
		\yhat_i (\bt_g) = \bC_g \E_q (\numu) + \sum_{l=1}^L \E_q (\zeta_{il}) \bC_g \E_q (\nupsi{l}), \quad i = 1, \dots, n,
	\]
	
	\noindent where $\E_q (\numu)$ is the posterior estimate of $\numu$ from the VMP algorithm and similarly for
	$\E_q (\zeta_{il})$ and $\E_q (\nupsi{l})$. In summary, the post processing steps simply realign the mean function,
	orthogonalize and normalize the eigenfunctions and uncorrelate the scores, but do not affect the fits to the
	observed data.
	
\end{remark}

\begin{proposition}
	
	The vectors $\bzetahat_1, \dots, \bzetahat_N$ are independent and satisfy:
	
	\[
		\frac{1}{n} \sum_{i=1}^n \bzetahat_i = \bzero \quad
		\text{and} \quad
		\frac{1}{n-1} \sum_{i=1}^n \bzetahat_i \bzetahat^{\intercal}_i = \diag \left(
			\left|\left| \psitilde_1 \right|\right|^2, \dots, \left|\left| \psitilde_L \right|\right|^2
		\right).
	\]
	
	\noindent Furthermore, the vectors $\psihat_1 (\bt_g), \dots, \psihat_L (\bt_g)$ are eigenvectors of the sample
	covariance matrix of $\yhat_1 (\bt_g), \dots, \yhat_n (\bt_g)$.
	
\label{prop:bi_orthogonal}
\end{proposition}

\begin{remark}
	
	Proposition \ref{prop:bi_orthogonal} shows that the sample properties of the posterior estimates for the scores
	obey the assumptions of the scores in the Karhunen-Lo\`{e}ve decomposition in \eqref{kl_expansion}.
	Furthermore, the vectors $\psihat_1 (\bt_g), \dots, \psihat_L (\bt_g)$ respect the orthogonality conditions in $\ell^2$.
	Therefore, \eqref{post_curve_est} may be interpreted as a vector version of the truncated
	Karhunen-Lo\`{e}ve decomposition. As a consequence, the numerical estimates of
	$|| \psitilde_l ||^2$, $l = 1, \dots, L$ are the posterior estimates of the eigenvalues of the covariance operator
	$\Sigma$ (see the first paragraph of Section \ref{sec:fpca}).
	
\end{remark}

\noindent The proof of Lemma \ref{lem:response_est} is presented in Appendix \ref{app:proof_lem_response_est},
and the proof of Proposition \ref{prop:bi_orthogonal} is presented in Appendix \ref{app:proof_prop_bi_eigenfunctions}.


\section{Simulations}
\label{sec:sims}

We illustrate the use of Algorithms \ref{alg:fpca_gauss_lik_frag} and \ref{alg:mean_fpc_gauss_pen_frag}
through a series of simulations of model \eqref{bayes_fpca_mod}. Pseudocode for the VMP algorithm is
provided in Algorithm \ref{alg:vmp_alg}.

\begin{algorithm}
	\caption{
		Generic VMP algorithm for the Gaussian response FPCA model \eqref{bayes_fpca_mod} with
		mean field restriction \eqref{fpca_mf_restrn}.
	}
	\label{alg:vmp_alg}
	\begin{algorithmic}[1]
		\Inputs All hyperparameters and observed data
		\Initialize All factor to stochastic node messages.
			\Comment{see \eqref{msg_fact_sn}}
		\Updates
			\While{$\lpyq$ has not converged}
				\State Update all stochastic node to factor messages.
					\Comment{see \eqref{msg_sn_fact}}
				\State Update the fragment for $p (\by | \bnu, \bzeta_1, \dots, \bzeta_n, \sigsqeps)$
					\Comment{see Algorithm \ref{alg:fpca_gauss_lik_frag}}
				\State Update the fragment for $p (\sigsqeps | \aeps)$
					\Comment{see Algorithm 2 of \citeA{maestrini20}}
				\State Update the fragment for $p (\aeps)$
					\Comment{see Algorithm 1 of \citeA{maestrini20}}
				\State Update the fragment for $p (\bnu | \sigsqmu, \sigsqpsi{1}, \dots, \sigsqpsi{L})$
					\Comment{see Algorithm \ref{alg:mean_fpc_gauss_pen_frag}}
				\For{$i = 1, \dots, n$}
					\State Update the fragment for $p (\bzeta_i)$
						\Comment{see Section 4.1.1 of \citeA{wand17}}
				\EndFor
				\State Update the fragment for $p (\sigsqmu | \amu)$
					\Comment{see Algorithm 2 of \citeA{maestrini20}}
				\State Update the fragment for $p (\amu)$
					\Comment{see Algorithm 1 of \citeA{maestrini20}}
				\For{$i = 1, \dots, n$}
					\State Update the fragment for $p (\sigsqpsi{l} | \apsi{l})$
						\Comment{see Algorithm 2 of \citeA{maestrini20}}
					\State Update the fragment for $p (\apsi{l})$
						\Comment{see Algorithm 1 of \citeA{maestrini20}}
				\EndFor
			\EndWhile
			\State Rotate, translate and re-scale $\bPsi$ and $\bXi$.
				\Comment{see Section \ref{sec:biorthogonal}}
		\Outputs $\muhat (\bt_g)$, $\psihat_1 (\bt_g), \dots, \psihat_L (\bt_g)$ and $\bzetahat_1, \dots, \bzetahat_n$.
	\end{algorithmic}
\end{algorithm}

\begin{figure}[t!]
\centering
\tabskip=0pt
\halign{#\cr
	\hbox{%
		\begin{subfigure}{\textwidth}
			\centering
			\includegraphics[width=\textwidth]{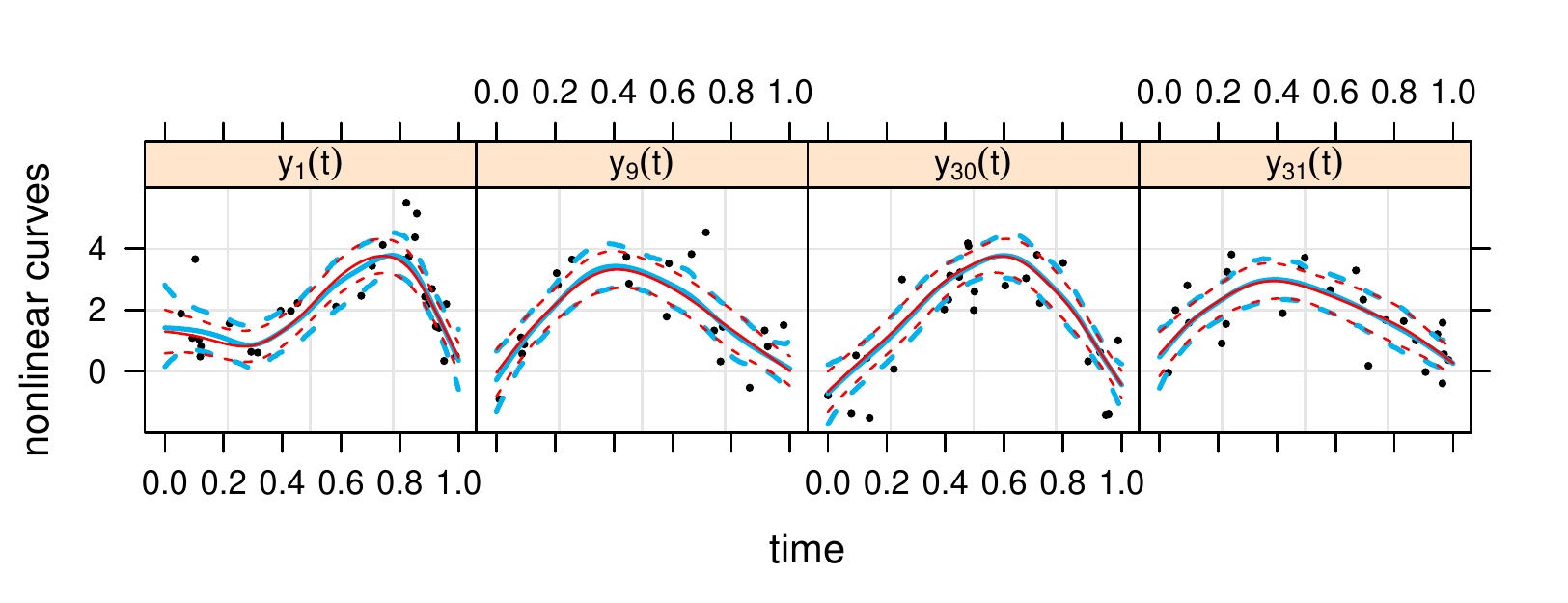}
		\caption{}
		\label{subfig:gauss_response_curves}
		\end{subfigure}%
	}\cr
	\noalign{\vfill}
	\hbox{%
		\begin{subfigure}{\textwidth}
			\centering
			\includegraphics[width=\textwidth]{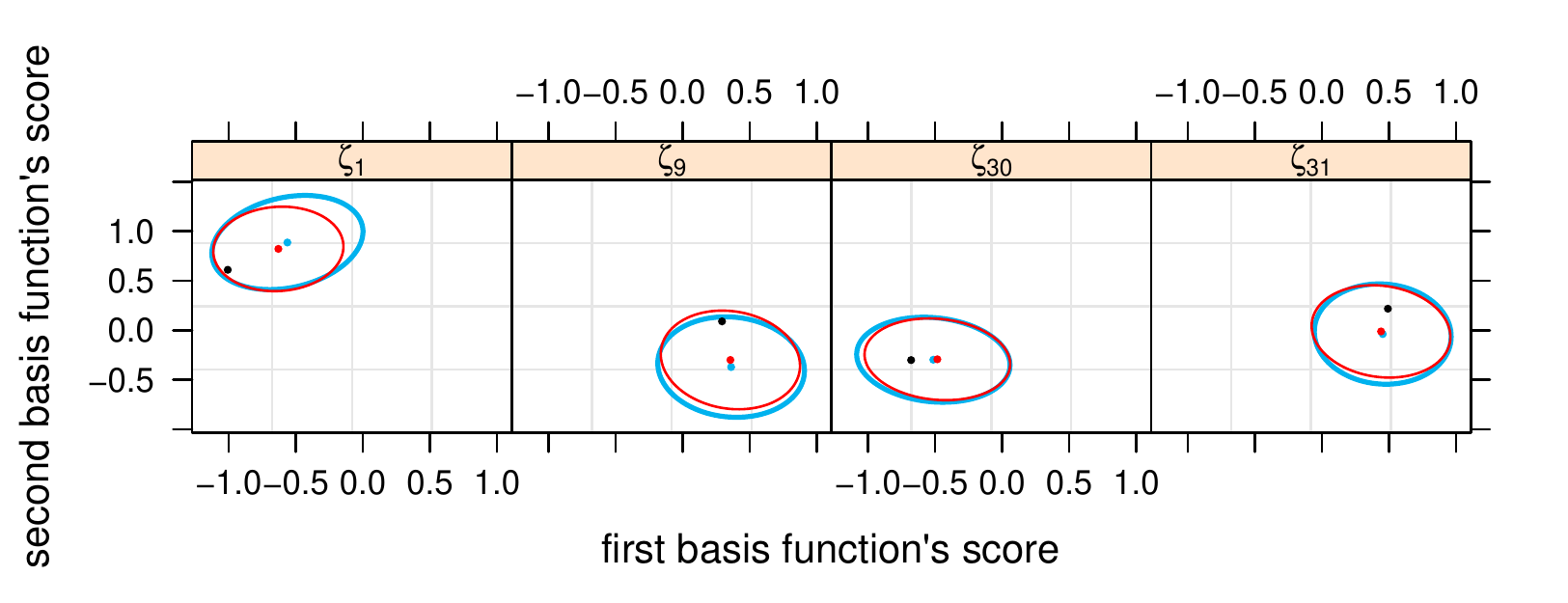}
		\caption{}
		\label{subfig:gauss_response_scores}
		\end{subfigure}%
	}\cr
}
\caption{
	The results from one simulation of the Gaussian response FPCA model in \eqref{bayes_fpca_mod}. The
	simulation parameters are outlined in Section \ref{sec:sims}.
	In \subref{subfig:gauss_response_curves}, the simulated data are shown in black, while the VMP-based
	variational Bayes posterior estimates are presented in red and the corresponding MCMC estimates
	are shown in blue. In each panel, the solid lines represent
	the posterior mean, while the dashed line represents the 95\% pointwise credible sets for the mean.
	In \subref{subfig:gauss_response_scores}, we present the vector of scores for each of the
	randomly selected response curves,
	shown in black, as well as the VMP-based variational Bayes posterior estimates, shown in red, and the
	MCMC-based posterior estimates, shown in blue. The red and blue dots represent the VMP-based
	variational Bayes posterior means and the MCMC-based posterior means, respectively.
	The ellipses represent the 95\% credible contours.
}
\label{fig:gauss_resp_sim}
\end{figure}


\subsection{Accuracy Assessment}
\label{sec:acc_ass}

For model \eqref{bayes_fpca_mod}, we simulated 36 response curves with the number
of observations $T_i$ for the $i$th curve sampled uniformly over $\{ 20, 21, \dots, 30 \}$. Furthermore, the time
observations within the $i$th curve $\{ t_{i1}, \dots, t_{i T_i} \}$ were sampled uniformly over the interval $(0, 1)$,
while the residual variance $\sigsqeps$ was set to 1. The mean function was

\begin{equation}
	\mu (t) = 3 \sin (\pi t),
\label{mean_func_gauss_sim}
\end{equation}

\noindent and the eigenfunctions were

\begin{equation}
	\psi_1 (t) = \sqrt{2} \sin (2 \pi t) \quad
	\text{and} \quad
	\psi_2 (t) = \sqrt{2} \cos (2 \pi t),
\label{bf_gauss_sim}
\end{equation}

\noindent Each vector of principal component scores were simulated according to

\begin{equation}
	\bzeta_i = \begin{bmatrix}
		\zeta_{i1} \\
		\zeta_{i2}
	\end{bmatrix} \indsim \normal \left(
		\begin{bmatrix}
			0 \\
			0
		\end{bmatrix},
		\begin{bmatrix}
			1 & 0 \\
			0 & 0.25
		\end{bmatrix}
	\right), \quad i = 1, \dots, n.
\label{score_sims}
\end{equation}

\noindent Nonparameteric regression with O'Sullivan penalized splines for the nonlinear curves was performed
with $K = 10$. Finally, the simulations were conducted by setting $L = 3$, rather
than 2 (the number of eigenfunctions), to assess the flexibility of the VMP algorithm under slight model misspecification.

The results from the simulation are presented in Figure \ref{fig:gauss_resp_sim}, where a random sample of
four of the functional responses are selected for visual clarity. In addition, we have included the results from an MCMC
treatment of model \eqref{bayes_fpca_mod} in blue for comparison with
the VMP-based variational Bayes fits in red.
MCMC simulations were conducted through \textsf{Rstan}, the \textsf{R} \cite{r20} interface to the probabilistic
programming language \textsf{Stan} \cite{rstan20}.
The variational Bayes fits have good agreement with their MCMC
counterparts, as well as the simulated data.
In particular, the post-VMP procedures that are outlined in Section \ref{sec:post_vmp_steps} neatly
complement the standard VMP algorithm.

We then incorporated five settings for the number of response curves: $n \in \{ 10, 50, 100, 250, 500 \}$. For each of these
settings, we conducted 100 simulations of model \eqref{bayes_fpca_mod} with the aim of analysing the error of
the posterior mean estimates of the mean curve in \eqref{mean_func_gauss_sim} and the functional principal
component eigenfunctions in \eqref{bf_gauss_sim}. The error of each simulation was determined via the
integrated squared error:

\begin{equation}
	\text{ISE} (f, \fhat) = \int_0^1 \left| f (x) - \fhat(x) \right|^2 dx,
\label{ise}
\end{equation}

\noindent where, in our simulations, $f (\cdot)$ represents the true function that generated the data, while $\fhat (\cdot)$
represents the VMP-based variational Bayes posterior mean curve.

\begin{figure}[t!]
\centering
\tabskip=0pt
\halign{#\cr
	\hbox{%
		\begin{subfigure}{\textwidth}
		\centering
			\includegraphics[width=\textwidth]{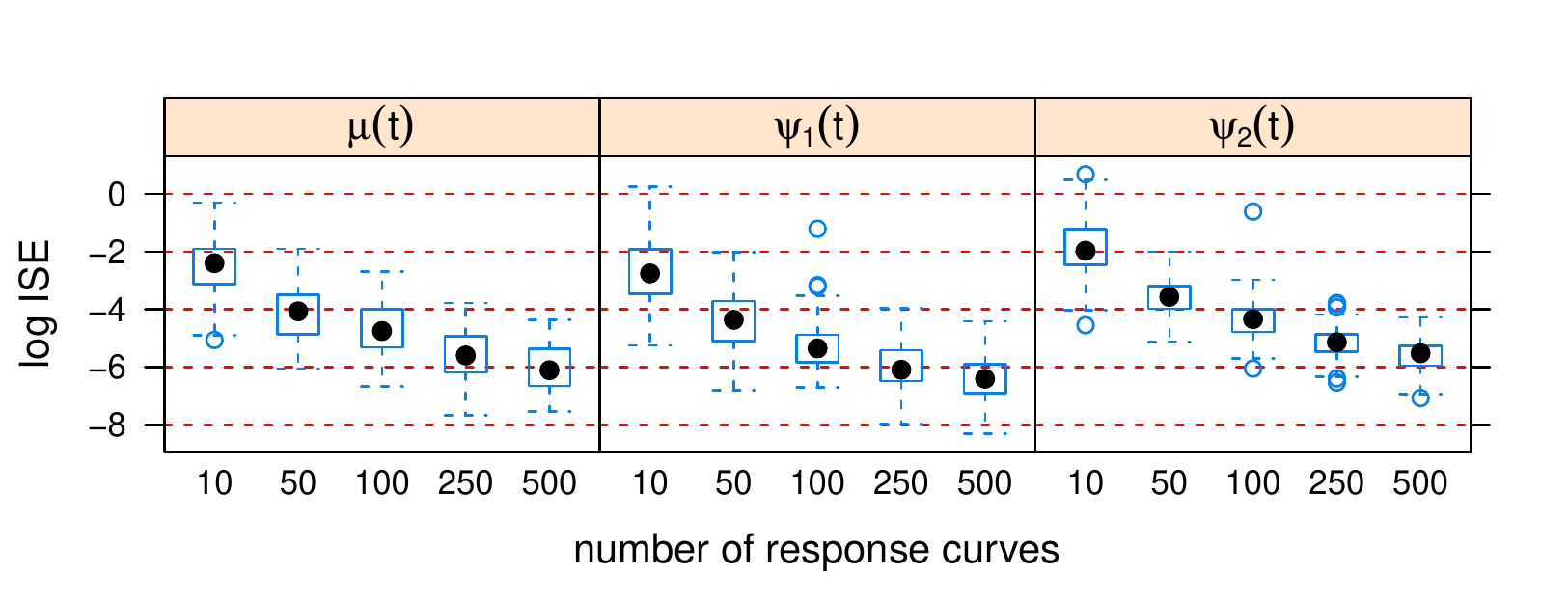}
		\caption{}
		\label{subfig:bf_accs}
		\end{subfigure}%
	}\cr
	\noalign{\vfill}
	\hbox{%
		\begin{subfigure}{\textwidth}
		\centering
			\includegraphics[width=\textwidth]{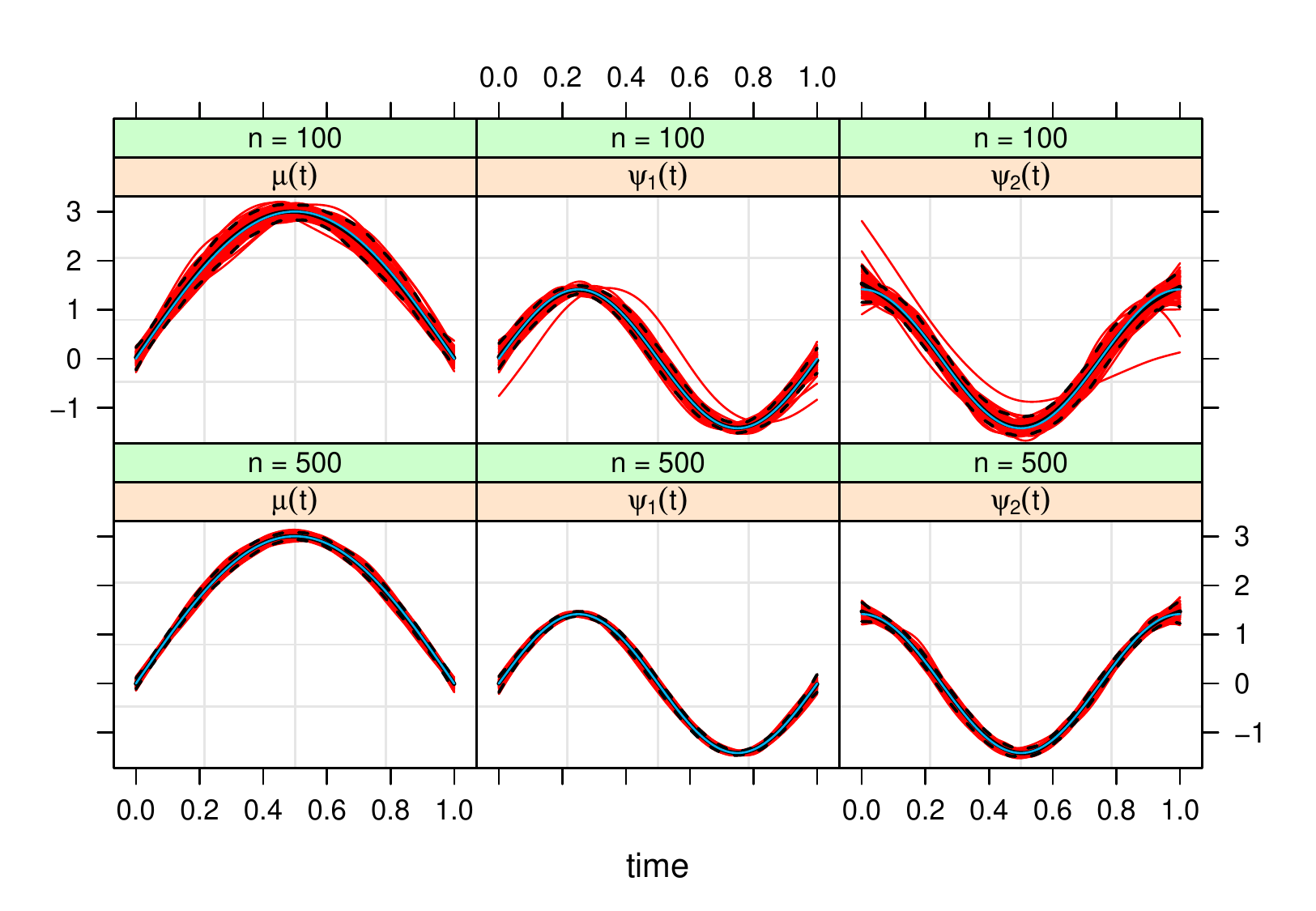}
		\caption{}
		\label{subfig:bf_sims}
		\end{subfigure}%
	}\cr
}
\caption{
	The results from a simulation study of the Gaussian response FPCA model in \eqref{bayes_fpca_mod}. The
	simulation parameters are outlined in Section \ref{sec:acc_ass}.
	The box plots in \subref{subfig:bf_accs} are a summary of the logarithm of the
	integrated squared error values in \eqref{ise}
	for 100 simulations of each of the settings $n \in \{ 10, 50, 100, 250, 500 \}$.
	In \subref{subfig:bf_sims}, we present the results for the mean function and the eigenfunctions when
	$n = 100$ (top row) and $n = 500$ (bottom row). The true functions are shown in blue in each panel,
	and the VMP-based posterior mean curve for each of the simulations is presented in red. In addition, we have
	included the pointwise mean curve and the pointwise 95\% confidence intervals resulting from the MCMC
	posterior estimates for each of the generated datasets in black. Note that, in each panel,
	the pointwise MCMC mean curve overlaps very tightly with the true curve,
	making it difficult to see.
}
\label{fig:gauss_resp_sim_st}
\end{figure}

The box plots for the logarithm of the integrated squared error values
in Figure \ref{fig:gauss_resp_sim_st} \subref{subfig:bf_accs} reflect the excellent results for the
settings where $n = 50, \ 100, \ 250 \ \text{and} \ 500$.
Overall, the results for the setting where $n = 10$ are good, however,
there are a few simulations where the posterior estimates of the second functional principal component
eigenfunction $\psi_2 (\cdot)$ are poor. This is to be expected because the scores
associated with this eigenfunction were generated from a $\normal (0, 0.25)$ distribution reflecting its weaker
contribution to the data generation process. Also, as expected, the error scores for all curves tend to decline with
increasing $n$. In Figure \ref{fig:gauss_resp_sim_st} \subref{subfig:bf_sims}, we present all of the simulated
posterior mean curves of
the mean function and the functional principal component eigenfunctions, for the case where $n = 100$
and $n = 500$, which are
overlaid with the true functions in blue. These plots demonstrate the strength of the VMP algorithm for estimating
the underlying curves that generate an observed set of functional data. Furthermore, the variability in the curve
estimates is drastically reduced with increasing $n$. In addition to each of the VMP posterior
estimates, we have included the pointwise mean curve and the pointwise 95\% confidence interval for the MCMC
simulations for each of the generated data sets. Evidently, there is strong agreement between the VMP simulations
and the MCMC simulations.


\subsection{Computational Speed Comparisons}
\label{sec:speed_comp}

In the previous section, we saw that the mean field product restriction in \eqref{fpca_mf_restrn} does not
compromise the accuracy of variational Bayesian inference for FPCA. However,
the major advantage offered by variational Bayesian inference via VMP is fast approximate inference in 
comparison to MCMC simulations. Several published articles have addressed the computational speed gains
from using variational Bayesian inference. \citeA{faes11} presented speed gains for parametric and nonparametric
regression with missing data, \citeA{luts15} presented timing comparisons for semiparametric regression
models with count responses, and
\citeA{lee16} and \citeA{nolanmw20} established speed gains for multilevel data models
with streamlined matrix algebraic results.
In all cases, the variational Bayesian inference algorithms had strong accuracy in
comparison to MCMC simulations and were far superior in computational speed.

In Table \ref{tab:speed_comp}, we present a similar set of results for the computational speed of VMP and MCMC
for model \eqref{bayes_fpca_mod}. The simulations were identical to those that were used to generate the
results in Figure \ref{fig:gauss_resp_sim_st}, where there were 100 simulations over five settings for the number
of response curves $n \in \{10, 50, 100, 250, 500\}$. In addition, the simulations were performed on a laptop computer
with 8 GB of random access memory and a 1.6 GHz processor. In Table \ref{tab:speed_comp},
we present the median elapsed computing time (in seconds),
with the first quartile and the third quartile shown in brackets.
Notice that most of the VMP simulations are completed within 1 minute, whereas the elapsed computing time
for the MCMC simulations tends to vary from approximately 1 minute, for $n = 10$, to over an hour, for $n = 500$.
The most impressive results are in the fourth column, where the median VMP simulation is 19.6 times faster
than the median MCMC simulation for $n = 10$, 34.3 times faster for $n = 50$, 37.9 times faster for $n = 100$,
49.0 times faster for $n = 250$ and 59.8 times faster for $n = 500$.

\begin{table}
\begin{center}
\begin{tcolorbox}[size=tight,on line,left=0mm,right=0mm,width=0.9\textwidth,bottom=0mm,top=1mm,arc=0mm,outer arc=0pt, box align=center,boxrule=1.5pt]
\captionsetup{width=0.9\textwidth}
\captionof{table}{
	Median (first quartile, third quartile) elapsed computing time in seconds for VMP and MCMC  with
	$n \in (10, 50, 100, 250, 500)$. The fourth column presents the ratio of the median elapsed time for MCMC
	to the median elapsed time for VMP.
}
\begin{tabularx}{\textwidth}{X c X | X c X | X c X | X c X}
  \rowcolor[gray]{.8}
  & $n$ & & & VMP & & & MCMC & & & MCMC/VMP & \\
  \rowcolor{white!50}
  & 10 & & & 2.1 (1.3, 2.8) & & & 41.2 (37.3, 45.6) & & & 19.6 & \\
  \rowcolor{white!50}
  & 50 & & & 5.5 (3.4, 9.9) & & & 188.4 (178.8, 213.9) & & & 34.3 & \\
  \rowcolor{white!50}
  & 100 & & & 11.8 (7.0, 18.2) & & & 446.7 (415.1, 475.7) & & & 37.9 & \\
  \rowcolor{white!50}
  & 250 & & & 33.1 (18.1, 48.6) & & & 1620.8 (1446.6, 1864.7) & & & 49.0 & \\
  \rowcolor{white!50}
  & 500 & & & 58.0 (32.1, 91.0) & & & 3471.2 (2832.9, 4497.8) & & & 59.8 & \\
\end{tabularx}
\label{tab:speed_comp}
\end{tcolorbox}
\end{center}
\end{table}

\begin{figure}[t!]
\centering
\tabskip=0pt
\halign{#\cr
	\hbox{%
		\begin{subfigure}{\textwidth}
		\centering
			\includegraphics[height=2.5in, width=6.5in]{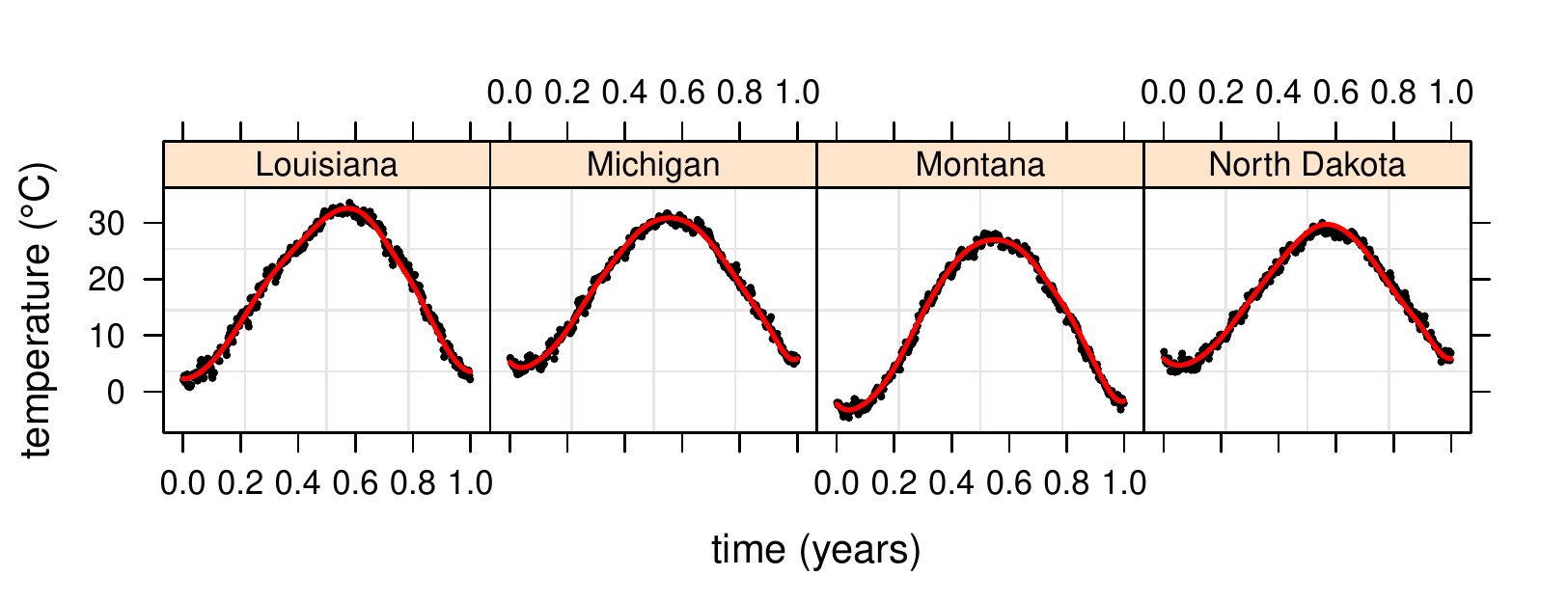}
		\caption{}
		\label{subfig:us_fits}
		\end{subfigure}%
	}\cr
	\noalign{\vfill}
	\hbox{%
		\begin{subfigure}{\textwidth}
		\centering
			\includegraphics[height=5in, width=6.5in]{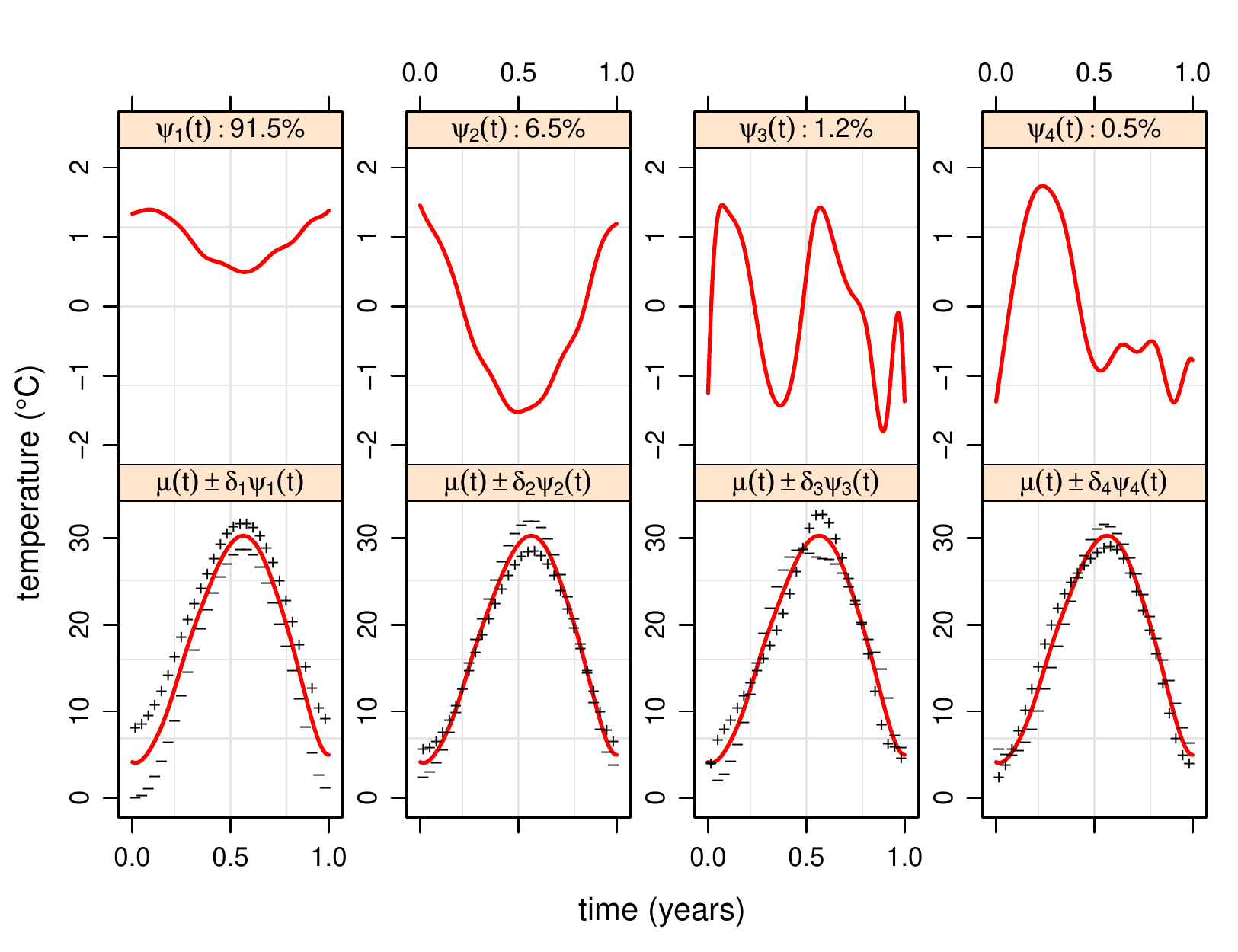}
		\caption{}
		\label{subfig:us_mean_shift}
		\end{subfigure}%
	}\cr
}
\caption{
	Application of the VMP algorithm for FPCA to the United States temperature data. The fits in \subref{subfig:us_fits}
	are for four randomly selected weather stations in the dataset.
	The plots in \subref{subfig:us_mean_shift} present the pointwise posterior mean estimates of the eigenfunctions
	(top panel) and
	show the estimated mean function with
	perturbations from each eigenfunction (bottom panel): $\muhat (t) \pm \delta_l \psihat_l (t)$,
	$l = 1, \dots, 4$.
}
\label{fig:us_weather_data}
\end{figure}


\section{Application: United States Temperature Data}
\label{sec:us_weather_data}

We now provide an illustration of our methodology with an application to temperature data collected from
various United
States weather stations, which is available from the \textsf{rnoaa} package \cite{rnoaa21} in \textsf{R}.
The \textsf{rnoaa}
package is an interface to the National Oceanic and Atmospheric Administration's climate data.
The function \textsf{ghcnd\_stations()} provides access to all available global historical climatology
network daily weather data for each weather site from 1960 to 1994. The information includes the longitude
and latitude for each site, and this was used to determine the state or the federal district of the site.
Our analysis focused on
maximum daily temperature that was averaged over the 25 years of available data.

From this package, we collected full data sets (data available for every day
of the year) from 2837 weather stations, where 49 states and federal districts were represented. For each
state or federal district, we took a random sample of 3 of the available sites. In cases where there were
less than 3 sites available (Rhode Island and District of Columbia), we used all available sites. This resulted
in 145 sites used in our application, with 365 observations for each site.

Chapter 8 of \citeA{ramsay05} conducts a similar analysis of Canadian temperature data from various weather
stations. In their application, they uncovered four functional principal component eigenfunctions.
Similarly, we conducted VMP simulations with $L = 4$.
The results are presented in Figure \ref{fig:us_weather_data}. In Figure \ref{fig:us_weather_data}
\subref{subfig:us_fits}, we display the results of four randomly selected weather stations, from four different
states. There is relatively small residual variability in the observed dataset because we are using long-term averages.
As a consequence, the pointwise 95\% credible sets would not be visible in the plots,
so we have only included the pointwise variational Bayesian posterior means.
In Figure \ref{fig:us_weather_data} \subref{subfig:us_mean_shift}, we present the pointwise posterior estimates
for each eigenfunction (top panel) and the
the effect of perturbing the estimated mean function with each eigenfunction (bottom panel):
$\muhat (t) \pm \delta_l \psihat_l (t)$,
$l = 1, \dots, 4$. The plus (minus) signs indicate the shift that each eigenfunction makes to the mean function
with a positive (negative) perturbation. In addition, the value of $\delta_l$ was simply selected such that the effect
of the perturbation would be visibly apparent. Note that the top and bottom panels in Figure
\ref{fig:us_weather_data} \subref{subfig:us_mean_shift} should be analysed concurrently when determining
the effect of each eigenfunction.

The bottom panel for the first eigenfunction (which accounts for 91.5\% of the total variation)
shows that it is a mean shift that perturbs the mean function in the positive
(negative) direction when it is added (subtracted). The top panel shows that this effect is stronger in the Winter months
than the Summer months, indicating that US temperature is most variable in Winter. Similar analysis of the
second eigenfunction (which accounts for 6.5\% of the total variation) shows that it represents uniformity in the measured
temperatures. It perturbs the mean function in the negative (positive) direction in the Summer (Winter) months when
it is added. As a consequence, weather stations at locations with larger discrepancies
between Winter and Summer temperatures will have a strong and negative score for this eigenfunction.
The third and fourth eigenfunctions are harder to interpret given their weak contributions to the
total variation. The scores associated with the first two eigenfunctions for the displayed weather stations are
(0.21, -1.26) for the weather station in Louisiana, (0.32, -0.30) for the weather station in Michigan, (-5.24, -0.04)
for the weather station in Montana and (-0.66, 0.85) for the weather station in North Dakota.
The scores for the first eigenfunction indicate that slightly higher than average temperatures
are to be expected in Louisiana and Michigan, slightly lower than average temperatures in North Dakota and well
below average temperatures in Montana. Furthermore, the scores for the second eigenfunction show that the
greatest variability between Summer and Winter months can be found
in Louisiana, whereas Montana may appear to be more uniform in comparison to Louisiana. In addition, Michigan
experiences more than average differences in temperature between Summer and Winter, while
the difference in Winter and Summer temperatures in Montana are close to the average differences in the
United States.


\section{Closing Remarks}
\label{sec:closing_remarks}

We have provided a comprehensive overview of Bayesian FPCA with a VMP-based mean field variational Bayes
approach. Our coverage has focused on the Gaussian likelihood specification for the observed data, and it
includes the introduction of two new fragments for VMP:

\begin{enumerate}
	\item the functional principal component Gaussian likelihood fragment (Algorithm \ref{alg:fpca_gauss_lik_frag});
	\item and the functional principal component Gaussian penalization fragment (Algorithm \ref{alg:mean_fpc_gauss_pen_frag}).
\end{enumerate}

\noindent These are directly compatible with the fragment-based computational constructions of VMP outlined in
\citeA{wand17}. This is, to our knowledge, the first VMP construction of a Bayesian FPCA model. In addition,
we have outlined a sequence of post-VMP steps that are necessary for producing orthonormal functional
principal component eigenfunctions and uncorrelated scores.

Simulations were conducted to assess the speed and accuracy of the VMP simulations against MCMC
counterparts. The approximate variational  posterior density functions were in good agreement with the
MCMC estimations, and the VMP algorithm was approximately 20 to 50 times faster than the MCMC
algorithm depending on the number of response curves. An application to a large US temperature dataset
showed that the VMP-based FPCA algorithm can be used for strong inference in big data applications.

This study could be extended to other
functional data models, such as function on scalar or vector regression models, that are yet to be treated
under a VMP-based mean field variational Bayes approach. In addition, extending the likelihood specification to
generalized outcomes would also satisfy a popular area of research in functional data analysis.


\bibliography{bibliography}

\begin{thebibliography}{}

\bibitem [\protect \citeauthoryear {%
Benko%
, H\"{a}rdle%
\BCBL {}\ \BBA {} Kneip%
}{%
Benko%
\ \protect \BOthers {.}}{%
{\protect \APACyear {2009}}%
}]{%
benko09}
\APACinsertmetastar {%
benko09}%
\begin{APACrefauthors}%
Benko, M.%
, H\"{a}rdle, W.%
\BCBL {}\ \BBA {} Kneip, A.%
\end{APACrefauthors}%
\unskip\
\newblock
\APACrefYearMonthDay{2009}{}{}.
\newblock
{\BBOQ}\APACrefatitle {Common functional principal components} {Common
  functional principal components}.{\BBCQ}
\newblock
\APACjournalVolNumPages{The Annals of Statistics}{37}{}{1--34}.
\PrintBackRefs{\CurrentBib}

\bibitem [\protect \citeauthoryear {%
Bishop%
}{%
Bishop%
}{%
{\protect \APACyear {1999}}%
}]{%
bishop99}
\APACinsertmetastar {%
bishop99}%
\begin{APACrefauthors}%
Bishop, C\BPBI M.%
\end{APACrefauthors}%
\unskip\
\newblock
\APACrefYearMonthDay{1999}{}{}.
\newblock
{\BBOQ}\APACrefatitle {Variational Pincipal Components} {Variational pincipal
  components}.{\BBCQ}
\newblock
\BIn{} \APACrefbtitle {Proceedings of the Ninth International Conference on
  Artificial Neural Networks.} {Proceedings of the ninth international
  conference on artificial neural networks.}
\newblock
\APACaddressPublisher{}{Institute of Electrical and Electronics Engineers}.
\PrintBackRefs{\CurrentBib}

\bibitem [\protect \citeauthoryear {%
Bishop%
}{%
Bishop%
}{%
{\protect \APACyear {2006}}%
}]{%
bishop06}
\APACinsertmetastar {%
bishop06}%
\begin{APACrefauthors}%
Bishop, C\BPBI M.%
\end{APACrefauthors}%
\unskip\
\newblock
\APACrefYear{2006}.
\newblock
\APACrefbtitle {{Pattern Recognition and Machine Learning}} {{Pattern
  Recognition and Machine Learning}}.
\newblock
\APACaddressPublisher{New York}{Springer}.
\PrintBackRefs{\CurrentBib}

\bibitem [\protect \citeauthoryear {%
Blei%
, Kucukelbir%
\BCBL {}\ \BBA {} McAuliffe%
}{%
Blei%
\ \protect \BOthers {.}}{%
{\protect \APACyear {2017}}%
}]{%
blei17}
\APACinsertmetastar {%
blei17}%
\begin{APACrefauthors}%
Blei, D\BPBI M.%
, Kucukelbir, A.%
\BCBL {}\ \BBA {} McAuliffe, J\BPBI D.%
\end{APACrefauthors}%
\unskip\
\newblock
\APACrefYearMonthDay{2017}{}{}.
\newblock
{\BBOQ}\APACrefatitle {Variational inference: {A} review for statisticians}
  {Variational inference: {A} review for statisticians}.{\BBCQ}
\newblock
\APACjournalVolNumPages{Journal of the {American} Statistical
  Association}{112}{}{859--877}.
\PrintBackRefs{\CurrentBib}

\bibitem [\protect \citeauthoryear {%
Chamberlain%
\ \protect \BOthers {.}}{%
Chamberlain%
\ \protect \BOthers {.}}{%
{\protect \APACyear {2021}}%
}]{%
rnoaa21}
\APACinsertmetastar {%
rnoaa21}%
\begin{APACrefauthors}%
Chamberlain, S.%
, Anderson, B.%
, Salmon, M.%
, Erickson, A.%
, Potter, N.%
, Stachelek, J.%
\BDBL {}rOpenSci%
\end{APACrefauthors}%
\unskip\
\newblock
\APACrefYearMonthDay{2021}{}{}.
\newblock
\APACrefbtitle {{'NOAA'} Weather Data from {\textsf{R}}.} {{'NOAA'} weather
  data from {\textsf{r}}.}
\newblock
\begin{APACrefURL} \url{https://docs.ropensci.org/rnoaa/} \end{APACrefURL}
\newblock
\APACrefnote{\textsf{R} package version 1.3.0}
\PrintBackRefs{\CurrentBib}

\bibitem [\protect \citeauthoryear {%
Di%
, Crainiceanu%
, Caffo%
\BCBL {}\ \BBA {} Punjabi%
}{%
Di%
\ \protect \BOthers {.}}{%
{\protect \APACyear {2009}}%
}]{%
di09}
\APACinsertmetastar {%
di09}%
\begin{APACrefauthors}%
Di, C\BPBI Z.%
, Crainiceanu, C\BPBI M.%
, Caffo, B\BPBI S.%
\BCBL {}\ \BBA {} Punjabi, N\BPBI M.%
\end{APACrefauthors}%
\unskip\
\newblock
\APACrefYearMonthDay{2009}{}{}.
\newblock
{\BBOQ}\APACrefatitle {Multilevel functional principal component analysis}
  {Multilevel functional principal component analysis}.{\BBCQ}
\newblock
\APACjournalVolNumPages{The Annals of Applied Statistics}{3}{}{458--488}.
\PrintBackRefs{\CurrentBib}

\bibitem [\protect \citeauthoryear {%
Durb\'{a}n%
, Harezlak%
, Wand%
\BCBL {}\ \BBA {} Carroll%
}{%
Durb\'{a}n%
\ \protect \BOthers {.}}{%
{\protect \APACyear {2005}}%
}]{%
durban05}
\APACinsertmetastar {%
durban05}%
\begin{APACrefauthors}%
Durb\'{a}n, M.%
, Harezlak, J.%
, Wand, M\BPBI P.%
\BCBL {}\ \BBA {} Carroll, R\BPBI J.%
\end{APACrefauthors}%
\unskip\
\newblock
\APACrefYearMonthDay{2005}{}{}.
\newblock
{\BBOQ}\APACrefatitle {Simple fitting of subject specific curves for
  longitudinal data} {Simple fitting of subject specific curves for
  longitudinal data}.{\BBCQ}
\newblock
\APACjournalVolNumPages{Statistics in Medicine}{24}{}{1153--1167}.
\PrintBackRefs{\CurrentBib}

\bibitem [\protect \citeauthoryear {%
Faes%
, Ormerod%
\BCBL {}\ \BBA {} Wand%
}{%
Faes%
\ \protect \BOthers {.}}{%
{\protect \APACyear {2011}}%
}]{%
faes11}
\APACinsertmetastar {%
faes11}%
\begin{APACrefauthors}%
Faes, C.%
, Ormerod, J\BPBI T.%
\BCBL {}\ \BBA {} Wand, M\BPBI P.%
\end{APACrefauthors}%
\unskip\
\newblock
\APACrefYearMonthDay{2011}{}{}.
\newblock
{\BBOQ}\APACrefatitle {Variational {Bayesian} inference for parametric and
  nonparametric regression with missing data} {Variational {Bayesian} inference
  for parametric and nonparametric regression with missing data}.{\BBCQ}
\newblock
\APACjournalVolNumPages{Journal of the American Statistical
  Association}{106}{}{959--971}.
\PrintBackRefs{\CurrentBib}

\bibitem [\protect \citeauthoryear {%
Gelman%
}{%
Gelman%
}{%
{\protect \APACyear {2006}}%
}]{%
gelman06}
\APACinsertmetastar {%
gelman06}%
\begin{APACrefauthors}%
Gelman, A.%
\end{APACrefauthors}%
\unskip\
\newblock
\APACrefYearMonthDay{2006}{}{}.
\newblock
{\BBOQ}\APACrefatitle {Prior distributions for variance parameters in
  hierarchical models (comment on article by {B}rowne and {D}raper)} {Prior
  distributions for variance parameters in hierarchical models (comment on
  article by {B}rowne and {D}raper)}.{\BBCQ}
\newblock
\APACjournalVolNumPages{Bayesian Analysis}{1}{}{515--534}.
\PrintBackRefs{\CurrentBib}

\bibitem [\protect \citeauthoryear {%
Gentle%
}{%
Gentle%
}{%
{\protect \APACyear {2007}}%
}]{%
gentle07}
\APACinsertmetastar {%
gentle07}%
\begin{APACrefauthors}%
Gentle, J\BPBI E.%
\end{APACrefauthors}%
\unskip\
\newblock
\APACrefYear{2007}.
\newblock
\APACrefbtitle {{Matrix Algebra}} {{Matrix Algebra}}.
\newblock
\APACaddressPublisher{New York}{Springer}.
\PrintBackRefs{\CurrentBib}

\bibitem [\protect \citeauthoryear {%
Goldsmith%
, Greven%
\BCBL {}\ \BBA {} Crainiceanu%
}{%
Goldsmith%
\ \protect \BOthers {.}}{%
{\protect \APACyear {2013}}%
}]{%
goldsmith13}
\APACinsertmetastar {%
goldsmith13}%
\begin{APACrefauthors}%
Goldsmith, J.%
, Greven, S.%
\BCBL {}\ \BBA {} Crainiceanu, C.%
\end{APACrefauthors}%
\unskip\
\newblock
\APACrefYearMonthDay{2013}{}{}.
\newblock
{\BBOQ}\APACrefatitle {Corrected confidence bands for functional data using
  principal components} {Corrected confidence bands for functional data using
  principal components}.{\BBCQ}
\newblock
\APACjournalVolNumPages{Biometrics}{69}{}{41--51}.
\PrintBackRefs{\CurrentBib}

\bibitem [\protect \citeauthoryear {%
Goldsmith%
\ \BBA {} Kitago%
}{%
Goldsmith%
\ \BBA {} Kitago%
}{%
{\protect \APACyear {2016}}%
}]{%
goldsmith16}
\APACinsertmetastar {%
goldsmith16}%
\begin{APACrefauthors}%
Goldsmith, J.%
\BCBT {}\ \BBA {} Kitago, T.%
\end{APACrefauthors}%
\unskip\
\newblock
\APACrefYearMonthDay{2016}{}{}.
\newblock
{\BBOQ}\APACrefatitle {Assessing systematic effects of stroke on motorcontrol
  by using hierarchical function-on-scalar regression} {Assessing systematic
  effects of stroke on motorcontrol by using hierarchical function-on-scalar
  regression}.{\BBCQ}
\newblock
\APACjournalVolNumPages{Journal of the Royal Statistical Society. Series C,
  Applied statistics}{65}{}{215--236}.
\PrintBackRefs{\CurrentBib}

\bibitem [\protect \citeauthoryear {%
Goldsmith%
, Zippunnikov%
\BCBL {}\ \BBA {} Schrack%
}{%
Goldsmith%
\ \protect \BOthers {.}}{%
{\protect \APACyear {2015}}%
}]{%
Goldsmith15}
\APACinsertmetastar {%
Goldsmith15}%
\begin{APACrefauthors}%
Goldsmith, J.%
, Zippunnikov, V.%
\BCBL {}\ \BBA {} Schrack, J.%
\end{APACrefauthors}%
\unskip\
\newblock
\APACrefYearMonthDay{2015}{}{}.
\newblock
{\BBOQ}\APACrefatitle {Generalized multilevel function-on-scalar regression and
  principal component analysis} {Generalized multilevel function-on-scalar
  regression and principal component analysis}.{\BBCQ}
\newblock
\APACjournalVolNumPages{Biometrics}{71}{}{344--353}.
\PrintBackRefs{\CurrentBib}

\bibitem [\protect \citeauthoryear {%
Greven%
, Crainiceanu%
, Caffo%
\BCBL {}\ \BBA {} Reich%
}{%
Greven%
\ \protect \BOthers {.}}{%
{\protect \APACyear {2011}}%
}]{%
greven2011}
\APACinsertmetastar {%
greven2011}%
\begin{APACrefauthors}%
Greven, S.%
, Crainiceanu, C.%
, Caffo, B.%
\BCBL {}\ \BBA {} Reich, D.%
\end{APACrefauthors}%
\unskip\
\newblock
\APACrefYearMonthDay{2011}{}{}.
\newblock
{\BBOQ}\APACrefatitle {Longitudinal functional principal component analysis}
  {Longitudinal functional principal component analysis}.{\BBCQ}
\newblock
\BIn{} \APACrefbtitle {Recent Advances in Functional Data Analysis and Related
  Topics} {Recent advances in functional data analysis and related topics}\
  (\BPGS\ 149--154).
\newblock
\APACaddressPublisher{}{Springer}.
\PrintBackRefs{\CurrentBib}

\bibitem [\protect \citeauthoryear {%
Jaakkola%
\ \BBA {} Jordan%
}{%
Jaakkola%
\ \BBA {} Jordan%
}{%
{\protect \APACyear {2000}}%
}]{%
jaakkola00}
\APACinsertmetastar {%
jaakkola00}%
\begin{APACrefauthors}%
Jaakkola, T\BPBI S.%
\BCBT {}\ \BBA {} Jordan, M\BPBI I.%
\end{APACrefauthors}%
\unskip\
\newblock
\APACrefYearMonthDay{2000}{}{}.
\newblock
{\BBOQ}\APACrefatitle {Bayesian parameter estimation via variational methods}
  {Bayesian parameter estimation via variational methods}.{\BBCQ}
\newblock
\APACjournalVolNumPages{Statistics and Computing}{10}{}{25--37}.
\PrintBackRefs{\CurrentBib}

\bibitem [\protect \citeauthoryear {%
James%
, Hastie%
\BCBL {}\ \BBA {} Sugar%
}{%
James%
\ \protect \BOthers {.}}{%
{\protect \APACyear {2000}}%
}]{%
james2000}
\APACinsertmetastar {%
james2000}%
\begin{APACrefauthors}%
James, G\BPBI M.%
, Hastie, T\BPBI J.%
\BCBL {}\ \BBA {} Sugar, C\BPBI A.%
\end{APACrefauthors}%
\unskip\
\newblock
\APACrefYearMonthDay{2000}{}{}.
\newblock
{\BBOQ}\APACrefatitle {Principal component models for sparse functional data}
  {Principal component models for sparse functional data}.{\BBCQ}
\newblock
\APACjournalVolNumPages{Biometrika}{87}{3}{587--602}.
\PrintBackRefs{\CurrentBib}

\bibitem [\protect \citeauthoryear {%
Knowles%
\ \BBA {} Minka%
}{%
Knowles%
\ \BBA {} Minka%
}{%
{\protect \APACyear {2011}}%
}]{%
knowles11}
\APACinsertmetastar {%
knowles11}%
\begin{APACrefauthors}%
Knowles, D\BPBI A.%
\BCBT {}\ \BBA {} Minka, T.%
\end{APACrefauthors}%
\unskip\
\newblock
\APACrefYearMonthDay{2011}{}{}.
\newblock
{\BBOQ}\APACrefatitle {Non-conjugate variational message passing for
  multinomial and binary regression} {Non-conjugate variational message passing
  for multinomial and binary regression}.{\BBCQ}
\newblock
\BIn{} \APACrefbtitle {Advances in Neural Information Processing Systems}
  {Advances in neural information processing systems}\ (\BPGS\ 1701--1709).
\PrintBackRefs{\CurrentBib}

\bibitem [\protect \citeauthoryear {%
Lee%
\ \BBA {} Wand%
}{%
Lee%
\ \BBA {} Wand%
}{%
{\protect \APACyear {2016}}%
}]{%
lee16}
\APACinsertmetastar {%
lee16}%
\begin{APACrefauthors}%
Lee, C\BPBI Y\BPBI Y.%
\BCBT {}\ \BBA {} Wand, M\BPBI P.%
\end{APACrefauthors}%
\unskip\
\newblock
\APACrefYearMonthDay{2016}{}{}.
\newblock
{\BBOQ}\APACrefatitle {Streamlined mean field variational {Bayes} for
  longitudinal and multilevel data analysis.} {Streamlined mean field
  variational {Bayes} for longitudinal and multilevel data analysis.}{\BBCQ}
\newblock
\APACjournalVolNumPages{Biometrical Journal}{58}{}{868--895}.
\PrintBackRefs{\CurrentBib}

\bibitem [\protect \citeauthoryear {%
Luts%
\ \BBA {} Wand%
}{%
Luts%
\ \BBA {} Wand%
}{%
{\protect \APACyear {2015}}%
}]{%
luts15}
\APACinsertmetastar {%
luts15}%
\begin{APACrefauthors}%
Luts, J.%
\BCBT {}\ \BBA {} Wand, M\BPBI P.%
\end{APACrefauthors}%
\unskip\
\newblock
\APACrefYearMonthDay{2015}{}{}.
\newblock
{\BBOQ}\APACrefatitle {Variational inference for count response semiparametric
  regression.} {Variational inference for count response semiparametric
  regression.}{\BBCQ}
\newblock
\APACjournalVolNumPages{Bayesian Analysis}{10}{}{991--1023}.
\PrintBackRefs{\CurrentBib}

\bibitem [\protect \citeauthoryear {%
Maestrini%
\ \BBA {} Wand%
}{%
Maestrini%
\ \BBA {} Wand%
}{%
{\protect \APACyear {2018}}%
}]{%
maestrini18}
\APACinsertmetastar {%
maestrini18}%
\begin{APACrefauthors}%
Maestrini, L.%
\BCBT {}\ \BBA {} Wand, M\BPBI P.%
\end{APACrefauthors}%
\unskip\
\newblock
\APACrefYearMonthDay{2018}{}{}.
\newblock
{\BBOQ}\APACrefatitle {Variational message passing for skew $t$ regression}
  {Variational message passing for skew $t$ regression}.{\BBCQ}
\newblock
\APACjournalVolNumPages{Stat}{7}{}{e196}.
\PrintBackRefs{\CurrentBib}

\bibitem [\protect \citeauthoryear {%
Maestrini%
\ \BBA {} Wand%
}{%
Maestrini%
\ \BBA {} Wand%
}{%
{\protect \APACyear {2020}}%
}]{%
maestrini20}
\APACinsertmetastar {%
maestrini20}%
\begin{APACrefauthors}%
Maestrini, L.%
\BCBT {}\ \BBA {} Wand, M\BPBI P.%
\end{APACrefauthors}%
\unskip\
\newblock
\APACrefYearMonthDay{2020}{}{}.
\newblock
{\BBOQ}\APACrefatitle {The {Inverse G-Wishart} distribution and variational
  message passing} {The {Inverse G-Wishart} distribution and variational
  message passing}.{\BBCQ}
\newblock
\APACjournalVolNumPages{arXiv e-prints}{}{}{arXiv:2005.09876}.
\PrintBackRefs{\CurrentBib}

\bibitem [\protect \citeauthoryear {%
{McLean}%
\ \BBA {} Wand%
}{%
{McLean}%
\ \BBA {} Wand%
}{%
{\protect \APACyear {2019}}%
}]{%
mclean19}
\APACinsertmetastar {%
mclean19}%
\begin{APACrefauthors}%
{McLean}, M\BPBI W.%
\BCBT {}\ \BBA {} Wand, M\BPBI P.%
\end{APACrefauthors}%
\unskip\
\newblock
\APACrefYearMonthDay{2019}{}{}.
\newblock
{\BBOQ}\APACrefatitle {Variational message passing for elaborate response
  regression models} {Variational message passing for elaborate response
  regression models}.{\BBCQ}
\newblock
\APACjournalVolNumPages{Bayesian Analysis}{14}{}{371--398}.
\PrintBackRefs{\CurrentBib}

\bibitem [\protect \citeauthoryear {%
Menictas%
\ \BBA {} Wand%
}{%
Menictas%
\ \BBA {} Wand%
}{%
{\protect \APACyear {2013}}%
}]{%
menictas13}
\APACinsertmetastar {%
menictas13}%
\begin{APACrefauthors}%
Menictas, M.%
\BCBT {}\ \BBA {} Wand, M\BPBI P.%
\end{APACrefauthors}%
\unskip\
\newblock
\APACrefYearMonthDay{2013}{}{}.
\newblock
{\BBOQ}\APACrefatitle {Variational inference for marginal longitudinal
  semiparametric regression} {Variational inference for marginal longitudinal
  semiparametric regression}.{\BBCQ}
\newblock
\APACjournalVolNumPages{Stat}{2}{}{61--71}.
\PrintBackRefs{\CurrentBib}

\bibitem [\protect \citeauthoryear {%
Minka%
}{%
Minka%
}{%
{\protect \APACyear {2005}}%
}]{%
minka05}
\APACinsertmetastar {%
minka05}%
\begin{APACrefauthors}%
Minka, T.%
\end{APACrefauthors}%
\unskip\
\newblock
\APACrefYearMonthDay{2005}{}{}.
\newblock
\APACrefbtitle {Divergence measures and message passing} {Divergence measures
  and message passing}\ \APACbVolEdTR{}{\BTR{}}.
\newblock
\APACaddressInstitution{Cambridge, UK}{Microsoft Research Ltd.}
\PrintBackRefs{\CurrentBib}

\bibitem [\protect \citeauthoryear {%
Monahan%
\ \BBA {} Stefanski%
}{%
Monahan%
\ \BBA {} Stefanski%
}{%
{\protect \APACyear {1992}}%
}]{%
monahan92}
\APACinsertmetastar {%
monahan92}%
\begin{APACrefauthors}%
Monahan, J\BPBI F.%
\BCBT {}\ \BBA {} Stefanski, L\BPBI A.%
\end{APACrefauthors}%
\unskip\
\newblock
\APACrefYearMonthDay{1992}{}{}.
\newblock
{\BBOQ}\APACrefatitle {Normal scale mixture approximations to ${F}^* (z)$ and
  computation of the logistic-normal integral} {Normal scale mixture
  approximations to ${F}^* (z)$ and computation of the logistic-normal
  integral}.{\BBCQ}
\newblock
\BIn{} \APACrefbtitle {{Handbook of the Logistic Distribution}} {{Handbook of
  the Logistic Distribution}}\ (\BPGS\ 529--540).
\PrintBackRefs{\CurrentBib}

\bibitem [\protect \citeauthoryear {%
Nolan%
, Menictas%
\BCBL {}\ \BBA {} Wand%
}{%
Nolan%
\ \protect \BOthers {.}}{%
{\protect \APACyear {2020}}%
}]{%
nolanmw20}
\APACinsertmetastar {%
nolanmw20}%
\begin{APACrefauthors}%
Nolan, T\BPBI H.%
, Menictas, M.%
\BCBL {}\ \BBA {} Wand, M\BPBI P.%
\end{APACrefauthors}%
\unskip\
\newblock
\APACrefYearMonthDay{2020}{}{}.
\newblock
{\BBOQ}\APACrefatitle {Streamlined computing for variational inference with
  higher level random effects} {Streamlined computing for variational inference
  with higher level random effects}.{\BBCQ}
\newblock
\APACjournalVolNumPages{Journal of Machine Learning Research}{21}{}{1--62}.
\PrintBackRefs{\CurrentBib}

\bibitem [\protect \citeauthoryear {%
Nolan%
\ \BBA {} Wand%
}{%
Nolan%
\ \BBA {} Wand%
}{%
{\protect \APACyear {2017}}%
}]{%
nolan17}
\APACinsertmetastar {%
nolan17}%
\begin{APACrefauthors}%
Nolan, T\BPBI H.%
\BCBT {}\ \BBA {} Wand, M\BPBI P.%
\end{APACrefauthors}%
\unskip\
\newblock
\APACrefYearMonthDay{2017}{}{}.
\newblock
{\BBOQ}\APACrefatitle {Accurate logistic variational message passing:
  {A}lgebraic and numerical details} {Accurate logistic variational message
  passing: {A}lgebraic and numerical details}.{\BBCQ}
\newblock
\APACjournalVolNumPages{Stat}{6}{}{102--112}.
\PrintBackRefs{\CurrentBib}

\bibitem [\protect \citeauthoryear {%
Nolan%
\ \BBA {} Wand%
}{%
Nolan%
\ \BBA {} Wand%
}{%
{\protect \APACyear {2020}}%
}]{%
nolan19}
\APACinsertmetastar {%
nolan19}%
\begin{APACrefauthors}%
Nolan, T\BPBI H.%
\BCBT {}\ \BBA {} Wand, M\BPBI P.%
\end{APACrefauthors}%
\unskip\
\newblock
\APACrefYearMonthDay{2020}{}{}.
\newblock
{\BBOQ}\APACrefatitle {Streamlined Solutions to Multilevel Sparse Matrix
  Problems} {Streamlined solutions to multilevel sparse matrix
  problems}.{\BBCQ}
\newblock
\APACjournalVolNumPages{ANZIAM Journal}{62}{}{18--41}.
\PrintBackRefs{\CurrentBib}

\bibitem [\protect \citeauthoryear {%
Ormerod%
\ \BBA {} Wand%
}{%
Ormerod%
\ \BBA {} Wand%
}{%
{\protect \APACyear {2010}}%
}]{%
ormerod10}
\APACinsertmetastar {%
ormerod10}%
\begin{APACrefauthors}%
Ormerod, J\BPBI T.%
\BCBT {}\ \BBA {} Wand, M\BPBI P.%
\end{APACrefauthors}%
\unskip\
\newblock
\APACrefYearMonthDay{2010}{}{}.
\newblock
{\BBOQ}\APACrefatitle {Explaining variational approximations} {Explaining
  variational approximations}.{\BBCQ}
\newblock
\APACjournalVolNumPages{The {American} Statistician}{64}{}{140--153}.
\PrintBackRefs{\CurrentBib}

\bibitem [\protect \citeauthoryear {%
Parisi%
}{%
Parisi%
}{%
{\protect \APACyear {1988}}%
}]{%
parisi88}
\APACinsertmetastar {%
parisi88}%
\begin{APACrefauthors}%
Parisi, G.%
\end{APACrefauthors}%
\unskip\
\newblock
\APACrefYear{1988}.
\newblock
\APACrefbtitle {{Statistical Field Theory}} {{Statistical Field Theory}}.
\newblock
\APACaddressPublisher{Redwood City, CA}{Addison-Wesley}.
\PrintBackRefs{\CurrentBib}

\bibitem [\protect \citeauthoryear {%
Ramsay%
\ \BBA {} Silverman%
}{%
Ramsay%
\ \BBA {} Silverman%
}{%
{\protect \APACyear {2005}}%
}]{%
ramsay05}
\APACinsertmetastar {%
ramsay05}%
\begin{APACrefauthors}%
Ramsay, J\BPBI O.%
\BCBT {}\ \BBA {} Silverman, B\BPBI W.%
\end{APACrefauthors}%
\unskip\
\newblock
\APACrefYear{2005}.
\newblock
\APACrefbtitle {{Functional Data Analysis}} {{Functional Data Analysis}}.
\newblock
\APACaddressPublisher{New York}{Springer}.
\PrintBackRefs{\CurrentBib}

\bibitem [\protect \citeauthoryear {%
{\textsf{R} Core Team}%
}{%
{\textsf{R} Core Team}%
}{%
{\protect \APACyear {2020}}%
}]{%
r20}
\APACinsertmetastar {%
r20}%
\begin{APACrefauthors}%
{\textsf{R} Core Team}.%
\end{APACrefauthors}%
\unskip\
\newblock
\APACrefYearMonthDay{2020}{}{}.
\newblock
{\BBOQ}\APACrefatitle {\textsf{R}: A Language and Environment for Statistical
  Computing} {\textsf{R}: A language and environment for statistical
  computing}{\BBCQ}\ [\bibcomputersoftwaremanual].
\newblock
\APACaddressPublisher{Vienna, Austria}{}.
\newblock
\begin{APACrefURL} \url{https://www.R-project.org/} \end{APACrefURL}
\PrintBackRefs{\CurrentBib}

\bibitem [\protect \citeauthoryear {%
Rohde%
\ \BBA {} Wand%
}{%
Rohde%
\ \BBA {} Wand%
}{%
{\protect \APACyear {2016}}%
}]{%
rohde16}
\APACinsertmetastar {%
rohde16}%
\begin{APACrefauthors}%
Rohde, D.%
\BCBT {}\ \BBA {} Wand, M\BPBI P.%
\end{APACrefauthors}%
\unskip\
\newblock
\APACrefYearMonthDay{2016}{}{}.
\newblock
{\BBOQ}\APACrefatitle {Semiparametric mean field variational {B}ayes: {G}eneral
  principles and numerical issues} {Semiparametric mean field variational
  {B}ayes: {G}eneral principles and numerical issues}.{\BBCQ}
\newblock
\APACjournalVolNumPages{Journal of Machine Learning Research}{17}{}{1--47}.
\PrintBackRefs{\CurrentBib}

\bibitem [\protect \citeauthoryear {%
Ruppert%
, Wand%
\BCBL {}\ \BBA {} Carroll%
}{%
Ruppert%
\ \protect \BOthers {.}}{%
{\protect \APACyear {2003}}%
}]{%
ruppert03}
\APACinsertmetastar {%
ruppert03}%
\begin{APACrefauthors}%
Ruppert, D.%
, Wand, M\BPBI P.%
\BCBL {}\ \BBA {} Carroll, R\BPBI J.%
\end{APACrefauthors}%
\unskip\
\newblock
\APACrefYear{2003}.
\newblock
\APACrefbtitle {{Semiparametric Regression}} {{Semiparametric Regression}}.
\newblock
\APACaddressPublisher{}{Cambridge University Press}.
\PrintBackRefs{\CurrentBib}

\bibitem [\protect \citeauthoryear {%
Ruppert%
, Wand%
\BCBL {}\ \BBA {} Carroll%
}{%
Ruppert%
\ \protect \BOthers {.}}{%
{\protect \APACyear {2009}}%
}]{%
ruppert09}
\APACinsertmetastar {%
ruppert09}%
\begin{APACrefauthors}%
Ruppert, D.%
, Wand, M\BPBI P.%
\BCBL {}\ \BBA {} Carroll, R\BPBI J.%
\end{APACrefauthors}%
\unskip\
\newblock
\APACrefYearMonthDay{2009}{}{}.
\newblock
{\BBOQ}\APACrefatitle {Semiparametric regression during 2003--2007}
  {Semiparametric regression during 2003--2007}.{\BBCQ}
\newblock
\APACjournalVolNumPages{Electronic Journal of Statistics}{3}{}{1193--1256}.
\PrintBackRefs{\CurrentBib}

\bibitem [\protect \citeauthoryear {%
{Stan Development Team}%
}{%
{Stan Development Team}%
}{%
{\protect \APACyear {2020}}%
}]{%
rstan20}
\APACinsertmetastar {%
rstan20}%
\begin{APACrefauthors}%
{Stan Development Team}.%
\end{APACrefauthors}%
\unskip\
\newblock
\APACrefYearMonthDay{2020}{}{}.
\newblock
\APACrefbtitle {\textsf{RStan}: the \textsf{R} interface to \textsf{Stan}.}
  {\textsf{RStan}: the \textsf{R} interface to \textsf{Stan}.}
\newblock
\begin{APACrefURL} \url{http://mc-stan.org/} \end{APACrefURL}
\newblock
\APACrefnote{\textsf{R} package version 2.21.2}
\PrintBackRefs{\CurrentBib}

\bibitem [\protect \citeauthoryear {%
Tipping%
\ \BBA {} Bishop%
}{%
Tipping%
\ \BBA {} Bishop%
}{%
{\protect \APACyear {1999}}%
}]{%
tipping99}
\APACinsertmetastar {%
tipping99}%
\begin{APACrefauthors}%
Tipping, M\BPBI E.%
\BCBT {}\ \BBA {} Bishop, C\BPBI M.%
\end{APACrefauthors}%
\unskip\
\newblock
\APACrefYearMonthDay{1999}{}{}.
\newblock
{\BBOQ}\APACrefatitle {Probabilistic principal component analysis}
  {Probabilistic principal component analysis}.{\BBCQ}
\newblock
\APACjournalVolNumPages{Journal of the Royal Statistical Society, Series
  B}{3}{}{611--622}.
\PrintBackRefs{\CurrentBib}

\bibitem [\protect \citeauthoryear {%
{van der Linde}%
}{%
{van der Linde}%
}{%
{\protect \APACyear {2008}}%
}]{%
vanderlinde08}
\APACinsertmetastar {%
vanderlinde08}%
\begin{APACrefauthors}%
{van der Linde}, A.%
\end{APACrefauthors}%
\unskip\
\newblock
\APACrefYearMonthDay{2008}{}{}.
\newblock
{\BBOQ}\APACrefatitle {Variational {B}ayesian functional {PCA}} {Variational
  {B}ayesian functional {PCA}}.{\BBCQ}
\newblock
\APACjournalVolNumPages{Computational Statistics and Data
  Analysis}{53}{}{517--533}.
\PrintBackRefs{\CurrentBib}

\bibitem [\protect \citeauthoryear {%
Wand%
}{%
Wand%
}{%
{\protect \APACyear {2017}}%
}]{%
wand17}
\APACinsertmetastar {%
wand17}%
\begin{APACrefauthors}%
Wand, M\BPBI P.%
\end{APACrefauthors}%
\unskip\
\newblock
\APACrefYearMonthDay{2017}{}{}.
\newblock
{\BBOQ}\APACrefatitle {Fast approximate inference for arbitrarily large
  semiparametric regression models via message passing (with discussion)} {Fast
  approximate inference for arbitrarily large semiparametric regression models
  via message passing (with discussion)}.{\BBCQ}
\newblock
\APACjournalVolNumPages{Journal of the {American} Statistical
  Association}{112}{}{137--168}.
\PrintBackRefs{\CurrentBib}

\bibitem [\protect \citeauthoryear {%
Wand%
\ \BBA {} Ormerod%
}{%
Wand%
\ \BBA {} Ormerod%
}{%
{\protect \APACyear {2008}}%
}]{%
wand08}
\APACinsertmetastar {%
wand08}%
\begin{APACrefauthors}%
Wand, M\BPBI P.%
\BCBT {}\ \BBA {} Ormerod, J\BPBI T.%
\end{APACrefauthors}%
\unskip\
\newblock
\APACrefYearMonthDay{2008}{}{}.
\newblock
{\BBOQ}\APACrefatitle {On semiparametric regression with {O'Sullivan} penalized
  splines} {On semiparametric regression with {O'Sullivan} penalized
  splines}.{\BBCQ}
\newblock
\APACjournalVolNumPages{{Australian} \& {New Zealand} Journal of
  Statistics}{50}{}{179--198}.
\PrintBackRefs{\CurrentBib}

\bibitem [\protect \citeauthoryear {%
Wang%
, Chiou%
\BCBL {}\ \BBA {} M\"{u}ller%
}{%
Wang%
\ \protect \BOthers {.}}{%
{\protect \APACyear {2016}}%
}]{%
wang16}
\APACinsertmetastar {%
wang16}%
\begin{APACrefauthors}%
Wang, J\BPBI L.%
, Chiou, J\BPBI M.%
\BCBL {}\ \BBA {} M\"{u}ller, H\BPBI G.%
\end{APACrefauthors}%
\unskip\
\newblock
\APACrefYearMonthDay{2016}{}{}.
\newblock
{\BBOQ}\APACrefatitle {Functional data analysis} {Functional data
  analysis}.{\BBCQ}
\newblock
\APACjournalVolNumPages{Annual Review of Statistics and Its
  Applications}{3}{}{257--295}.
\PrintBackRefs{\CurrentBib}

\bibitem [\protect \citeauthoryear {%
Winn%
\ \BBA {} Bishop%
}{%
Winn%
\ \BBA {} Bishop%
}{%
{\protect \APACyear {2005}}%
}]{%
winn05}
\APACinsertmetastar {%
winn05}%
\begin{APACrefauthors}%
Winn, J.%
\BCBT {}\ \BBA {} Bishop, C\BPBI M.%
\end{APACrefauthors}%
\unskip\
\newblock
\APACrefYearMonthDay{2005}{}{}.
\newblock
{\BBOQ}\APACrefatitle {Variational message passing} {Variational message
  passing}.{\BBCQ}
\newblock
\APACjournalVolNumPages{Journal of Machine Learning Research}{6}{}{661--694}.
\PrintBackRefs{\CurrentBib}

\bibitem [\protect \citeauthoryear {%
Xiao%
, Zipunnikov%
, Ruppert%
\BCBL {}\ \BBA {} Crainiceanu%
}{%
Xiao%
\ \protect \BOthers {.}}{%
{\protect \APACyear {2016}}%
}]{%
xiao16}
\APACinsertmetastar {%
xiao16}%
\begin{APACrefauthors}%
Xiao, L.%
, Zipunnikov, V.%
, Ruppert, D.%
\BCBL {}\ \BBA {} Crainiceanu, C.%
\end{APACrefauthors}%
\unskip\
\newblock
\APACrefYearMonthDay{2016}{}{}.
\newblock
{\BBOQ}\APACrefatitle {Fast covariance estimation for high-dimensional
  functional data} {Fast covariance estimation for high-dimensional functional
  data}.{\BBCQ}
\newblock
\APACjournalVolNumPages{Statistics and computing}{26}{1-2}{409--421}.
\PrintBackRefs{\CurrentBib}

\bibitem [\protect \citeauthoryear {%
Yao%
, M\"{u}ller%
\BCBL {}\ \BBA {} Wang%
}{%
Yao%
\ \protect \BOthers {.}}{%
{\protect \APACyear {2005}}%
}]{%
yao05}
\APACinsertmetastar {%
yao05}%
\begin{APACrefauthors}%
Yao, F.%
, M\"{u}ller, H\BPBI G.%
\BCBL {}\ \BBA {} Wang, J\BPBI L.%
\end{APACrefauthors}%
\unskip\
\newblock
\APACrefYearMonthDay{2005}{}{}.
\newblock
{\BBOQ}\APACrefatitle {Functional data analysis for sparse longitudinal data}
  {Functional data analysis for sparse longitudinal data}.{\BBCQ}
\newblock
\APACjournalVolNumPages{Journal of the {American} Statistical
  Association}{100}{}{577--590}.
\PrintBackRefs{\CurrentBib}

\end{thebibliography}


\appendix


\section{Proof of Theorem \ref{thm:orth_basis}}
\label{app:proof_thm_orth_basis}

We first note that

\begin{equation}
	y_i (t) - \mu (t) = \sum_{l=1}^L \zeta_{il} \psi_l (t), \quad i = 1, \dots, n.
\label{centered_kl_expansion}
\end{equation}

\noindent The existence of an orthonormal eigenbasis $\psi^*_1, \dots, \psi^*_L$ can be established via Gram-Schmidt
orthogonalization. We first set

\[
	\phi_1 \equiv \psi_1, \quad
	\phi_j \equiv \psi_j - \sum_{l=1}^{j-1} \frac{\langle \phi_l, \psi_j \rangle}{|| \phi_l ||^2} \phi_l, \quad j = 2, \dots, L.
\]

\noindent Next, set

\[
	\phi^*_j = \frac{\phi_j}{|| \phi_j ||}, \quad j = 1, \dots, L.
\]

\noindent Then $\phi^*_1, \dots, \phi^*_L$ form an orthonormal basis for the span of $\psi_1, \dots, \psi_L$. Therefore,
\eqref{centered_kl_expansion} can be re-written as

\[
	y_i (t) - \mu (t) = \sum_{l=1}^L \iota_{il} \phi^*_l (t), \quad i = 1, \dots, n,
\]

\noindent where

\[
	\iota_{il} \equiv \zeta_{il} ||\phi_l|| + \sum_{j=l+1}^L \zeta_{ij} \frac{\langle \phi_l , \psi_j \rangle}{||\phi_l||}, \quad
	l = 1, \dots, L-1, \quad
	\iota_{iL} \equiv \zeta_{iL} ||\phi_L||.
\]

\noindent Note that $\iota_{i1}, \dots, \iota_{iL}$ are correlated.

Now, define $\biota_i \equiv \T{(\iota_{i1}, \dots, \iota_{iL})}$, $i = 1, \dots, n$. Since the curves $y_1, \dots, y_n$ are
random observations of a Gaussian process, we have

\[
	\biota_i \indsim \normal (0, \bSigma_\iota), \quad i = 1, \dots, n.
\]

\noindent Next, establish the eigendecomposition of $\Sigma_\iota$, such
that $\bSigma_\iota = \bQ_\iota \bLambda_\iota \T{\bQ_\iota}$, where $\bLambda_\iota$
is a diagonal matrix consisting of the eigenvalues of
$\bSigma_\iota$ in descending order, and the columns of $\bQ_\iota$
are the corresponding eigenvectors. Then, it can be easily seen that

\[
	\bzeta^*_i \equiv \T{\bQ_\iota} \iota_i \indsim \normal (0, \bLambda_\iota), \quad i = 1, \dots, n.
\]

\noindent That is, the elements of $\bzeta^*_i$ are uncorrelated and $\bzeta^*_1, \dots, \bzeta^*_n$ are independent.

Next, define the eigenvectors of $\bSigma_\iota$ as $\bq_1, \dots, \bq_L$, such that $\bQ = \bigl[
\begin{smallmatrix} \bq_1 & \cdots & \bq_L \end{smallmatrix} \bigr]$. Furthermore, define the elements of each of
the eigenvectors such that $\bq_l = \T{( q_{l1}, \dots, q_{lL} )}$, $l = 1, \dots, L$. Then, set

\[
	\psi^*_l \equiv \sum_{j=1}^L q_{lj} \phi^*_j, \quad l = 1, \dots, L.
\]

\noindent The orthonormality of $\psi^*_1, \dots, \psi^*_L$ is easily verified:

\begin{align*}
	\langle \psi^*_l , \psi^*_j \rangle
		&= \biggl\langle \sum_{m=1}^L q_{lm} \phi^*_m , \sum_{k=1}^L q_{jk} \phi^*_k \biggr\rangle
		= \sum_{m=1}^L \sum_{k=1}^L q_{lm} q_{jk} \langle \phi^*_m , \phi^*_k \rangle
		= \sum_{k=1}^L q_{lk} q_{jk}
		= \T{\bq_l} \bq_j
		= \ind (l = j),
\end{align*}

\noindent where $\ind (\cdot)$ is the indicator function.

Finally, we have

\[
	y_i (t) - \mu (t)
		= \sum_{l=1}^L \iota_{il} \phi^*_l (t)
		= \sum_{l=1}^L \sum_{j=1}^L \zeta^*_{ij} q_{jl} \phi^*_l (t)
		= \sum_{j=1}^L \zeta^*_{ij} \sum_{l=1}^L q_{jl} \phi^*_l (t)
		= \sum_{j=1}^L \zeta^*_{ij} \ \psi^*_j (t).
\]

\noindent Assumptions \ref{asspn:scores} and \ref{asspn:signs} ensure that this decomposition is unique.


\section{Proof of Lemma \ref{lem:response_est}}
\label{app:proof_lem_response_est}

To prove \eqref{post_curve_est}, we first note that posterior curve estimates from the VMP algorithm satisfy

\begin{align}
\begin{split}
	\yhat_i (\bt_g)
		&= \bC_g \E_q (\numu) + \sum_{l=1}^L \E_q (\zeta_{il}) \bC_g \E_q (\nupsi{l}) \\
		&= \E_q \{ \mu (\bt_g) \} + \sum_{l=1}^L \E_q (\zeta_{il}) \E_q \{ \psi_l (\bt_g) \} \\
		&= \E_q \{ \mu (\bt_g) \} + \bPsi \E_q (\bzeta_i) \\
		&= \E_q \{ \mu (\bt_g) \} + \bU_\psi \bD_\psi \T{\bV_\psi} \E_q (\bzeta_i) \\
		&= [ \E_q \{ \mu (\bt_g) \} + \bU_\psi \bm_\zeta ]
			+ \bU_\psi \{ \bD_\psi \T{\bV_\psi} \E_q (\bzeta_i) - \bm_\zeta \} \\
		&= \muhat (\bt_g) + \bU_\psi \bQ \bLambda^{1/2} \bLambda^{-1/2} \T{\bQ} \{
			\bD_\psi \T{\bV_\psi} \E_q (\bzeta_i) - \bm_\zeta
		\} \\
		&= \muhat (\bt_g) + \bPsitilde \bzetatilde_i,
\end{split}
\label{E_q_y}
\end{align}

\noindent where $\bzetatilde_i \equiv \T{(\zetatilde_{i1}, \dots, \zetatilde_{iL})}$, $i = 1, \dots, n$. Next, define

\[
	\bY \equiv \begin{bmatrix} \yhat_1 (\bt_g) & \dots & \yhat_n (\bt_g) \end{bmatrix}
\]

\noindent Then, \eqref{E_q_y} implies

\[
	\bY - \mu^* (\bt_g) \T{\bone_N} = \bPsitilde \T{\bXitilde}.
\]

\noindent Now, let $\bc$ be the $L \times 1$ vector, with $|| \psitilde_l ||$ as the $l$th entry, $l = 1, \dots, L$.
Furthermore, let $1/\bc$ be the $L \times 1$ vector, with $1/|| \psitilde_l ||$ as the $l$th entry, $l = 1, \dots, L$.
Recall that we can approximate these values through numerical integration. Then,

\[
	\bY - \mu^* (\bt_g) \T{\bone_N} = \bPsitilde \diag (1/\bc) \diag (\bc) \T{\bXitilde}.
\]

\noindent It is easy to see that this implies \eqref{post_curve_est}.


\section{Proof of Proposition \ref{prop:bi_orthogonal}}
\label{app:proof_prop_bi_eigenfunctions}

The independence of $\bzetahat_1, \dots, \bzetahat_n$ is a consequence of the independence assumption in
\eqref{fpca_mf_min_restrn}. Let $\bc$ and $1/\bc$ retain their definitions from Appendix
\ref{app:proof_lem_response_est}. Then, note that

\[
	\bzetahat_i
		= \diag (\bc) \bzetatilde_i
		= \diag (\bc) \bLambda^{-1/2} \T{\bQ} \{ \bD_\psi \T{\bV_\psi} \E_q (\bzeta_i) - \bm_\zeta \}.
\]

\noindent Recall that $\bm_\zeta$ is the mean vector of the columns of $\bD_\psi \T{\bV_\psi} \T{\bXi}$.
Then, it is easy to see that $\sum_{i=1}^n \bzetahat_i = \bzero$. Next,

\begin{align*}
	\sum_{i=1}^n \bzetahat_i \bzetahat^{\intercal}_i
		&= \diag (\bc) \bLambda^{-1/2} \T{\bQ}
			\sum_{i=1}^n \left[
				\{ \bD_\psi \T{\bV_\psi} \E_q (\bzeta_i) - \bm_\zeta \}
					\T{\{ \bD_\psi \T{\bV_\psi} \E_q (\bzeta_i) - \bm_\zeta \}}
			\right]
			\bQ \bLambda^{-1/2} \diag (\bc) \\
		&= (n - 1) \diag (\bc) \bLambda^{-1/2} \T{\bQ} \bC_\zeta \bQ \bLambda^{-1/2} \diag (\bc) \\
		&= (n - 1) \diag (\bc) \bLambda^{-1/2} \T{\bQ} \bQ \bLambda \T{\bQ} \bQ \bLambda^{-1/2} \diag (\bc) \\
		&= (n - 1) \diag (\bc^2),
\end{align*}

\noindent which proves the results for the estimated scores.

From Lemma \ref{lem:response_est}, we have

\[
	\sum_{i=1}^n \yhat_i (\bt_g)
		= \sum_{i=1}^n \left\{ \muhat (\bt_g) + \sum_{l=1}^L \zetahat_{il} \psihat_l (\bt_g) \right\}
		= \sum_{i=1}^n \left\{ \muhat (\bt_g) + \bPsihat \bzetahat_i \right\}
		= n \muhat (\bt_g),
\]

\noindent where $\bPsihat \equiv \bigl[ \begin{smallmatrix} \psihat_1 (\bt_g) & \cdots & \psihat_L (\bt_g)
\end{smallmatrix} \bigr]$.
Therefore, the sample covariance matrix of $\yhat_1 (\bt_g), \dots, \yhat_n (\bt_g)$ is such that

\begin{align*}
	\sum_{i=1}^n \left[ \yhat_i (\bt_g) - \muhat (\bt_g) \right] & \T{\left[ \yhat_i (\bt_g) - \muhat (\bt_g) \right]} \\
		&= \sum_{i=1}^n \left(
			\sum_{l=1}^L \zetahat_{il} \psihat_l (\bt_g)
		\right) \T{\left(
			\sum_{l=1}^L \zetahat_{il} \psihat_l (\bt_g)
		\right)} \\
		&= \sum_{i=1}^n \left( \bPsihat \bzetahat_i \right) \T{\left( \bPsihat \bzetahat_i \right)} \\
		&= \bPsihat \left\{ \sum_{i=1}^n \left( \bzetahat_i \bzeta^{*\intercal}_i \right) \right\} \bPsihat^{\intercal} \\
		&= (n - 1) \bPsihat \diag (\bc^2) \bPsihat^{\intercal}.
\end{align*}

\noindent Simple rearrangement confirms that this is the eigenvalue decomposition of
the sample covariance matrix of $\yhat_1 (\bt_g), \dots, \yhat_n (\bt_g)$, proving the result for the vectors
$\psihat_1 (\bt_g), \dots, \psihat_L (\bt_g)$.


\section{Derivation of the Functional Principal Component Gaussian Likelihood Fragment}
\label{app:fpca_gauss_lik_frag}

From \eqref{fpc_gauss_lik_factor}, we have, for $i = 1, \dots, n$, 

\begin{equation}
	\log p (\by_i | \bnu, \bzeta_i, \sigsqeps) =
		-\frac{T_i}{2} \log (\sigsqeps)
		- \frac{1}{2\sigsqeps} \left|\left|
			\by_i - \bC_i \left( \numu - \sum_{l=1}^L \zeta_{il} \nupsi{l} \right)
		\right|\right|^2
		+ \const
\label{log_fpc_gauss_lik_factor}
\end{equation}

First, we establish the natural parameter vector for each of the optimal posterior density functions. These natural
parameter vectors are essential for determining expectations with respect to the optimal posterior distribution.
According to \eqref{npbf}, the natural parameter vector for $q (\bnu)$ is

\[
	\npbf{p (\by | \bnu, \bzeta_1, \dots, \bzeta_n, \sigsqeps)}{\bnu} =
		\np{p (\by | \bnu, \bzeta_1, \dots, \bzeta_n, \sigsqeps)}{\bnu}
		+ \np{\bnu}{p (\by | \bnu, \bzeta_1, \dots, \bzeta_n, \sigsqeps)},
\]

\noindent the natural parameter vector for $q (\bzeta_i)$, $i = 1, \dots, n$, is

\[
	\npbf{p (\by | \bnu, \bzeta_1, \dots, \bzeta_n, \sigsqeps)}{\bzeta_i} =
		\np{p (\by | \bnu, \bzeta_1, \dots, \bzeta_n, \sigsqeps)}{\bzeta_i}
		+ \np{\bzeta_i}{p (\by | \bnu, \bzeta_1, \dots, \bzeta_n, \sigsqeps)},
\]

\noindent and the natural parameter vector for $q(\sigsqeps)$ is

\[
	\npbf{p (\by | \bnu, \bzeta_1, \dots, \bzeta_n, \sigsqeps)}{\sigsqeps} =
		\np{p (\by | \bnu, \bzeta_1, \dots, \bzeta_n, \sigsqeps)}{\sigsqeps}
		+ \np{\sigsqeps}{p (\by | \bnu, \bzeta_1, \dots, \bzeta_n, \sigsqeps)}.
\]

Next, we consider the updates for standard expectations that occur for each of
the random variables and random vectors in
\eqref{log_fpc_gauss_lik_factor}. For $\bnu$, we need to determine the mean vector $\E_q (\bnu)$
and the covariance matrix $\Cov_q (\bnu)$. The expectations are taken with respect to the normalization
of

\[
	\msg{p (\by | \bnu, \bzeta_1, \dots, \bzeta_n, \sigsqeps)}{\bnu} (\bnu)
	\msg{\bnu}{p (\by | \bnu, \bzeta_1, \dots, \bzeta_n, \sigsqeps)} (\bnu),
\]

\noindent which is a multivariate normal density function with natural parameter vector $\npbf{p (\by | \bnu, \bzeta_1,
\dots, \bzeta_n, \sigsqeps)}{\bnu}$. From \eqref{gauss_vec_comm_params}, we have

\begin{align}
\begin{split}
	\E_q (\bnu)
		&\longleftarrow
			-\frac12 \left[
				\vect^{-1} \left\{
					\left( \npbf{p (\by | \bnu, \bzeta_1, \dots, \bzeta_n, \sigsqeps)}{\bnu} \right)_2
				\right\}
			\right]^{-1} \left( \npbf{p (\by | \bnu, \bzeta_1, \dots, \bzeta_n, \sigsqeps)}{\bnu} \right)_1 \\
	\text{and} \quad
	\Cov_q (\bnu)
		&\longleftarrow
			-\frac12 \left[
				\vect^{-1} \left\{
					\left( \npbf{p (\by | \bnu, \bzeta_1, \dots, \bzeta_n, \sigsqeps)}{\bnu} \right)_2
				\right\}
			\right]^{-1}.
\end{split}
\label{mom_lik_nu}
\end{align}

\noindent Furthermore, the mean vector has the form

\begin{equation}
	\E_q (\bnu) \equiv \T{
		\left\{ \T{\E_q (\numu)}, \T{\E_q (\nupsi{1})}, \dots, \T{\E_q (\nupsi{L})} \right\}
	},
\label{exp_lik_nu}
\end{equation}

\noindent and the covariance matrix has the form

\begin{equation}
	\Cov_q (\bnu) \equiv \begin{bmatrix}
		\Cov_q (\numu) & \Cov_q (\numu, \nupsi{1}) & \dots & \Cov_q (\numu, \nupsi{L}) \\
		\Cov_q (\nupsi{1}, \numu) & \Cov_q (\nupsi{1}) & \dots & \Cov_q (\nupsi{1}, \nupsi{L}) \\
		\vdots & \vdots & \ddots & \vdots \\
		\Cov_q (\nupsi{L}, \numu) & \Cov_q (\nupsi{L}, \nupsi{1}) & \dots & \Cov_q (\nupsi{L}) \\
	\end{bmatrix}.
\label{cov_lik_nu}
\end{equation}

\noindent Similarly, for each $i = 1, \dots, n$, we need to determine the optimal mean vector and covariance matrix
for $\bzeta_i$, which are $\E_q (\bzeta_i)$ and $\Cov_q (\bzeta_i)$, respectively. The expectations are taken with
respect to the normalization of

\[
	\msg{p (\by | \bnu, \bzeta_1, \dots, \bzeta_n, \sigsqeps)}{\bzeta_i} (\bzeta_i)
	\msg{\bzeta_i}{p (\by | \bnu, \bzeta_1, \dots, \bzeta_n, \sigsqeps)} (\bzeta_i),
\]

\noindent which is a multivariate normal density function with natural parameter vector $\npbf{p (\by | \bnu, \bzeta_1,
\dots, \bzeta_n, \sigsqeps)}{\bzeta_i}$. According to \eqref{gauss_vech_comm_params},

\begin{align}
\begin{split}
	\E_q (\bzeta_i)
		&\longleftarrow
			-\frac12 \left[
				\vect^{-1} \left\{
					\bD_L^{+ \intercal}
					\left( \npbf{p (\by | \bnu, \bzeta_1, \dots, \bzeta_n, \sigsqeps)}{\bzeta_i} \right)_2
				\right\}
			\right]^{-1} \left( \npbf{p (\by | \bnu, \bzeta_1, \dots, \bzeta_n, \sigsqeps)}{\bzeta_i} \right)_1 \\
	\text{and} \quad
	\Cov_q (\bzeta_i)
		&\longleftarrow
			-\frac12 \left[
				\vect^{-1} \left\{
					\bD_L^{+ \intercal}
					\left( \npbf{p (\by | \bnu, \bzeta_1, \dots, \bzeta_n, \sigsqeps)}{\bzeta_i} \right)_2
				\right\}
			\right]^{-1}, \quad \text{for $i = 1, \dots, n$.}
\end{split}
\label{exp_lik_zeta}
\end{align}

\noindent Finally, for $\sigsqeps$, we need to determine $\E_q (1/\sigsqeps)$, with the expectation taken with
respect to the normalization of

\[
	\msg{p (\by | \bnu, \bzeta_1, \dots, \bzeta_n, \sigsqeps)}{\sigsqeps} (\sigsqeps)
	\msg{\sigsqeps}{p (\by | \bnu, \bzeta_1, \dots, \bzeta_n, \sigsqeps)} (\sigsqeps).
\]

\noindent This is an inverse-$\chi^2$ density function, with natural parameter vector $\npbf{p (\by | \bnu, \bzeta_1, \dots,
\bzeta_n, \sigsqeps)}{\sigsqeps}$. According to Result 6 of \citeA{maestrini20},

\[
	\E_q (1/\sigsqeps)
		\longleftarrow
			\frac{
				\left( \npbf{p (\by | \bnu, \bzeta_1, \dots, \bzeta_n, \sigsqeps)}{\sigsqeps} \right)_1 + 1
			}{
				\left( \npbf{p (\by | \bnu, \bzeta_1, \dots, \bzeta_n, \sigsqeps)}{\sigsqeps} \right)_2
			}.
\]

Now, we turn our attention to the derivation of the message passed from $p (\by | \bnu, \bzeta_1, \dots, \bzeta_n,
\sigsqeps)$ to $\bnu$. Notice that

\begin{equation}
	\bC_i \left( \numu - \sum_{l=1}^L \zeta_{il} \nupsi{l} \right) = (\T{\bzetatilde_i} \otimes \bC_i) \bnu.
\label{theta_simplification}
\end{equation}

\noindent Therefore, as a function of $\bnu$, \eqref{log_fpc_gauss_lik_factor} can be re-written as

\begin{align*}
	\log p (\by_i | \bnu, \bzeta_i, \sigsqeps)
		& = -\frac{1}{2\sigsqeps} \left|\left|
			\by_i - (\T{\bzetatilde_i} \otimes \bC_i) \bnu
		\right|\right|^2 + \tni{\bnu} \\
		& = \T{\begin{bmatrix}
			\bnu \\
			\vect (\bnu \T{\bnu})
		\end{bmatrix}} \begin{bmatrix}
			\frac{1}{\sigsqeps} \T{(\T{\bzetatilde_i} \otimes \bC_i)} \by_i \\
			-\frac{1}{2\sigsqeps} \vect \left\{
				(\bzetatilde_i \T{\bzetatilde_i}) \otimes (\T{\bC_i} \bC_i)
			\right\}
		\end{bmatrix} + \tni{\bnu}.
\end{align*}

\noindent where, for each $i = 1, \dots, n$, $\zetatilde_i$ is defined in \eqref{zeta_tilde}.
From \eqref{msg_fact_sn} and \eqref{fpc_gauss_lik_factorized}, the message from the factor $p (\by | \bnu,
\bzeta_1, \dots, \bzeta_n, \sigsqeps)$ to $\bnu$ is as given in \eqref{msg_lik_nu}, which is proportional to a
multivariate normal density function. The update for the message's natural parameter vector,
in \eqref{np_lik_nu}, is dependent upon
the mean vector and covariance matrix of $\bzetatilde_i$, which are

\begin{equation}
	\E_q (\zetatilde_i) = \T{\{ 1, \T{\E_q (\bzeta_i)} \}} \quad
	\text{and} \quad
	\Cov_q (\zetatilde_i) = \blockdiag \left\{ 0, \Cov_q (\bzeta_i) \right\}, \quad
	\text{for $i = 1, \dots, n$,}
\label{exp_lik_zeta_tilde}
\end{equation}

\noindent where $\E_q (\bzeta_i)$ and $\Cov_q (\bzeta_i)$ are defined in \eqref{exp_lik_zeta}. Note that
a standard statistical result allows us to write

\begin{equation}
	\E_q (\bzetatilde_i \T{\bzetatilde_i}) =
		\Cov_q (\bzetatilde_i) + \E_q (\bzetatilde_i) \T{\E_q (\bzetatilde_i)}, \quad \text{for $i = 1, \dots, n$.}
\label{E_q_outer_zeta}
\end{equation}

Next, notice that

\begin{equation}
	\sum_{l=1}^L \zeta_{il} \nupsi{l} = \Vpsi \bzeta_i,
\label{zeta_simplification}
\end{equation}

\noindent where $\Vpsi$ is defined in \eqref{psi_related_defns}. Then, for each $i = 1, \dots, n$, the log-density function
in \eqref{log_fpc_gauss_lik_factor} can be represented as a function of $\bzeta_i$ by

\begin{align*}
	\log p (\by_i | \bnu, \bzeta_i, \sigsqeps)
		& = -\frac{1}{2\sigsqeps} \left|\left|
			\by_i - \bC_i \numu - \bC_i \Vpsi \bzeta_i
		\right|\right|^2 + \tni{\bzeta_i} \\
		& = \T{
			\begin{bmatrix}
				\bzeta_i \\
				\vech (\bzeta_i \T{\bzeta_i})
			\end{bmatrix}
		} \begin{bmatrix}
			\frac{1}{\sigsqeps} (\T{\Vpsi} \T{\bC_i} \by_i - \hmupsi{i}) \\
			-\frac{1}{2 \sigsqeps} \T{\bD_L} \vect (\Hpsi{i})
		\end{bmatrix} + \tni{\bzeta_i},
\end{align*}

\noindent where $\hmupsi{i}$ and $\Hpsi{i}$ are also defined in \eqref{psi_related_defns}.
\noindent From \eqref{msg_fact_sn} and \eqref{fpc_gauss_lik_factorized}, the message from the factor
$p (\by | \bnu, \bzeta_1, \dots, \bzeta_n, \sigsqeps)$ to $\bzeta_i$ is as given in \eqref{msg_lik_zeta}, which
is proportional to a multivariate normal density function. The message's natural parameter vector update, in
\eqref{np_lik_zeta}, is dependant on the following expectations that are yet to be determined:

\[
	\E_q (\Vpsi) \quad \text{and} \quad \E_q (\Hpsi{i}), \quad \E_q (\hmupsi{i}), \quad i = 1, \dots, n.
\]

\noindent Now, from \eqref{psi_related_defns},

\begin{equation}
	\E_q (\Vpsi) = \begin{bmatrix}
		\E_q (\nupsi{1}) & \dots & \E_q (\nupsi{L})
	\end{bmatrix},
\label{E_q_Vpsi}
\end{equation}

\noindent where, for $l = 1, \dots, L$, $\E_q (\nupsi{l})$ is defined by \eqref{mom_lik_nu} and \eqref{exp_lik_nu}.
Next, $\E_q (\hmupsi{i})$ is an $L \times 1$ vector, with $l$th component being

\begin{equation}
	\E_q (\hmupsi{i})_l =
		\tr \{ \Cov_q (\numu, \nupsi{l}) \T{\bC_i} \bC_i \}
		+ \T{\E_q (\nupsi{l})} \T{\bC_i} \bC_i \E_q (\numu), \quad
	l = 1, \dots, L,
\label{exp_lik_hmupsi}
\end{equation}

\noindent which depends on sub-vectors of $\E_q (\bnu)$ and sub-blocks of $\Cov_q (\bnu)$ that are defined
in \eqref{exp_lik_nu} and \eqref{cov_lik_nu}, respectively. Finally, $\E_q (\Hpsi{i})$ is an $L \times L$ matrix,
with $(l, l')$ component being

\begin{equation}
	\E_q (\Hpsi{i})_{l, l'} =
		\tr \{ \Cov_q (\nupsi{l'}, \nupsi{l}) \T{\bC_i} \bC_i \}
		+ \T{\E_q (\nupsi{l})} \T{\bC_i} \bC_i \E_q (\nupsi{l'}), \quad
	1 \le l, l' \le L.
\label{exp_lik_Hpsi}
\end{equation}

The final message to consider is the message from $p (\by | \bnu, \bzeta_1, \dots, \bzeta_n, \sigsqeps)$ to
$\sigsqeps$. As a function of $\sigsqeps$, \eqref{log_fpc_gauss_lik_factor} takes the form

\begin{align*}
	\log p (\by_i | \bnu, \bzeta_i, \sigsqeps)
		& = -\frac{T_i}{2} \log (\sigsqeps) - \frac{1}{2 \sigsqeps} \left|\left|
			\by_i - \bC_i \bV \bzetatilde_i
		\right|\right|^2 + \tni{\sigsqeps} \\
		& = \T{
			\begin{bmatrix}
				\log (\sigsqeps) \\
				\frac{1}{\sigsqeps}
			\end{bmatrix}
		} \begin{bmatrix}
			-\frac{T_i}{2} \\
			-\frac12 \left|\left| \by_i - \bC_i \bV \bzetatilde_i \right|\right|^2
		\end{bmatrix} + \tni{\sigsqeps},
\end{align*}

\noindent where $\bV$ is defined in \eqref{V_mat} and, for each $i = 1, \dots, n$, $\zetatilde_i$ is defined in
\eqref{zeta_tilde}.
From \eqref{msg_fact_sn} and \eqref{fpc_gauss_lik_factorized}, the message from $p (\by | \bnu,
\bzeta_1, \dots, \bzeta_n, \sigsqeps)$ to $\sigsqeps$ is as given in \eqref{msg_lik_sigsqeps}, which is proportional
to a inverse-$\chi^2$ density function. The message's natural parameter vector, in \eqref{np_lik_sigsqeps}, depends
on the mean of the square norm $|| \by_i - \bC_i \bV \bzetatilde_i ||^2$, for $i = 1, \dots, n$.
This expectation takes the form

\begin{align*}
	\E_q \left(
		\left|\left| \by_i - \bC_i \bV \bzetatilde_i \right|\right|^2
	\right) =
		& \T{\by_i} \by_i - 2 \T{\E_q (\bzetatilde_i)} \T{\E_q (\bV)} \T{\bC_i} \by_i \\
		& + \tr \left[
			\left\{ \Cov_q (\bzetatilde_i) + \E_q (\bzetatilde_i) \T{\E_q (\bzetatilde_i)} \right\} \E_q (\bH_i)
		\right],
\end{align*}

\noindent where we introduce the matrices

\begin{equation}
	\bH_i \equiv \begin{bmatrix}
		\hmu{i} & \T{\hmupsi{i}} \\
		\hmupsi{i} & \Hpsi{i}
	\end{bmatrix}, \quad
	\text{for $i = 1, \dots, n$},
\label{H_mat}
\end{equation}

\noindent and vectors

\begin{equation}
	\hmu{i} \equiv \T{\numu} \bC_i \bC_i \numu, \quad
	\text{for $i = 1, \dots, n$}.
\label{hmu_vec}
\end{equation}

\noindent For each $i = 1, \dots, n$, the mean vector $\E_q (\bzetatilde_i)$ and $\Cov_q (\bzetatilde_i)$ are defined in
\eqref{exp_lik_zeta_tilde}. However, $\E_q (\bV)$ and $\E_q (\bH_i)$, $i = 1, \dots, n$, are yet to be determined.
From \eqref{V_mat},

\[
	\E_q (\bV) = \begin{bmatrix}
		\E_q (\numu) & \E_q (\nupsi{1}) & \dots & \E_q (\nupsi{L})
	\end{bmatrix},
\]

\noindent where the component mean vectors are defined by \eqref{exp_lik_nu}.
For each $i = 1, \dots, n$, the expectation of $\bH_i$, defined in \eqref{H_mat},
with respect to the optimal posterior distribution is

\[
	\E_q (\bH_i) \equiv \begin{bmatrix}
		\E_q (\hmu{i}) & \T{\E_q (\hmupsi{i})} \\
		\E_q (\hmupsi{i}) & \E_q (\Hpsi{i})
	\end{bmatrix},
\]

\noindent where $\hmu{i}$ is defined in \eqref{hmu_vec} with expected value

\[
	\E_q (\hmu{i}) \equiv
		\tr \{ \Cov_q (\numu) \T{\bC_i} \bC_i \}
		+ \T{\E_q (\numu)} \T{\bC_i} \bC_i \E_q (\numu).
\]

\noindent Furthermore, $\E_q (\hmupsi{i})$ and $\E_q (\Hpsi{i})$ are defined in \eqref{exp_lik_hmupsi} and
\eqref{exp_lik_Hpsi}, respectively.

The FPCA Gaussian likelihood fragment, summarized in Algorithm \ref{alg:fpca_gauss_lik_frag}, is a
proceduralization of these results.


\section{Derivation of the Functional Principal Component Gaussian Penalization Fragment}
\label{app:mean_fpc_gauss_pen_frag}

From \eqref{mean_fpc_gauss_pen_factor}, we have, for $l = 1, \dots, L$,

\begin{align}
\begin{split}
	\log p (\numu, \nupsi{l} | \sigsqmu, \sigsqpsi{l}) =
		& -\frac{K}{2} \log (\sigsqmu) - \frac{K}{2} \log (\sigsqpsi{l})
			- \frac12 \T{(\betamu - \bmu_{\betamu})} \bSigma_{\betamu}^{-1} (\betamu - \bmu_{\betamu}) \\
		& - \frac{1}{2 \sigsqmu} \T{\umu} \umu
			- \frac12 \T{(\betapsi{l} - \bmu_{\betapsi{l}})} \bSigma_{\betapsi{l}}^{-1} (\betapsi{l} - \bmu_{\betapsi{l}})
			- \frac{1}{2 \sigsqpsi{l}} \T{\upsi{l}} \upsi{l}.
\end{split}
\label{log_mean_fpc_gauss_pen_factor}
\end{align}

First, we establish the natural parameter vector for each of the optimal posterior density functions. As explained
in Appendix \ref{app:fpca_gauss_lik_frag}, these natural
parameter vectors are essential for determining expectations with respect to the optimal posterior distribution.
According to \eqref{npbf}, the natural parameter vector for $q (\bnu)$ is

\[
	\npbf{p(\bnu | \sigsqmu, \sigsqpsi{1}, \dots, \sigsqpsi{L})}{\bnu} =
		\np{p(\bnu | \sigsqmu, \sigsqpsi{1}, \dots, \sigsqpsi{L})}{\bnu}
		+ \np{\bnu}{p(\bnu | \sigsqmu, \sigsqpsi{1}, \dots, \sigsqpsi{L})},
\]

\noindent the natural parameter vector for $q (\sigsqmu)$ is

\[
	\npbf{p(\bnu | \sigsqmu, \sigsqpsi{1}, \dots, \sigsqpsi{L})}{\sigsqmu} =
		\np{p(\bnu | \sigsqmu, \sigsqpsi{1}, \dots, \sigsqpsi{L})}{\sigsqmu}
		+ \np{\sigsqmu}{p(\bnu | \sigsqmu, \sigsqpsi{1}, \dots, \sigsqpsi{L})},
\]

\noindent and, for $l = 1, \dots, L$, the natural parameter vector for $q(\sigsqpsi{l})$ is

\[
	\npbf{p(\bnu | \sigsqmu, \sigsqpsi{1}, \dots, \sigsqpsi{L})}{\sigsqpsi{l}} =
		\np{p(\bnu | \sigsqmu, \sigsqpsi{1}, \dots, \sigsqpsi{L})}{\sigsqpsi{l}}
		+ \np{\sigsqpsi{l}}{p(\bnu | \sigsqmu, \sigsqpsi{1}, \dots, \sigsqpsi{L})}.
\]

Next, we consider the updates for standard expectations of each of the random variables and random vectors
that appear in \eqref{log_mean_fpc_gauss_pen_factor}. For $\bnu$, we require the mean vector $\E_q (\bnu)$
and covariance matrix $\Cov_q (\bnu)$ under the optimal posterior distribution. The expectations are taken with
respect to the normalization of

\[
	\msg{p(\bnu | \sigsqmu, \sigsqpsi{1}, \dots, \sigsqpsi{L})}{\bnu} (\bnu)
	\msg{\bnu}{p(\bnu | \sigsqmu, \sigsqpsi{1}, \dots, \sigsqpsi{L})} (\bnu),
\]

\noindent which is a multivariate normal density function with natural parameter vector $\npbf{p(\bnu | \sigsqmu,
\sigsqpsi{1}, \dots, \sigsqpsi{L})}{\bnu}$. From \eqref{gauss_vec_comm_params}, we have

\begin{align}
\begin{split}
	\E_q (\bnu)
		&\longleftarrow
			-\frac12 \left[
				\vect^{-1} \left\{
					\left( \npbf{p(\bnu | \sigsqmu, \sigsqpsi{1}, \dots, \sigsqpsi{L})}{\bnu} \right)_2
				\right\}
			\right]^{-1} \left( \npbf{p(\bnu | \sigsqmu, \sigsqpsi{1}, \dots, \sigsqpsi{L})}{\bnu} \right)_1 \\
	\text{and} \quad
	\Cov_q (\bnu)
		&\longleftarrow
			-\frac12 \left[
				\vect^{-1} \left\{
					\left( \npbf{p(\bnu | \sigsqmu, \sigsqpsi{1}, \dots, \sigsqpsi{L})}{\bnu} \right)_2
				\right\}
			\right]^{-1}.
\end{split}
\label{mom_pen_nu}
\end{align}

\noindent The sub-vectors and sub-matrices of $\E_q (\bnu)$ and $\Cov_q (\bnu)$ are identical to those in
\eqref{exp_lik_nu} and \eqref{cov_lik_nu}, respectively. For the mean and FPC Gaussian penalization fragment,
however, we need to note further sub-vectors and sub-matrices. First,

\begin{equation}
	\E_q (\numu) \equiv \T{\left\{ \T{\E_q (\betamu)}, \T{\E_q (\umu)} \right\}} \quad
	\text{and} \quad
	\E_q (\nupsi{l}) \equiv \T{\left\{ \T{\E_q (\betapsi{l})}, \T{\E_q (\upsi{l})} \right\}}, \quad
	\text{for $l = 1, \dots, L$}
\label{exp_betau}
\end{equation}

\noindent and, second,

\begin{equation}
	\Cov_q (\numu) \equiv \begin{bmatrix}
		\Cov_q (\betamu) & \Cov_q (\betamu, \umu) \\
		\Cov_q (\umu, \betamu) & \Cov_q (\umu)
	\end{bmatrix}
\label{cov_betau_mu}
\end{equation}

\noindent and

\begin{equation}
	\Cov_q (\nupsi{l}) \equiv \begin{bmatrix}
		\Cov_q (\betapsi{l}) & \Cov_q (\betapsi{l}, \upsi{l}) \\
		\Cov_q (\upsi{l}, \betapsi{l}) & \Cov_q (\upsi{l})
	\end{bmatrix}, \quad
	\text{for $l = 1, \dots, L$.}
\label{cov_betau_psi}
\end{equation}

\noindent For $\sigsqmu$, we need to determine
$\E_q (1/\sigsqmu)$, with expectation taken with respect to the normalization of

\[
	\msg{p(\bnu | \sigsqmu, \sigsqpsi{1}, \dots, \sigsqpsi{L})}{\sigsqmu} (\sigsqmu)
	\msg{\sigsqmu}{p(\bnu | \sigsqmu, \sigsqpsi{1}, \dots, \sigsqpsi{L})} (\sigsqmu),
\]

\noindent which is an inverse-$\chi^2$ density function with natural parameter vector
$\npbf{p(\bnu | \sigsqmu, \sigsqpsi{1}, \dots, \sigsqpsi{L})}{\sigsqmu}$. According to Result 6 of \citeA{maestrini20},

\begin{equation}
	\E_q (1/\sigsqmu)
		\longleftarrow
			\frac{
				\left( \npbf{p(\bnu | \sigsqmu, \sigsqpsi{1}, \dots, \sigsqpsi{L})}{\sigsqmu} \right)_1 + 1
			}{
				\left( \npbf{p(\bnu | \sigsqmu, \sigsqpsi{1}, \dots, \sigsqpsi{L})}{\sigsqmu} \right)_2
			}.
\label{exp_pen_sigsqmu}
\end{equation}

\noindent Similar arguments can be used to show that

\begin{equation}
	\E_q (1/\sigsqpsi{l})
		\longleftarrow
			\frac{
				\left( \npbf{p(\bnu | \sigsqpsi{l}, \sigsqpsi{1}, \dots, \sigsqpsi{L})}{\sigsqpsi{l}} \right)_1 + 1
			}{
				\left( \npbf{p(\bnu | \sigsqpsi{l}, \sigsqpsi{1}, \dots, \sigsqpsi{L})}{\sigsqpsi{l}} \right)_2
			}, \quad \text{for $l = 1, \dots, L$.}
\label{exp_pen_sigsqpsi}
\end{equation}

Now, we turn our attention to the derivation of the messages passed from the factor. As a function of $\bnu$,
\eqref{log_mean_fpc_gauss_pen_factor} this can be re-written as

\begin{align*}
	\log p(\bnu | \sigsqmu, \sigsqpsi{1}, \dots, \sigsqpsi{L})
		& = -\frac12 \T{\bnu} \Sigmanu^{-1} \bnu + \T{\bnu} \Sigmanu^{-1} \munu + \tni{\bnu} \\
		& = \T{\begin{bmatrix}
			\bnu \\
			\vect (\bnu \T{\bnu})
		\end{bmatrix}} \begin{bmatrix}
			\Sigmanu^{-1} \munu \\
			-\frac12 \vect (\Sigmanu^{-1})
		\end{bmatrix} + \tni{\bnu},
\end{align*}

\noindent where $\munu$ and $\Sigmanu$ are defined in \eqref{munu_Sigmanu}.
From \eqref{msg_fact_sn}, the message from the factor $p(\bnu | \sigsqmu, \sigsqpsi{1}, \dots, \sigsqpsi{L})$
to $\bnu$ is as given in \eqref{msg_pen_nu}, which is proportional to a multivariate normal density function.
The update for the message's natural parameter vector, in \eqref{np_pen_nu},
is dependant upon the expectation of $\Sigmanu^{-1}$, which is given by

\[
	\E_q (\Sigmanu^{-1}) =
		\blockdiag \left\{
			\begin{bmatrix}
				\bSigma_{\betamu} & \T{\textbf{O}} \\
				\textbf{O} & \E_q (1/\sigsqmu) \bI_K
			\end{bmatrix},
			\blockdiag_{l = 1, \dots, L} \left(
				\begin{bmatrix}
					\bSigma_{\betapsi{l}} & \T{\textbf{O}} \\
					\textbf{O} & \E_q (1/\sigsqpsi{l}) \bI_K
				\end{bmatrix}
			\right)
		\right\},
\]

\noindent where $\E_q (1/\sigsqmu)$ and, for $l = 1, \dots, L$, $\E_q (1/\sigsqpsi{l})$ are defined in
\eqref{exp_pen_sigsqmu} and \eqref{exp_pen_sigsqpsi}, respectively.

As a function of $\sigsqmu$, \eqref{log_mean_fpc_gauss_pen_factor} can be re-written as

\begin{align*}
	\log p(\bnu | \sigsqmu, \sigsqpsi{1}, \dots, \sigsqpsi{L})
		& = -\frac{K}{2} \log (\sigsqmu) - \frac{1}{2\sigsqmu} \T{\umu} \umu + \tni{\sigsqmu} \\
		& = \T{\begin{bmatrix}
			\log (\sigsqmu) \\
			1/\sigsqmu
		\end{bmatrix}} \begin{bmatrix}
			-\frac{K}{2} \\
			-\frac12 \T{\umu} \umu
		\end{bmatrix} + \tni{\sigsqmu}.
\end{align*}

\noindent From \eqref{msg_fact_sn}, the message from the factor $p(\bnu | \sigsqmu, \sigsqpsi{1}, \dots,
\sigsqpsi{L})$ to $\sigsqmu$ is as given in \eqref{msg_pen_sigsqmu}, which is an inverse-$\chi^2$
density function upon normalization. The message's natural parameter vector update in \eqref{np_pen_sigsqmu}
depends on $\E_q (\T{\umu} \umu)$. Standard statistical results and sub-vector and sub-matrix definitions in
\eqref{exp_betau} and \eqref{cov_betau_mu} can be employed to show that

\[
	\E_q (\T{\umu} \umu) = \T{\E_q (\umu)} \E_q (\umu) + \tr \left\{ \Cov_q (\umu) \right\}.
\]

As a function of $\sigsqpsi{l}$, for $l = 1, \dots, L$, \eqref{log_mean_fpc_gauss_pen_factor} can be re-written as

\begin{align*}
	\log p(\bnu | \sigsqmu, \sigsqpsi{1}, \dots, \sigsqpsi{L})
		& = -\frac{K}{2} \log (\sigsqpsi{l}) - \frac{1}{2\sigsqpsi{l}} \T{\upsi{l}} \upsi{l} + \tni{\sigsqpsi{l}} \\
		& = \T{\begin{bmatrix}
			\log (\sigsqpsi{l}) \\
			1/\sigsqpsi{l}
		\end{bmatrix}} \begin{bmatrix}
			-\frac{K}{2} \\
			-\frac12 \T{\upsi{l}} \upsi{l}
		\end{bmatrix} + \tni{\sigsqpsi{l}}.
\end{align*}

\noindent From \eqref{msg_fact_sn}, the message from the factor $p(\bnu | \sigsqmu, \sigsqpsi{1}, \dots,
\sigsqpsi{L})$ to $\sigsqpsi{l}$ is as given in \eqref{msg_pen_sigsqpsi}, which is an inverse-$\chi^2$
density function upon normalization. The message's natural parameter vector update in \eqref{np_pen_sigsqpsi}
depends on $\E_q (\T{\upsi{l}} \upsi{l})$. Standard statistical results and sub-vector and sub-matrix definitions in
\eqref{exp_betau} and \eqref{cov_betau_psi} can be employed to show that

\[
	\E_q (\T{\upsi{l}} \upsi{l}) = \T{\E_q (\upsi{l})} \E_q (\upsi{l}) + \tr \left\{ \Cov_q (\upsi{l}) \right\}.
\]

The mean and FPC Gaussian penalization fragment, summarized in Algorithm \ref{alg:mean_fpc_gauss_pen_frag},
is a proceduralization of these results.

\end{document}